\documentclass[12pt]{article}

\usepackage{amsmath,amssymb}
\usepackage{bm,braket,ascmac}
\usepackage[dvipdfmx]{graphicx}
\usepackage{cite}
\usepackage{authblk}

\newcommand{\argmin}{\mathop{\rm min}\limits}

\begin{document}

\title{Transient Dynamics of Double Quantum Dots Coupled to Two Reservoirs}
\author[1]{Takahisa Fukadai}
\author[1]{Tomohiro Sasamoto}
\affil[1]{{\it Department of Physics, Tokyo Institute of Technology, 2-12-1 Ookayama, Meguro-ku, Tokyo 152-8550, Japan}}
\date{25 April 2018}
\maketitle

\begin{abstract}
We study the time-dependent properties of double quantum dots coupled to two
 reservoirs using the nonequilibrium Green function method. For an arbitrary time-dependent bias, we derive an expression for the
 time-dependent electron
 density of a dot and several currents, including the current between
 the dots in the wide-band-limit approximation. For the special case of a constant bias, we calculate
 the electron density and the currents numerically. As a result, we find
 that these quantities oscillate and that the number of
 crests in a single period of the current from a dot changes with the bias voltage. We also obtain an analytical
 expression for the relaxation time, which expresses how fast the system
 converges to its steady state. From the expression, we find that
 the relaxation time becomes constant when the coupling strength
 between the dots is sufficiently large in comparison with the difference of coupling strength between the dots and the reservoirs.
\end{abstract}


\section{Introduction}
Electron transport is a typical example of nonequilibrium phenomena
and has been useful for developing the understanding of nonequilibrium statistical
physics. Outstanding work on the electron transport was accomplished by
Landauer\cite{landauer} and B$\mathrm{\ddot{u}}$ttiker\cite{buttiker}. They derived a
simple formula in which the conductance between leads is expressed in terms
of simple physical quantities such as the transmission coefficient. Although the
formula was derived for noninteracting electrons with several
assumptions, it has been used widely
due to its applicability and simplicity. The Landauer-B$\mathrm{\ddot{u}}$ttiker (LB) formula was extended to interacting
case in Refs. \cite{jwm1992,jwm1993,iwm1994} with the nonequilibrium
Green function method\cite{schwinger1961,keldysh1964}. In these studies, the time-dependent current of a system coupled to two leads was
expressed in terms of interacting Green functions, even in the case where
arbitrary time-dependent biases are applied to leads. For the special case where
the system and reservoirs are in a nonequilibrium steady state, the
expression for the current reproduces the LB formula.

The approach used in these studies is based on the partitioned approach, which means that a system and
reservoirs are in independent equilibrium states characterized by a
chemical potential $\mu_{\alpha}$ and an inverse temperature
$\beta_{\alpha}$ with $\alpha$ representing the system and the reservoir at
an initial time $t_0$. After time $t_0$, a bias
voltage is applied to each reservoir and couplings between the system
and the reservoirs are added. When we only consider the steady state, this
approach is justified because all effects of the initial condition
disappear in the steady state\cite{stefalmb2004}. However, this approach
has a problem when we focus on the transient dynamics, where effects of the
initial condition cannot be ignored, and in a real
experiment we only switch on the bias, not the bias and the
coupling. Therefore, the approach is not
suitable for investigating the dynamics or time-dependent nonequilibrium
properties of nanosystems. The appropriate treatment of this problem may be to set the system and the reservoirs in the same equilibrium state
at the initial time, where the couplings between the system and the
reservoirs have already been added. This approach is
called the partition-free approach. 
This improvement was introduced in Refs. \cite{cini1980,stefalmb2004} and
applied in Refs. \cite{myohanen2009,tuovinen2013,tuovinen2014,Ridley2015}. In the study\cite{myohanen2009}, the authors calculated
the Green functions including the effects of the Coulomb interaction
with an approximation of the self-energy to conserve physical
quantities properly such as energy with the partition-free approach. Without the Coulomb interaction, transport
properties were calculated exactly\cite{tuovinen2013} in the wide-band-limit approximation (WBLA). The method of calculation in
\cite{tuovinen2013} has been employed to study the case where an arbitrary time-dependent bias is applied to
reservoirs\cite{Ridley2015} and to investigate graphene
nanoribbons\cite{tuovinen2014}.

Although the authors in the studies above assumed general scatterers and therefore their results are applicable to many systems, we need to consider concrete settings to answer specific questions, such as how the fact that a scatterer consists of several subsystems affects physical quantities. 
An application of the extended LB formula is the quantum dot. In a quantum dot, many interesting phenomena have been observed such as the Kondo effect\cite{sara1998} due to its low dimensionality.
For quantum transport through several quantum dots, one can expect a variety of phenomena to arise from the couplings between the system and the reservoirs or their geometrical configuration. From this viewpoint, many studies on quantum transport through double quantum dots (DQDs) have been conducted experimentally\cite{ono2002,hayashi2003,chang2004,petta2005} and theoretically\cite{colemon1984,hewson1993,aguado2000,greoges1999,lopez2002,tanaka2005,konik2007}.
Although there have been many analytical studies of DQDs, the analysis has been
mainly carried out in the stationary and special cases. For example, the slave boson
approach\cite{colemon1984,hewson1993}, which has been used widely in
the studies of DQDs, assumes that
the Coulomb interactions between electrons are very large and that only one
electron can exist in each dot. With this approach and the mean-field
approximation, many studies on the steady state of DQDs with a parallel, serial, or T-shaped
geometry have been made\cite{greoges1999,aguado2000,lopez2002,tanaka2005}. For the special case where the parameters satisfy the Yang-Baxter
relations, the Hamiltonian of DQDs including the Coulomb interaction is
exactly solvable with the Bethe ansatz\cite{konik2007}. 
Understanding the transient dynamics of DQDs themselves is also important for several
reasons. One is that DQDs
can be applied to quantum computation\cite{nielsen2000} and studying how the
decoherence occurs is important in this context. Moreover, the transient dynamics is
interesting as a nonequilibrium phenomenon because of the competition between the
initial correlation effects and the existence of coupling between the dots. Until now, theoretical
investigations of the transient dynamics of DQDs have mainly been conducted with the master
equation\cite{ziegler2000} or quantum master equation\cite{harbola2006}, which cannot treat quantum effects such as coherence properly. Some analysis including quantum effect has been conducted numerically\cite{marvlje2006,myohanen2008,matisse2008,tanaka2012,hartle2014}. 
 
In this paper, to understand the dynamics of DQDs analytically, we study
the transient dynamics of DQDs
in a serial geometry with the nonequilibrium Green function method and
the partition-free approach in the WBLA. We obtain
an exact expression for the electron density of a dot and the current
between DQDs with arbitrary parameters in the case where the Coulomb
interaction is irrelevant. Experimentally, the case in which high bias
voltages are applied to reservoirs or the total system is set at a high
temperature may be an example of our analysis. For the quenched case, where a constant bias
voltage is suddenly applied to a reservoir, we calculate these
quantities numerically and clarify the dependence of these quantities
on the parameters. From these expressions, we also obtain the relaxation
time from any initial condition to the steady state, which is useful for
understanding how the decoherence is affected by certain parameters. 

The paper is organized as follows. We first introduce the nonequilibrium
Green
function method in Sect. 2, then we briefly review the previous study
of the single dot case\cite{Ridley2015} in Sect. 3 to highlight the
difference between the analysis of the previous case and our case. In Sect. 4, we
show the analysis of DQDs, where the second-order partial
differential equations for Green functions are derived. We solve the
equations and obtain the expressions for the Green functions. We numerically
calculate the electron density and the currents in Sect. 5 for the
quenched case where a bias voltage is suddenly applied to a reservoir.


\section{Nonequilibrium Green Function}
The nonequilibrium Green function approach\cite{schwinger1961,keldysh1964} is a method to calculate physical
quantities in nonequilibrium many-body quantum systems. It is
especially useful for calculating nonequilibrium  physical quantities
perturbatively.  First, we consider a general case and explain the
formalism of the nonequilibrium Green function.

We consider the case where there is a central region surrounded by
reservoirs. In the central region, there are some subsystems. The total
Hamiltonian is represented as $H_0$ at initial time
$t_0$. For the initial state, we assume that the subsystems in the central
region and the reservoirs are
already coupled and that they
are in the same thermal-equilibrium state characterized by an inverse temperature $\beta$ and a chemical potential
$\mu$. After the initial time, the perturbative term $V(t)$ is added to the initial Hamiltonian $H(t) = H_0
+ V(t)$. A bias voltage applied to a reservoir is an example of the perturbation. We assume that the total Hamiltonian can be expressed in the
matrix form $H(t)=\sum_{ij} h_{ij}(t) d^{\dagger}_{i} d_{j}$. The operators
$d^{\dagger}_i, d_i$ express the creation and annihilation operators of the
system or the reservoirs. 
\begin{figure}
\begin{center}
 \includegraphics[width=0.5\linewidth]{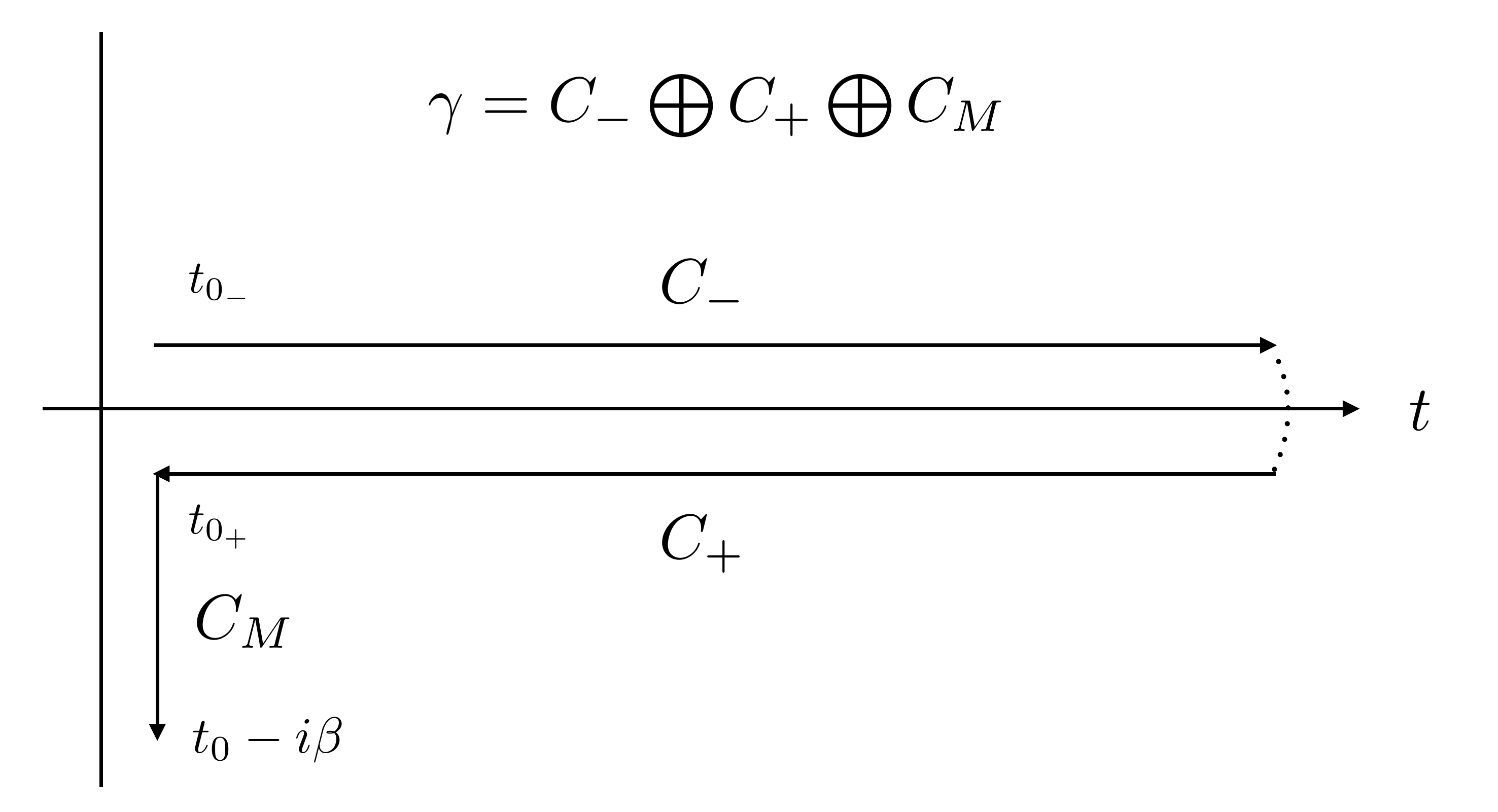}
\end{center}
\caption{(Color Online) Konstantinov-Perel' contour $\gamma$.}
\label{fig:1}
\end{figure}
Here, we define the Konstantinov-Perel' contour $\gamma$, which is an extension of the
Keldysh contour to include the effects of the initial state, as shown in
Fig. \ref{fig:1}. The contour consists of three paths: $C_{-}:\{t_0 \rightarrow \infty \}, C_{+}:\{ \infty \rightarrow
t_0 \}$ and, 
$C_{M}:\{t_0 \rightarrow t_0 - i \beta \}$. The nonequilibrium Green function on the
contour is defined as
\begin{equation}
 G_{ij}(z_1,z_2) := -i \frac{1}{Z} \mathrm{Tr}[e^{-\beta H_0}
  \mathcal{T}_c [d_{i}(z_1) d^{\dagger}_{j}(z_2) ]], \label{eq:1}
\end{equation}
where $Z=\mathrm{Tr}e^{-\beta H_0}$ is the partition function at the initial
time. Here $z=t_{-},t_{+},t_0-i\tau$ represents a position on the
contour and the time. $t_{-}$ and $t_{+}$ express time $t$ on the paths
$C_{-}$ and $C_{+}$ respectively. $\tau(0<\tau<\beta)$ is an imaginary time on path
$C_{M}$. The operators appearing in Eq. (\ref{eq:1}) are defined as
\begin{equation*}
 d_{i}(z) :=
\begin{cases}
 d_{H,i}(t), & z \in C_{-} \bigoplus C_{+}, \\
d^M_{i}(\tau), & z \in C_{M},
 \end{cases}
\end{equation*}
where $d_{H,i}(t):=e^{i\int^t_{t_0} ds H(s)} d e^{-i\int^t_{t_0} ds
 H(s)}$ is the operator in the Heisenberg picture and $d^M(\tau)$ is the
 Matsubara operator $d^M(\tau):= e^{ H_0
 \tau} d e^{-H_0 \tau}$.
$\mathcal{T}_c$ is the contour-ordering operator which depends on the
 direction of the arrow on the contour. The nonequilibrium Green
 function Eq. (\ref{eq:1}) takes several forms according to the positions of the two
 arguments $z_1$ and $z_2$. We represent these Green functions in Table I.
\begin{table}[htb]
  \begin{center}
    \begin{tabular}{|c|c|c|} \hline
      Symbol & Position on the contour $\gamma$ & Name \\ \hline \hline
 $G^{--}_{ij}(t_1,t_2)$ & $ z_1,z_2 \in C_{-}$ &   \\ \hline 
 $G^{++}_{ij}(t_1,t_2)$ & $ z_1,z_2 \in C_{+}$ &   \\ \hline 
$G^{>}_{ij}(t_1,t_2)$   & $(z_1 > z_2) \land (z_1,z_2 \in C_{-}
  \bigoplus C_{+})$ & Greater  \\ \hline
    $G^{<}_{ij}(t_1,t_2)$   & $(z_1 < z_2) \land (z_1,z_2 \in
  C_{-} \bigoplus C_{+})$  & Lesser \\ \hline
      $G^{\rceil}_{ij}(t_1,\tau_2)$ & $(z_1
 \in C_{-} \bigoplus C_{+}) \land (z_2 \in C_M)$ & Right　\\ \hline
 $G^{\lceil}_{ij}(\tau_1,t_2)$ &  $(z_1
 \in C_M) \land (z_2 \in C_{-} \bigoplus C_{+})$ & Left \\ \hline
 $G^M_{ij}(\tau_1,\tau_2)$ & $z_1,z_2 \in C_M$  & Matsubara \\ \hline
    \end{tabular}
  \end{center}
\caption{Definitions of the Green functions.}
\end{table}
As an example, let us consider the case where these two arguments appearing in the Eq. (1)
are in part of the contour  $z_1,z_2 \in C_{-} \bigoplus
C_{+}$ and $z_1$ is in front of $z_2$ on the contour. In this case, since the argument
$z_1$ is always in front of $z_2$, $\mathcal{T}_c$
exchanges these two operators and then $(-1)$ is multiplied. Therefore,
the nonequilibrium Green function takes the form $G^{<}_{ij}(t_1,t_2) = i/Z \mathrm{Tr}[e^{-\beta H_0}
  d_{H,i}^{\dagger}(t_2) d_{H,j}(t_1) ]$ for $(z_1 < z_2) \land (z_1,z_2 \in
  C_{-} \bigoplus C_{+})$. This function is called the lesser Green function. We also define retarded/advanced Green functions as 
\begin{eqnarray}
G_{ij}^r(t_1,t_2) &:=& \theta (t_1-t_2) [G^{>}_{ij}(t_1,t_2) - G^{<}_{ij}(t_1,t_2)],
  \label{eq:2} \\ 
G_{ij}^a(t_1,t_2) &:=& - \theta (t_2-t_1) [G^{>}_{ij}(t_1,t_2) -
 G^{<}_{ij}(t_1,t_2)], \label{eq:3}
\end{eqnarray}
where $\theta(t)$ is the Heaviside function $\theta(t)=1(t>0),0(t<0)$.

We define the Hamiltonian on the contour $h_{ij}(z)$ as
\begin{equation*}
h_{ij}(z) :=
 \begin{cases}
  h_{ij}(t), & z \in C_{-} \bigoplus C_{+}, \\
  h_{ij}(t_0), & z \in C_{M}. 
 \end{cases}
\end{equation*}
Using the contour $\gamma$ and the nonequilibrium Green function on the
contour, we can use the diagram technique even in the nonequilibrium
case. For details, see \cite{zagoskin}.

\section{Single-Dot Case}
Before we discuss the DQD case, which is our main interest, we briefly review the case of a single dot that has only one energy
level and is coupled to two reservoirs. This is a special case of
\cite{Ridley2015}. In this case, we analyze the equation of motion for
the nonequilibrium Green function. The equation is self-consistent but
it becomes solvable after employing the WBLA. By solving the equation of motion, we can obtain a
closed-form expression for the nonequilibrium Green function. This section will be useful for understanding the differences
between the analysis of the single-dot case and that of the double-dots
case, where we cannot obtain a closed-form expression for the
nonequilibrium Green function by solving the equation of motion with the WBLA.

\subsection{Definition of system}
We consider the case where the total system consists of a quantum dot in
the central region and two reservoirs. At an initial time
$t_0$, the dot and the reservoirs are in the
thermal-equilibrium state characterized by an inverse temperature $\beta$ and
a chemical potential $\mu$. After the time $t \geq t_0$, a bias voltage $V_{\alpha}(t)$ is applied to reservoir $\alpha$ and then the total system tends to a nonequilibrium state. 
The Hamiltonian of the total system is represented as 
\begin{equation}
 H(t) =
  \sum_{k \alpha} \epsilon_{k\alpha}(t) d^{\dagger}_{k \alpha}
  d_{k \alpha} + \epsilon_1(t) d^{\dagger}_1 d_1 +
  V,   \label{eq:4}
\end{equation}
where $\epsilon_{k \alpha}(t)$ and $\epsilon_1(t)$ are defined as $\epsilon_{k\alpha}(t)=\epsilon_{k \alpha}-\mu(t <
t_0),\epsilon_{k\alpha} +V_{\alpha}(t) (t \ge t_0)$ and $\epsilon_1(t)=\epsilon_1-\mu(t <
t_0),\epsilon_1 (t \ge t_0)$. $\epsilon_{k \alpha}$ is the $k$th
eigenvalue of the Hamiltonian of reservoir $\alpha$ and $\epsilon_1$ is the eigenvalue of the dot. $d_{k\alpha},d^{\dagger}_{k\alpha}$
are the creation and the annihilation operators of reservoir $\alpha$ and
$d_1,d^{\dagger}_1$ are those of the dot, respectively.
In this case, the particles are fermions and thus these operators satisfy
the anticommutation relations $\{d_i,d^{\dagger}_j\}=\delta_{ij},$ $\{d_i,d_j\} =
\{d^{\dagger}_i,d^{\dagger}_j\}=0$, where the index $i$ denotes the
index of a reservoir $\alpha$ or the dot $i=k\alpha,1$.
Each term of the total Hamiltonian has the following meaning: the first term
is the (diagonalized) Hamiltonian of the reservoirs, the second is the
Hamiltonian of the dot, and the third is the coupling between the
dot and the reservoirs. We assume that the coupling term takes the form
\begin{equation*}
 V = \sum_{k\alpha} T_{k\alpha,1} d^{\dagger}_{k \alpha}
  d_1 + T_{1, k\alpha} d^{\dagger}_1 d_{k \alpha}.
\end{equation*}
 We use a matrix
 representation of the Hamiltonian $h_{ij}(t)$ of the total system to match Ex. (\ref{eq:4}). For example, when the indices are $i=k \alpha$ and $j=k' \alpha$, the value of $h_{ij}(t)$ is 
\begin{equation}
 h_{k \alpha,k' \alpha}(t) =
\begin{cases}
 (\epsilon_{k \alpha} - \mu) \delta_{k,k'}, & t < t_0, \\
 (\epsilon_{k \alpha} + V_{\alpha}(t) ) \delta_{k,k'}, & t \geq t_0.
\end{cases} \notag 
\end{equation}
We define the matrix $\mathbf{h}_{\alpha \alpha}(t)$ by 
$[\mathbf{h}_{\alpha \alpha}(t)]_{k,k'} = h_{k\alpha,k'\alpha}(t)$. Similarly, we define other
matrices $\mathbf{h}_{\alpha \alpha}(t), \mathbf{h}_{\alpha 1},
\mathbf{h}_{1 \alpha}$, and $h_{11}(t)$.

\subsection{Equations of motion and previous result}

By differentiating the definition of each Green function and using
the commutation relations, we derive the equations of motion for the
nonequilibrium Green functions as
\begin{gather}
\left[ i \frac{d}{dz_1} -  h_{11}(z_1) \right]  G_{11}(z_1,z_2) = 
  \delta(z_1,z_2) + \sum_{\alpha} \mathbf{h}_{1\alpha}
 \mathbf{G}_{\alpha 1}(z_1,z_2),  \label{eq:5} \\ 
 G_{11}(z_1,z_2) \left[ -i \frac{\overleftarrow{d}}{dz_2} -h_{11}(z_2) \right] = 
  \delta(z_1,z_2) + \sum_{\alpha} \mathbf{G}_{1
 \alpha}(z_1,z_2) \mathbf{h}_{\alpha 1 }, \label{eq:6}
\end{gather}
\begin{gather}
\left[  i \frac{d}{dz_{1}} - \mathbf{h}_{\alpha \alpha}(z_1) \right]
 \mathbf{G}_{\alpha 1}(z_1,z_2) = \mathbf{h}_{\alpha 1}
 G_{11}(z_1,z_2), \label{eq:7} \\
 \mathbf{G}_{1 \alpha }(z_1,z_2)  \left[  -i
 \frac{\overleftarrow{d}}{dz_{2}} -  \mathbf{h}_{\alpha \alpha}(z_2) \right]
  = G_{11}(z_1,z_2) \mathbf{h}_{1 \alpha},  \label{eq:8}
\end{gather}
where we use the matrix representation of the nonequilibrium
Green function $[\mathbf{G}(z_1,z_2)]_{ij} = G_{ij}(z_1,z_2)$. The nonequilibirum Green function is defined  in the same way as Eq. (\ref{eq:1}). The
operator $\frac{\overleftarrow{d}}{dz}$ is the differential operator acting
from the right side.

We need an expression for the lesser
Green function to obtain an expression for one-particle quantities such as the electron density of the dot. To obtain an
expression for the
lesser Green function, we need to solve the equations of motion with the Kubo-Martin-Schwinger(KMS) boundary conditions\cite{kubo1957,matrinschwinger1959}. All Green functions must
satisfy the KMS conditions, which arise from the
condition that the initial state is in thermal equilibrium. The KMS
conditions are expressed as
\begin{equation}
\begin{cases}
\mathbf{G}(z_1,t_0-i \beta) = - \mathbf{G}(z_1,t_0), \\ 
\mathbf{G}(t_0-i \beta,z_2) = - \mathbf{G}(t_0,z_2),
\end{cases} \label{eq:9}
\end{equation} 
which are directly derived from definition (\ref{eq:1}). Note that we can choose another initial condition, as in the partitioned
approach, where the system and the reservoirs are in different
equilibrium states at the initial time. However this sometime
leads to unphysical behavior of physical quantities\cite{stefalmb2004}. Since
there are unknown functions in the 
equations of motion for $G_{11}(z_1,z_2)$, (\ref{eq:5}) and (\ref{eq:6}), we cannot obtain a closed-form expression for
$G_{11}(z_1,z_2)$. Now we express the solutions of Eqs. (\ref{eq:7}) and (\ref{eq:8}) as 
\begin{eqnarray}
 \mathbf{G}_{\alpha 1}(z_1,z_2) &=& \int_{\gamma} d \bar{z}
  \mathbf{g}_{\alpha \alpha}(z_1,\bar{z}) \mathbf{h}_{\alpha 1}(\bar{z})
  G_{11}(\bar{z},z_2), \label{eq:10} \\
  \mathbf{G}_{1 \alpha }(z_1,z_2) &=& \int_{\gamma} d \bar{z}
  G_{11}(z_1,\bar{z}) \mathbf{h}_{1 \alpha}(\bar{z})
  \mathbf{g}_{\alpha \alpha}(\bar{z},z_2), \label{eq:11}
\end{eqnarray}
where $\mathbf{g}_{\alpha\alpha}$ is the non-perturbative Green function
of reservoir $\alpha$, which obeys the equation of motion
\begin{equation*}
 \left[  i \frac{d}{dz_{1}} - \mathbf{h}_{\alpha \alpha}(z_1) \right]
  \mathbf{g}_{\alpha \alpha}(z_1,z_2) = \mathbf{1}\delta(z_1,z_2),
\end{equation*}
and $\int_{\gamma} d \bar{z}$ is the integration on contour
$\gamma$. This equation of motion is equal to that for the isolated
reservoir $\alpha$ since there is only the delta
function in the right-hand-side(RHS) of the equation. Actually, we can check that expressions (\ref{eq:10}) and (\ref{eq:11})
are solutions of the equations of motion (\ref{eq:7}) and (\ref{eq:8})
satisfying the KMS conditions. Then, by substituting Eqs. (\ref{eq:10}) and (\ref{eq:11}) into Eqs. (\ref{eq:5}) and
(\ref{eq:6}), finally we obtain the equation of motion for $G_{11}(z_1,z_2)$,
\begin{equation}
\begin{split}
 \left[ i \frac{d}{d z_1} - h_{11}(z_1) \right]
 G_{11}(z_1,z_2) &=  \delta(z_1,z_2) +
  \int_{\gamma} d \bar{z} \Sigma_{11}(z_1,\bar{z})
  G_{11}(\bar{z},z_2), \\
 G_{11}(z_1,z_2) \left[ -i \frac{\overleftarrow{d}}{dz_2} -h_{11}(z_2)
 \right] &=  \delta(z_1,z_2) +  \int_{\gamma} d \bar{z}
 G_{11}(z_1,\bar{z}) \Sigma_{11}(\bar{z},z_2)
\end{split}
  \label{eq:12}
\end{equation}
where $\Sigma_{11}$ is the embedded self-energy defined as
\begin{equation}
 \Sigma_{11}(z_1,z_2) = \mathbf{h}_{1 \alpha }(z_1)
 \mathbf{g}_{\alpha \alpha}(z_1,z_2) \mathbf{h}_{\alpha 1}(z_2).  \label{eq:13}
\end{equation}
At this point, we obtain a closed-form expression for the embedded self-energy because it is expressed in
terms of the known parameters and the non-perturbative Green function, which
is calculated easily. Now we use the WBLA. This approximation means that the energy bands of reservoirs are
very wide and that only the electrons on the Fermi
energy are transported. Mathematically, the WBLA is equivalent to the
substitution $\sum_{k}
\rightarrow \int \frac{dk}{2\pi} \rightarrow \rho(\epsilon_F) \int
d\epsilon_k $, where $\rho_{\epsilon_F}$ is the density of states in the
vicinity of the Fermi energy. Therefore, the level width
$\Gamma_{\alpha}(\omega)=2\pi \sum_{k} T_{1,k\alpha} T_{k\alpha,1}
\delta(\omega-\epsilon_{k\alpha})$ becomes a constant $\Gamma_{\alpha}$
and the embedded self-energy takes a simple form in this approximation. For example, the
retarded self-energy $\Sigma^r_{11}(t_1,t_2):=\theta(t_1-t_2)
(\Sigma^{>}_{11}(t_1,t_2)-\Sigma^<_{11}(t_1,t_2))$ in the WBLA is written as 
\begin{equation*}
 \Sigma^r_{11}(t_1,t_2) = -\frac{i}{2} \Gamma \delta(t_1-t_2),
\end{equation*}
where  $\Gamma=\sum_{\alpha} \Gamma_{\alpha}$ is the total level width. The lesser $\Sigma^<_{11}(t_1,t_2)$ and greater
$\Sigma^>_{11}(t_1,t_2)$ self-energies are defined similarly to the lesser and
greater Green functions in Sect. 2, respectively. For the derivation, see
Appendix A. The calculation of the self-energy in Appendix A is for the
case of the DQDs, but the self-energy for the single-dot
case is obtained in the same way.

From the equation of motion (\ref{eq:12}) and the embedded self-energy
in the WBLA, we can obtain the following equations of motion for the Matsubara,
right/left, and lesser Green functions in the WBLA:
\begin{equation}
 \left[-\frac{d}{d\tau_1} - h^M_{11} \right]
 G^{M}_{11}(\tau_1,\tau_2) = i 
  \delta(\tau_1-\tau_2) + \left( \Sigma^M_{11} *
  G^M_{11} \right)_{(\tau_1,\tau_2)}, \label{eq:14} 
\end{equation}
\begin{equation}
\begin{split}
  \left( i \frac{d}{dt_1} - h^{eff}_{11} \right)
  G^{\rceil}_{11}(t_1,\tau_2) &=
  \left( \Sigma^{\rceil}_{11}* G^{M}_{11}
  \right)_{(t_1,\tau_2)}, \\
G^{\lceil}_{11}(\tau_1,t_2) \left( -i \frac{\overleftarrow{d}}{dt_2} -
 h^{eff}_{11} \right) &= \left( G^{M}_{11}*\Sigma^{\lceil}_{11}
  \right)_{(\tau_1,t_2)},
\end{split}
 \label{eq:15} 
\end{equation}
\begin{equation}
\left( i \frac{d}{dt_1} - h^{eff}_{11} \right)
  G^{<}_{11}(t_1,t_2) = \left( \Sigma^{<}_{11} \cdot
 G^{a}_{11} + \Sigma^{\rceil}_{11}*
 G^{\lceil}_{11} \right)_{(t_1,t_2)}, \label{eq:16}
\end{equation}
where $h^M_{11}$ is the energy on the vertical part of the contour $\gamma$ and $h^{eff}_{11}$ is the effective energy of the system: $h^M_{11}=\epsilon_1 - \mu$, \ $h^{eff}_{11}:=h_{11}
- i \Gamma/2 $.  $(f \cdot g)_{(t,t')}$ is the convolution of $f$
and $g$ and $(f*g)_{(\tau,\tau')}$ is the imaginary time convolution
of $f$ and $g$: $(f \cdot g)_{(t,t')}=-i\int^{\infty}_{t_0} ds f(t,s)
g(s,t')$,  $(f * g)_{(\tau,\tau')}=\int^{\beta}_{0} ds f(\tau,s)
g(s,\tau')$. In this derivation, we use the Langreth
rule\cite{langreth1976} to convert the integration in the most RHS of
(\ref{eq:12}) to an integral along the real axis. Since we have an
expression of the Matsubara self-energy, we can obtain an expression of the Matsubara
Green function from Eq. (\ref{eq:14}) by applying the Fourier
transformation. Then, the resulting equations are inhomogeneous
first-order differential equations and thus we can solve these
equations. Details of these calculations are given in
\cite{Ridley2015}. Finally, we obtain the following expression of the lesser Green function:
\begin{multline}
 G^<_{11}(t_1,t_2) = e^{-ih^{eff}_{11}(t_1-t_0)} \int^{\infty}_{-\infty}
  d \omega f(\omega-\mu) \sum_{\alpha} [K_{\alpha}(t_1,t_0;\omega)
  \Gamma_{\alpha} G^a(\omega) -  G^r(\omega) \Gamma_{\alpha}
  K^{*}_{\alpha}(t_2,t_0;\omega) \\
 +iK_{\alpha}(t_1,t_0;\omega)
  \Gamma_{\alpha} K^{*}_{\alpha}(t_2,t_0;\omega) +
  iA_{\alpha}(\omega)] e^{i (h^{eff}_{11})^{*}(t_2-t_0)}, \label{eq:17}
\end{multline}
where $G^r(\omega) = \{\omega - h^{eff}_{11} \}^{-1}$ and $G^a(\omega) =
\{\omega - (h^{eff}_{11})^{*} \}^{-1}$ are Fourier transforms of the retarded/advanced Green functions, respectively. $f(\omega) = \left\{
e^{\beta \omega} +1 \right\}^{-1}$ is the Fermi
distribution function. Here, we introduce
the spectral function $A_{\alpha}(\omega) = G^r(\omega) \Gamma_{\alpha}
G^a(\omega) $ and the function 
\begin{equation*}
 K_{\alpha}(t,t_0;\omega) = \int^{t}_{t_0} ds
  e^{-i(\omega-h^{eff}_{11})(s- t_0)} e^{-i\psi_{\alpha}(s,t_0)},
\end{equation*}
which includes all the effects from the biased voltage $\psi_{\alpha}(t,t_0):=
\int^t_{t_0} ds V_{\alpha}(s)$. Since the electron density is expressed as
$\rho(t) = -i G^<_{11}(t,t)$, we calculate the electron
density almost directly. Using the concrete expression (\ref{eq:17}), we calculate the electron density
numerically for the quenched case where a constant bias voltage is
applied suddenly: $V_L=6,V_R=0$. We take other
parameters as $\epsilon_1=1, \Gamma_L=\Gamma_R=1/2$, and $\beta=100$.
\begin{figure}[ht]
 \begin{center}
  \includegraphics[width=0.5\hsize]{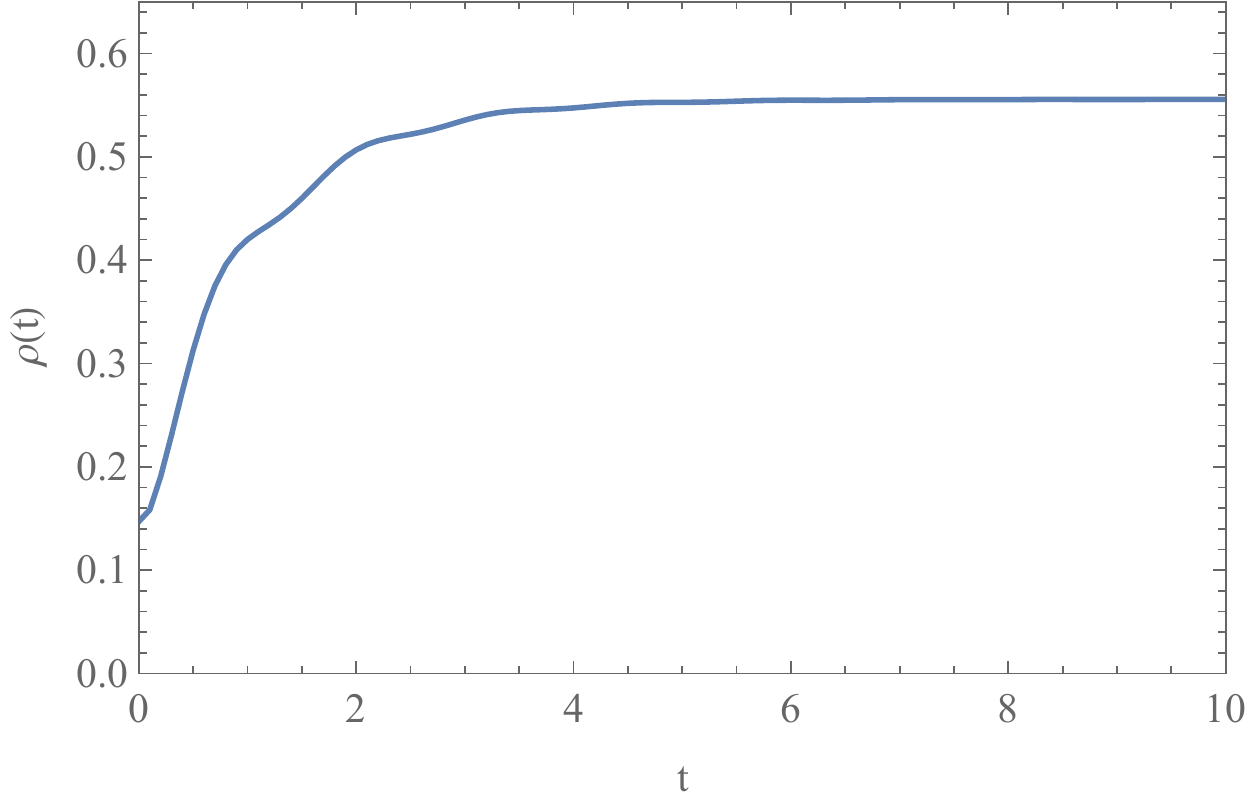}
\end{center}
\caption{(Color Online) Time dependence of electron density $\rho$ for $\beta=100$.}
\label{fig:2}
\end{figure}
We can see that the electron density approaches
the limiting value exponentially without oscillation (Fig. \ref{fig:2}). 
This is because we consider a dot having only one energy level. When we consider the case of double dots, however, an oscillation appears even though each dot has only one energy level. See Fig. \ref{fig:4} in Sect. 5 for this argument.

\section{Double-Dot Case}
\subsection{Definition of system}

\begin{figure}[tbp]
\begin{center}
 \includegraphics[width=0.5\linewidth]{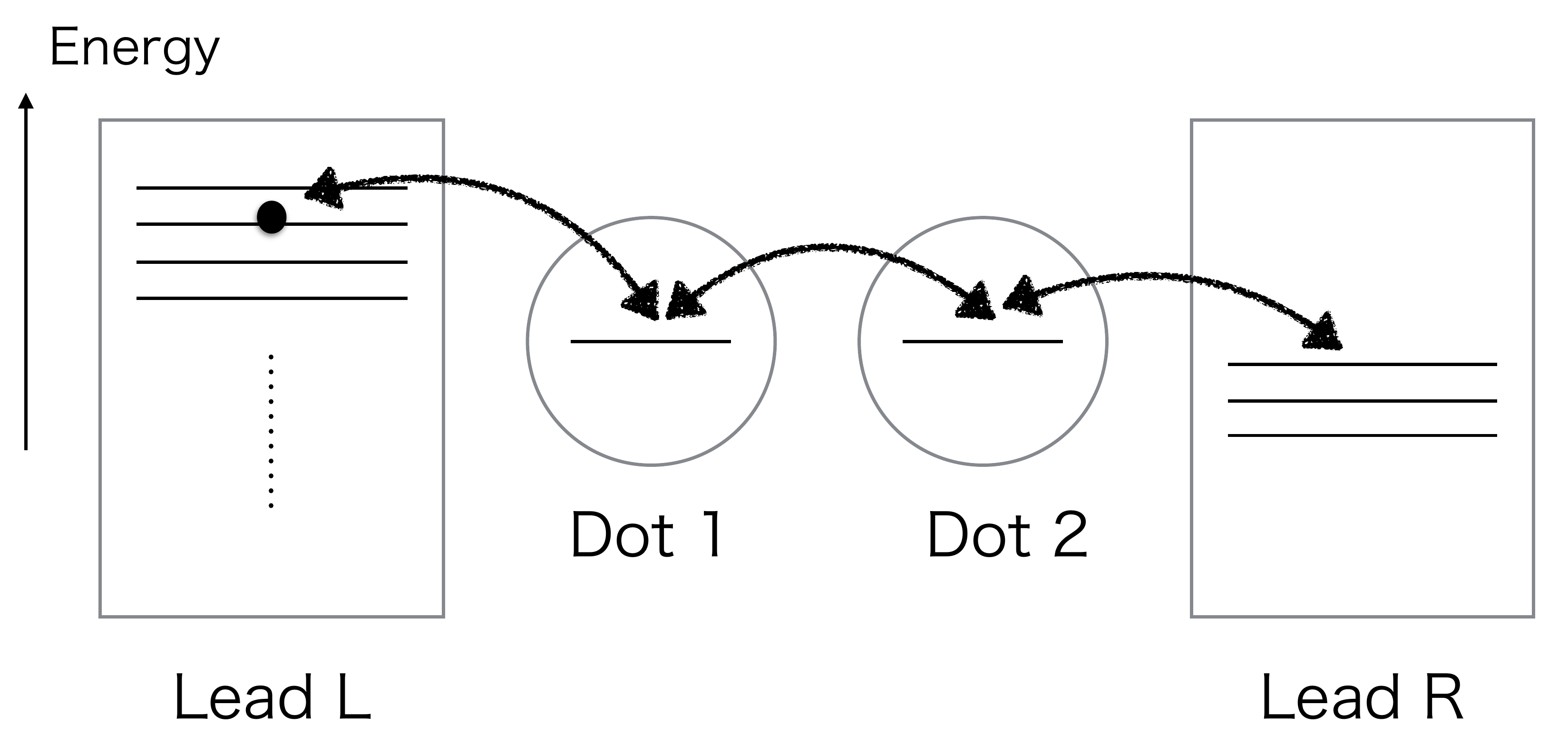}
\end{center}
\caption{(Color Online) Schematic diagram of system and reservoirs.}
\label{fig:3}
\end{figure}
Next, we consider the case of the DQDs, which is the main
target of our study. Figure \ref{fig:3} is a
schematic diagram of this case. In the case, there are two quantum
dots, not one, in the central region. Since there is coupling
between the dots, we can expect that richer phenomena will arise than those of the single dot. We assume that the initial
condition is the same as in the case of the single dot in Sect. 4. At
an initial time $t_0$, the dots and the reservoirs are in the thermal-equilibrium state characterized by an inverse temperature $\beta$ and
a chemical potential $\mu$. After the initial time $t \geq t_0$, a bias voltage $V_{\alpha}(t)$ is applied to
each reservoir $\alpha$ and the system enters a nonequilibrium state. 
The Hamiltonian of the total system is represented as 
\begin{multline}
 H(t) = \\
\begin{cases}
  \sum_{\alpha=\{L,R\},k\alpha} (\epsilon_{k\alpha} - \mu ) d^{\dagger}_{k\alpha}
  d_{k\alpha} + \sum_{n=\{1,2\}} (\epsilon_n - \mu) d^{\dagger}_n d_n +
  V_{1L} + V_{2R} + T_d,  & t < t_0 \\
  \sum_{\alpha=\{L,R\},k\alpha} (\epsilon_{k\alpha} + V_{\alpha}(t)) d^{\dagger}_{k\alpha}
  d_{k\alpha} + \sum_{n=\{1,2\}} \epsilon_n d^{\dagger}_n d_n +
  V_{1L} + V_{2R} + T_d, & t \geq t_0
\end{cases} 
\label{eq:18} 
\end{multline}
where $\epsilon_{k \alpha}$ is the $k$th eigenvalue of the Hamiltonian
of reservoir $\alpha$ and $\epsilon_{n}$ is the eigenvalue of dot $n$ $(n=1,2)$. We assume that the Hamiltonian of the reservoirs can be diagonalized
and that each dot has just one energy level. $d_{k\alpha},d^{\dagger}_{k\alpha}$
are the creation and  annihilation operators of reservoir $\alpha$ and
$d_n,d^{\dagger}_n$ are those of dot $n$, respectively.
The transported particles are fermions and thus the operators satisfy
the anticommutation relations $\{d_i,d^{\dagger}_j\}=\delta_{ij}$,
$\{d_i,d_j\} = \{d^{\dagger}_i,d^{\dagger}_j\}=0$.
Each term of the Hamiltonian has the following meaning: the first term
is the (diagonalized) Hamiltonian of the reservoirs, the second is the
Hamiltonian of the dots, the third is the coupling between dot 1 and the left reservoir,
the fourth is the coupling between dot 2 and the right reservoir and, the fifth, which does not appear in the case of a single dot, is the
coupling between the two dots. Here, we assume that the interactions are written
in the forms
\begin{eqnarray*}
 V_{i \alpha} &=& \sum_{k\alpha,n} T_{k\alpha,n} d^{\dagger}_{k \alpha}
  d_n + T_{n, k\alpha} d^{\dagger}_n d_{k \alpha}, \\
  T_d &=& t_{12}  d^{\dagger}_1 d_2 +  t_{21} d^{\dagger}_2 d_1.
\end{eqnarray*}
Note that we are interested in how the fact that
the central system consists of two subsystems affects physical
quantities. For this reason, we consider the region where the Coulomb
interaction can be ignored. If there was an effect of the Coulomb
interaction, there would be more interesting phenomena, but it would be
difficult to tell whether the cause of a phenomenon is from the fact
that the system consists of some subsystems or from the Coulomb
interaction. Experimentally, this approximation may be justified when the energy scale of electrons is larger than that of the Coulomb
interaction, i.e., when a high bias voltage is applied or when a system
is at a high temperature. We
represent the Hamiltonian of the total system (\ref{eq:18}) in the same
way as in the single-dot case,
\begin{equation}
  H(t)=\sum_{ij} h_{ij}(t) d^{\dagger}_i d_j, \notag
\end{equation}
where index $i$ can be $k \alpha$ or $n$ and the elements of $h_{ij}(t)$ are defined to match Eq. (\ref{eq:18}). We also define the
contour $\gamma$ and the nonequilibrium Green function in the same way
as in Sect. 3.

 \subsection{Equation of motion for nonequilibrium Green function}

By differentiating the definition of each Green function in Table I and using the anticommutation
relations for fermions, we can derive following equations of motion for the nonequilibrium Green functions:
\begin{gather}
\left[ i \frac{d}{dz_1} - h_{11}(z_1) \right] G_{11}(z_1,z_2) = 
 \delta(z_1,z_2) + \mathbf{h}_{1L} \cdot \mathbf{G}_{L1}(z_1,z_2) +
 t_{12} G_{21}(z_1,z_2),  \label{eq:19} \\ 
 G_{11}(z_1,z_2) \left[ -i \frac{\overleftarrow{d}}{dz_2} - h_{11}(z_2) \right] =  \delta(z_1,z_2) + \mathbf{G}_{1L}(z_1,z_2) \cdot \mathbf{h}_{L1} +
  G_{12}(z_1,z_2) t_{21},  \label{eq:20} \\ 
\left[ i \frac{d}{dz_2} - h_{22}(z_1) \right] G_{22}(z_1,z_2) = 
 \delta(z_1,z_2) + \mathbf{h}_{2R} \cdot \mathbf{G}_{R2}(z_1,z_2) +
 t_{21} G_{12}(z_1,z_2),  \label{eq:21} \\ 
 G_{22}(z_1,z_2) \left[ -i \frac{\overleftarrow{d}}{dz_2} - h_{22}(z_2) \right] =  \delta(z_1,z_2) + \mathbf{G}_{2R}(z_1,z_2) \cdot \mathbf{h}_{R2} +
  G_{21}(z_1,z_2) t_{12}.  \label{eq:22}
\end{gather}
In the case of double dots,
we must consider the equations of motion for not only $G_{11}$ and
$G_{22}$ but also $G_{12}$ and $G_{21}$ for analysis:
\begin{gather}
\left[ i \frac{d}{dz_1} - h_{11}(z_1) \right] G_{12}(z_1,z_2) = 
  \mathbf{h}_{1L} \cdot \mathbf{G}_{L2}(z_1,z_2) +
 t_{12} G_{22}(z_1,z_2),  \label{eq:23} \\
 G_{12}(z_1,z_2) \left[ -i \frac{\overleftarrow{d}}{dz_2} - h_{22}(z_2) \right] =  \mathbf{G}_{1R}(z_1,z_2) \cdot \mathbf{h}_{R2} +
  G_{11}(z_1,z_2) t_{12}, \label{eq:24}  \\
\left[ i \frac{d}{dz_1} - h_{22}(z_1) \right] G_{21}(z_1,z_2) = 
  \mathbf{h}_{2R} \cdot \mathbf{G}_{R1}(z_1,z_2) +
 t_{21} G_{11}(z_1,z_2), \label{eq:25} \\
 G_{21}(z_1,z_2) \left[ -i \frac{\overleftarrow{d}}{dz_2} - h_{11}(z_2) \right] =  \mathbf{G}_{2L}(z_1,z_2) \cdot \mathbf{h}_{L1} +
  G_{22}(z_1,z_2) t_{21},  \label{eq:26}
\end{gather}
\begin{gather}
\left[  i \frac{d}{dz_{1}} - \mathbf{h}_{LL}(z_1) \right]
 \mathbf{G}_{L1}(z_1,z_2) = \mathbf{h}_{L 1} G_{11}(z_1,z_2),
 \label{eq:27} \\
\left[ i \frac{d}{dz_{1}} - \mathbf{h}_{RR}(z_1) \right]
 \mathbf{G}_{R1}(z_1,z_2) = \mathbf{h}_{R 2} G_{21}(z_1,z_2), \label{eq:28}
\end{gather}
where we use the matrix representation of the nonequilibrium
Green function $ \{ \mathbf{G}(z_1,z_2) \}_{ij} = G_{ij}(z_1,z_2)$.
As in the case of the single dot in Sect. 3, all Green functions must satisfy the KMS conditions
\begin{equation}
\begin{cases}
\mathbf{G}(z_1,t_0-i \beta) = - \mathbf{G}(z_1,t_0), \\ 
\mathbf{G}(t_0-i \beta,z_2) = - \mathbf{G}(t_0,z_2). \label{eq:29}
\end{cases}
\end{equation}
We can construct solutions of equations (\ref{eq:27}) and
(\ref{eq:28}) satisfying the KMS conditions in the form
\begin{eqnarray}
 \mathbf{G}_{L1} &=& \int_{\gamma} d \bar{z} \mathbf{g}_{LL}(z_1,\bar{z}) \mathbf{h}_{L1}
  G_{11}(\bar{z},z_2), \label{eq:30} \\
  \mathbf{G}_{R1} &=& \int_{\gamma} d \bar{z}\mathbf{g}_{RR}(z_1,\bar{z}) \mathbf{h}_{R2}
  G_{21}(\bar{z},z_2), \label{eq:31}
\end{eqnarray}
where $\mathbf{g}_{\alpha\alpha}$ is the non-perturbative Green function
of reservoir $\alpha$, which obeys the equation of motion,
\begin{equation*}
 \left[  i \frac{d}{dz_{1}} - \mathbf{h}_{\alpha \alpha}(z_1) \right]
  \mathbf{g}_{\alpha \alpha}(z_1,z_2) = \mathbf{1}\delta(z_1,z_2)
\end{equation*}
and $\int_{\gamma} d \bar{z}$ is the integration on contour $\gamma$.
Finally, by substituting Eqs. (\ref{eq:30}) and (\ref{eq:31}) into Eqs. (\ref{eq:19}) and (\ref{eq:25}), we obtain
\begin{multline}
 \left[ i \frac{d}{d z_1} - h_{11}(z_1) \right]
 G_{11}(z_1,z_2) =   1 \delta(z_1,z_2) \\ +
  \int_{\gamma} d \bar{z} \Sigma_{11}(z_1,\bar{z})
  G_{11}(\bar{z},z_2) + t_{12} G_{21}(z_1,z_2),  \label{eq:32} 
\end{multline}
\begin{eqnarray}
  & \left[ i \frac{d}{d z_1} - h_{22}(z_1) \right]
 G_{21}(z_1,z_2) = \int_{\gamma} d \bar{z} \Sigma_{22}(z_1,\bar{z})
  G_{21}(\bar{z},z_2) + t_{21} G_{11}(z_1,z_2), & \label{eq:33} \\
 &\Sigma_{11}(z_1,\bar{z}) = \mathbf{h}_{1L} \cdot
 \mathbf{g}_{LL}(z_1,\bar{z}) \mathbf{h}_{L1}, \  \Sigma_{22}(z_1,\bar{z})
 = \mathbf{h}_{2R} \cdot \mathbf{g}_{RR}(z_1,\bar{z}) \mathbf{h}_{R2}. \label{eq:34} & 
\end{eqnarray}
In comparison to the single-dot case, we find that new terms appear in
the RHSs of the equations of motion for the system (\ref{eq:32}) and (\ref{eq:33}). Hence, we cannot solve the equations in the same way 
as for the single-dot case, where we obtained the solution by
integrating the equation of motion. However, there are only two unknown
functions, $G_{11}$ and $G_{21}$, in the two differential equations (\ref{eq:32}) and (\ref{eq:33}). Therefore, in
principle, by substituting one equation into the other, we can
derive a second-order differential equation for the
nonequilibrium Green function in a closed form. We can expect to obtain a concrete expression of the nonequilibrium Green function by
integrating the second-order differential equation. This is the basic
idea of our analysis. Although this idea is very simple, a phenomenon that cannot be observed in
the analysis of the single dot appears. In the second-order partial
differential equation, a new quantity called the pseudo
self-energy appears. The problem is that divergence appears in a
second-order partial differential equation when one
computes the new quantity straightforwardly. We solve this
problem by changing the method of calculating the quantity. We explain
this problem and how we solve it in Appendix E.

 \subsection{Second-order differential equations of the nonequilibrium
  Green functions and the pseudo self-energy}

We can obtain the following second-order differential
equations for the lesser Green function of the system from Eqs. (\ref{eq:32}) and
(\ref{eq:33}). See Appendix D for details of the derivation
\begin{multline}
\bigg[-\frac{d^2}{dt^2_1} - i (\epsilon_1 + \epsilon_2) \frac{d}{dt_1} +
(\epsilon_1 \epsilon_2 -t_{12} t_{21} )\bigg] G^{<}_{11} (t_1,t_2) =
\int_{\gamma} d\bar{z} \tilde{\Sigma}^{-\ }_{11}(t_1, \bar{z})
G^{\ +}_{11}(\bar{z},t_2)  \\
+ i\left(\Sigma^r_{22}(t_1,t_0) G^{<}_{11}(t_0,t_2) +
      \Sigma^{<}_{22}(t_1,t_0) G^{a}_{11}(t_0,t_2) \right) +
\Sigma^{<}_{22}(t_1,t_2), \label{eq:35}
\end{multline}
\begin{multline}
G^{<}_{11} (t_1,t_2) \bigg[-\frac{\overleftarrow{d}^2}{dt^2_2} + i (\epsilon_1 + \epsilon_2) \frac{\overleftarrow{d}}{dt_2} +
(\epsilon_1 \epsilon_2 -t_{12} t_{21} )\bigg]  =
\int_{\gamma} d\bar{z} G^{-\ }_{11}(t_1,\bar{z}) \bar{\Sigma}^{\ +}_{11}(\bar{z},t_2)  \\
+ i\left( G^{r}_{11}(t_1,t_0) \Sigma^{<}_{22}(t_0,t_2) +
      G^{<}_{11}(t_1,t_0) \Sigma^{a}_{22}(t_0,t_2) \right) - \Sigma^{<}_{22}(t_1,t_2). \label{eq:36} 
\end{multline}
Here we define new quantities $\tilde{\Sigma}_{11}(z_1,z_2)$ and $\bar{\Sigma}_{11}(z_1,z_2)$ as
\begin{multline}
 \tilde{\Sigma}_{11}(z_1,z_2) := \left( i \frac{d}{dz_1} - h_2(z_1)
 \right) \Sigma_{11}(z_1,z_2) + \Sigma_{22}(z_1,z_2)
 \left( -i\frac{\overleftarrow{d}}{dz_2} -h_1(z_2) \right) \\
 - \int_{\gamma} d \bar{z} \Sigma_{22}(z_1,\bar{z})
 \Sigma_{11}(\bar{z},z_2), \label{eq:37}
\end{multline}
\begin{multline}
 \bar{\Sigma}_{11}(z_1,z_2) := \left( i \frac{d}{dz_1} - h_1(z_2)
 \right) \Sigma_{22}(z_1,z_2) + \Sigma_{11}(z_1,z_2)
 \left( -i\frac{\overleftarrow{d}}{dz_2} -h_2(z_2) \right)  \\
 - \int_{\gamma} d \bar{z} \Sigma_{11}(z_1,\bar{z})
 \Sigma_{22}(\bar{z},z_2). \label{eq:38}
\end{multline}
By comparing Eqs. (\ref{eq:35}) and (\ref{eq:36}) with Eq. (\ref{eq:12}), we see that
$\tilde{\Sigma}_{11}(z_1,z_2)$ and $\bar{\Sigma}_{11}(z_1,z_2)$ appear in
place of the self-energy in the RHS of (\ref{eq:12}). Although they are
similar in this sense,
$\tilde{\Sigma}_{11}(z_1,z_2)$ and $\bar{\Sigma}_{11}(z_1,z_2)$ include terms whose dimensions are [${\epsilon}^2$], such as $h_2(z_1)
\Sigma_{11}(z_1,z_2)$, not [$\epsilon$]. Thus, we simply call them
 pseudo self-energies here. We do not yet understood their physical
meaning. Note that the pseudo self-energies do not satisfy the
Langreth rule, which the usual self-energy satisfies and which we
used for the integration of the equation of motion in the single-dot
case. For the pseudo self-energies, we have to apply
a modified Langreth rule. We explain the rule in Appendix B.

For the analysis of equation (\ref{eq:35}) or second-order partial
differential equations for other Green functions that will appear later,
we require expressions of the embedded self-energy
and the pseudo self-energy. We calculate these quantities with the
WBLA. Within the WBLA, the
self-energy takes a simple form that is the same as in the single-dot case\cite{Ridley2015}. For example, the retarded/advanced self-energies
are represented as
\begin{equation}
\begin{matrix}
 \Sigma^r_{11}(t_1,t_2) = - \frac{i}{2}\Gamma_{L}\delta(t_1-t_2), &
 \Sigma^r_{22}(t_1,t_2) = - \frac{i}{2}\Gamma_{R} \delta(t_1-t_2), \\
 \Sigma^a_{11}(t_1,t_2) =  \frac{i}{2}\Gamma_{L} \delta(t_1-t_2),& 
 \Sigma^a_{22}(t_1,t_2) =  \frac{i}{2}\Gamma_{R} \delta(t_1-t_2). \\ \label{eq:39}
\end{matrix} 
\end{equation}
The expressions of other self-energies are given in Appendix A. The new
quantities, the pseudo self-energies, are calculated from the expressions of
the self-energy and definition (\ref{eq:37}). However, we must pay
attention when we calculate the
retarded/advanced parts of the pseudo self-energy. This is because a
diverging term appears
in the second-order differential equations for the retarded/advanced part if
we calculate the pseudo self-energies directly. This technical
problem is explained in Appendix E. Finally, the retarded pseudo
self-energy is expressed as
\begin{eqnarray}
 \tilde{\Sigma}^r_{11}(t_1,t_2) &:=& \theta(t_1-t_2)
  \left( \tilde{\Sigma}^{>}_{11}(t_1,t_2)-\tilde{\Sigma}^{<}_{11}(t_1,t_2) \right) \notag \\
 &=& \frac{1}{2}(\Gamma_{L}  + \Gamma_{R} ) \frac{d}{d(t_1-t_2)}
  \delta(t_1-t_2) \notag  + \frac{i}{2}
(\epsilon_2 \Gamma_{L} + \epsilon_1 \Gamma_{R})
\delta(t_1-t_2) \notag \\
 && \quad -(\mathbf{h}_{1L}\mathbf{h}_{L1} + \mathbf{h}_{2R}\mathbf{h}_{R2})
 \delta(t_1-t_2). \label{eq:40} 
\end{eqnarray}
In the WBLA, $\bar{\Sigma}$ takes the same form as
$\tilde{\Sigma}$. Details of the derivation of the retarded pseudo-self energy and other
pseudo-self energies are given in Appendix D.

 \subsection{Integration of second-order partial differential equations}

 \subsubsection{Matsubara Green function}

Using Eqs. (\ref{eq:32}) and (\ref{eq:33}) on the vertical part of the contour, we
 can derive a second-order partial differential equation for the Matsubara Green
 function $G^M_{11}(\tau_1,\tau_2)$ in the same way as for Eq. (\ref{eq:35}), 
 \begin{multline}
 \left[ \frac{d^2}{d \tau^2_{1}} +(\epsilon_1 -\mu + \epsilon_2 + \mu)
  \frac{d}{d\tau_1} + \left((\epsilon_1-\mu)(\epsilon_2 -\mu) -t_{12}
  t_{21} \right)\right] G^M_{11}(\tau_1,\tau_2) \\
 = \left( \tilde{\Sigma}^{M}_{11}*G^M_{11} \right)_{(\tau_1,\tau_2)} -
  \Sigma^M_{22}(\tau_1,\tau_2) - i (\epsilon_2
  -\mu)\delta(\tau_1-\tau_2) - \frac{d}{d(\tau_1-\tau_2)}
  \delta(\tau_1-\tau_2).
\label{eq:41}
 \end{multline}
 By expanding the Matsubara Green function $G^M_{11}(\tau_1,\tau_2)$, the Matsubara self-energy, and $\delta(\tau_1-\tau_2)$ into the Matsubara sum, we obtain
 \begin{equation}
  G^M_{11}(\omega_q) =
 \begin{cases}
  \frac{\omega_q - \epsilon_2^{M,eff}}{\omega^2_q -(\epsilon_1^{M,eff}
  +\epsilon^{M,eff}_2) \omega_{q} +\left( \epsilon^{M,eff}_1
  \epsilon^{M,eff}_2 -t_{12} t_{21} \right)}, & \mathrm{Im}\omega_q >0, \\
 \frac{\omega_q - (\epsilon_2^{M,eff})^{*}}{\omega^2_q -((\epsilon_1^{M,eff})^{*}
  +(\epsilon^{M,eff}_2)^{*} ) \omega_{q} +\left( (\epsilon^{M,eff}_1)^{*}
  (\epsilon^{M,eff}_2)^{*} -t_{12} t_{21} \right)}, & \mathrm{Im}\omega_q <0,
 \end{cases}
 \label{eq:42}
 \end{equation}
 \begin{equation}
  G^M_{11}(\tau_1,\tau_2) = \frac{i}{\beta}\sum_q
   e^{-\omega_q(\tau_1-\tau_2)} G^M_{11}(\omega_q), \label{eq:43}
 \end{equation}
 where $\epsilon^{M,eff}_1 = \epsilon_1-\mu-i/2 \Gamma_{L}$ and
 $\epsilon^{M,eff}_2 = \epsilon_2-\mu-i/2 \Gamma_{R}$. When we
 consider the case where dot 2 does not exist and therefore $\epsilon^{M,eff}_2 = t_{12} = t_{21} =0$,
 the Matsubara Green function becomes $G^M_{11}(\omega_q) = \{ \omega_q
 + \epsilon^{M,eff}_1 \}^{-1}$, which is the same result as in the single-dot case\cite{Ridley2015}.
 Other Matsubara Green functions are calculated similarly and the results are
 \begin{equation}
  G^M_{22}(\omega_q) =
 \begin{cases}
  \frac{\omega_q - \epsilon_1^{M,eff}}{\omega^2_q -(\epsilon_1^{M,eff}
  +\epsilon^{M,eff}_2) \omega_{q} +\left( \epsilon^{M,eff}_1
  \epsilon^{M,eff}_2 -t_{12} t_{21} \right)}, & \mathrm{Im}\omega_q >0, \\
 \frac{\omega_q - (\epsilon_1^{M,eff})^{*}}{\omega^2_q -((\epsilon_1^{M,eff})^{*}
  +(\epsilon^{M,eff}_2)^{*} ) \omega_{q} +\left( (\epsilon^{M,eff}_1)^{*}
  (\epsilon^{M,eff}_2)^{*} -t_{12} t_{21} \right)}, & \mathrm{Im}\omega_q <0,
 \end{cases}
 \label{eq:44}
 \end{equation}
\begin{equation}
  G^M_{12}(\omega_q) =
 \begin{cases}
  \frac{t_{12}}{\omega^2_q -(\epsilon_1^{M,eff}
  +\epsilon^{M,eff}_2) \omega_{q} +\left( \epsilon^{M,eff}_1
  \epsilon^{M,eff}_2 -t_{12} t_{21} \right)}, & \mathrm{Im}\omega_q >0, \\
 \frac{t_{12}}{\omega^2_q-((\epsilon_1^{M,eff})^{*}
  +(\epsilon^{M,eff}_2)^{*} ) \omega_{q} +\left( (\epsilon^{M,eff}_1)^{*}
  (\epsilon^{M,eff}_2)^{*} -t_{12} t_{21} \right)}, & \mathrm{Im}\omega_q <0,
 \end{cases} \label{eq:45}
 \end{equation}
\begin{equation}
  G^M_{21}(\omega_q) =
 \begin{cases}
  \frac{t_{21}}{\omega^2_q -(\epsilon_1^{M,eff}
  +\epsilon^{M,eff}_2) \omega_{q} +\left( \epsilon^{M,eff}_1
  \epsilon^{M,eff}_2 -t_{12} t_{21} \right)}, & \mathrm{Im}\omega_q >0, \\
 \frac{t_{21}}{\omega^2_q -((\epsilon_1^{M,eff})^{*}
  +(\epsilon^{M,eff}_2)^{*} ) \omega_{q} +\left( (\epsilon^{M,eff}_1)^{*}
  (\epsilon^{M,eff}_2)^{*} -t_{12} t_{21} \right)}, & \mathrm{Im}\omega_q <0.
 \end{cases} \label{eq:46}
 \end{equation}
The lead-dot Matsubara Green functions, $\mathbf{G}^M_{L1}$ and
$\mathbf{G}^M_{1L}$, are determined from the solutions (\ref{eq:30}) and (\ref{eq:31}) as 
\begin{eqnarray}
 \mathbf{G}^M_{L1}(\tau_1,\tau_2) &=& \left( \mathbf{g}^M_{LL} * \mathbf{h}_{L1}(\tau)
		    G^M_{11} \right)_{(\tau_1,\tau_2)}, \label{eq:47} \\
\mathbf{G}^M_{1L}(\tau_1,\tau_2) &=& \left( G^M_{11} \mathbf{h}_{1L}(\tau) *\mathbf{g}^M_{LL}
		     \right)_{(\tau_1,\tau_2)}, \label{eq:48}
\end{eqnarray}
\begin{eqnarray}
 \mathbf{G}^M_{L2}(\tau_1,\tau_2) &=& \left( \mathbf{g}^M_{LL} * \mathbf{h}_{L1}(\tau)
		    G^M_{12} \right)_{(\tau_1,\tau_2)}, \label{eq:49} \\
\mathbf{G}^M_{2L}(\tau_1,\tau_2) &=& \left( G^M_{21} \mathbf{h}_{1L}(\tau) *\mathbf{g}^M_{LL}
		     \right)_{(\tau_1,\tau_2)}, \label{eq:50}
\end{eqnarray}
where we use the Langreth rule.
 \subsubsection{Retarded/advanced Green function}
 By differentiating the definition of the retarded/advanced Green functions
 and using Eqs. (\ref{eq:32}) and (\ref{eq:33}) with the modified Langreth rule, we can derive the second-order differential
equations for the retarded/advanced Green functions as
\begin{multline}
 \bigg[- \frac{d^2}{d{t_1}^2} -i(\epsilon^{eff}_1 + \epsilon^{eff}_2)
  \frac{d}{dt_1} + (\epsilon^{eff}_1 \epsilon^{eff}_2 -t_{12} t_{21}) \bigg] G^{r}_{11}(t_1,t_2) \\
  =i \frac{d}{d(t_1-t_2)}\delta(t_1-t_2) - \epsilon^{eff}_2
 \delta(t_1-t_2), \label{eq:51}
\end{multline}
\begin{multline}
 \bigg[- \frac{d^2}{d{t_1}^2} -i((\epsilon^{eff}_1)^{*} +
 (\epsilon^{eff}_2)^{*})
  \frac{d}{dt_1} + ((\epsilon^{eff}_1)^{*}
 (\epsilon^{eff}_2)^{*}  -t_{12} t_{21}) \bigg] G^{a}_{11}(t_1,t_2) \\
  =i \frac{d}{d(t_1-t_2)}\delta(t_1-t_2) - (\epsilon^{eff}_2)^{*} \delta(t_1-t_2), \label{eq:52}
\end{multline}
where $\epsilon^{eff}_1:=\epsilon_1-i/2 \Gamma_{L}$ and
$\epsilon^{eff}_2:=\epsilon_2-i/2 \Gamma_{R}$ are the effective
energies of the dots. In this case, the
additional terms, which are the second and third terms in Eqs. (\ref{eq:35})
and (\ref{eq:36}), disappear. The derivations of Eqs. (\ref{eq:51}) and (\ref{eq:52}) are in Appendix D.

From now on, we focus on equation Eq. (\ref{eq:51}), as we can
solve Eq. (\ref{eq:52}) in the same way. The general solution of
the homogeneous equation (\ref{eq:51}) is a linear
combination of $e^{-ik_1(t_1-t_2)}$ and $e^{-ik_2(t_1-t_2)}$, where $k_1$ and $k_2$
are the solutions of the characteristic equation and
are defined as 
\begin{align}
 k_1 = p-\sqrt{p^2-q}, && k_2 =
 p+\sqrt{p^2-q}, \notag \\
p= \frac{{\epsilon}^{eff}_1 + {\epsilon}^{eff}_2}{2}, &&
q= {\epsilon}^{eff}_1 {\epsilon}^{eff}_2 -t_{12} t_{21}. \notag 
\end{align} We assume that $G^{r}_{11}(t_1,t_2)$ takes the following form:
\begin{equation}
 G^{r}_{11}(t_1,t_2) = -i \theta(t_1-t_2) \left( C_1 e^{-i k_1(t_1-t_2)} + C_2
  e^{-i k_2(t_1-t_2)} \right  ). \label{eq:53}
\end{equation}
By substituting Eq. (\ref{eq:53}) into Eq. (\ref{eq:51}), we can
confirm that the function is
actually a particular solution when $C_1$ and $C_2$ satisfy the conditions
\begin{align}
 C_1 = \frac{k_2-\epsilon^{eff}_1 }{k_2-k_1}, && C_2 = -
  \frac{k_1-\epsilon^{eff}_1}{k_2-k_1} \notag.
\end{align}
When we consider the case of only dot 1 and reservoir $L$ and therefore $\epsilon_{2}=t_{12}=t_{21}=0$, Eq. (\ref{eq:51}) takes the following form:
\begin{equation*} 
 \bigg[- \frac{d^2}{d{t_1}^2} -i   \epsilon^{eff}_1
  \frac{d}{dt_1}  \bigg] G^{r}_{11}(t_1,t_2) 
  =i \frac{d}{d(t_1-t_2)}\delta(t_1-t_2). 
\end{equation*}
In this case, the solution is $G^r_{11}(t_1,t_2)=-i\theta(t_1-t_2)
e^{-i\epsilon^{eff}_1(t_1-t_2)}$, which is the same expression as for the
single-dot case\cite{Ridley2015}.
Similarly, we obtain the solution of Eq. (\ref{eq:52}) as
\begin{equation}
 G^{a}_{11}(t_1,t_2) = i \theta(t_2-t_1) \left( {C_1}^{*} e^{-i
					  k^{*}_1(t_1-t_2)} + {C_2}^{*}
					  e^{-i k^{*}_2(t_1-t_2)}
					  \right ). \label{eq:54}
\end{equation}

 \subsubsection{Right/left Green function}
 By differentiating the definition of the right/left Green functions of dot 1 and using Eqs. (\ref{eq:32}) and (\ref{eq:33}) and the modified Langreth rule, we can derive the second-order differential
equations for the right/left Green functions as
\begin{multline}
\bigg[-\frac{d^2}{dt^2} - i (\epsilon^{eff}_1 + \epsilon^{eff}_2) \frac{d}{dt} +
(\epsilon^{eff}_1 \epsilon^{eff}_2 -t_{12} t_{21} )\bigg] G^{\rceil}_{11} (t,\tau) =
\left(\tilde{\Sigma}^{\rceil}_{11} * G^M_{11} \right)_{(t,\tau)} -
 \Sigma^{\rceil}_{22}(t,\tau), \label{eq:55}
\end{multline}
\begin{multline}
G^{\lceil}_{11} (\tau,t) \bigg[-\frac{\overleftarrow{d}^2}{dt^2} + i ((\epsilon^{eff}_1)^{*} + (\epsilon^{eff}_2)^{*}) \frac{\overleftarrow{d}}{dt} +
((\epsilon^{eff}_1)^{*} (\epsilon^{eff}_2)^{*} -t_{12} t_{21} )\bigg]  =
\left(G^M_{11}*\bar{\Sigma}^{\lceil}_{11} \right)_{(\tau,t)} -
 \Sigma^{\lceil}_{22}(\tau,t). \label{eq:56}
\end{multline}
Treating this equation as a conventional second-order differential equation, Eq. (\ref{eq:55}) is solved with the boundary conditions
$G^{\rceil}_{11}(t_0,\tau) = G^M_{11}(0^+,\tau)$ and
$i\frac{d}{dt}G^{\rceil}_{11}(t,\tau)|_{t=t_0}=\epsilon^{eff}_1
G^M_{11}(0^+,\tau) + t_{12}G^{M}_{21}(0^+,\tau)$ as
\begin{multline}
 G^{\rceil}_{11} (t,\tau) = b_1(\tau) e^{-ik_1(t-t_0)} + b_2(\tau)
  e^{-ik_2(t-t_0)} \\ + \frac{i}{k_2-k_1} \left[\int^t_{t_0} ds
 \{ (\tilde{\Sigma}^{\rceil}_{11}*G^M_{11})_{(s,\tau)}
 -\Sigma^{\rceil}_{22}(s,\tau) \} (e^{-ik_1(t-s)}-e^{-ik_2(t-s)})
 \right], \label{eq:57}
\end{multline}
where
\begin{align*}
 b_1(\tau) = -\frac{1}{k_2-k_1}\{(\epsilon^{eff}_1-k_2) G^M_{11}(0^+,\tau) +
 (\Sigma^M_{11}*G^M_{11})_{(0^+,\tau)} + t_{12}G^M_{12}(0^+,\tau) \}, \\ 
 b_2(\tau) = \frac{1}{k_2-k_1}\{(\epsilon^{eff}_1-k_1) G^M_{11}(0^+,\tau) +
 (\Sigma^M_{11}*G^M_{11})_{(0^+,\tau)} + t_{12}G^M_{12}(0^+,\tau) \}.
\end{align*}
We can calculate other right and left Green functions similarly and the results are
\begin{multline}
 G^{\lceil}_{11} (\tau,t) = c_1(\tau) e^{i(k_1)^{*}(t-t_0)} + c_2(\tau)
  e^{i(k_2)^{*}(t-t_0)} \\ - \frac{i}{(k_2-k_1)^{*}} \left[\int^t_{t_0} ds
 \{ (G^M_{11}*\bar{\Sigma}^{\lceil}_{11})_{(\tau,s)}
 +\Sigma^{\lceil}_{22}(\tau,s) \} (e^{i(k_1)^{*}(t-s)}-e^{i(k_2)^{*}(t-s)})
 \right], \label{eq:58}
\end{multline}
\begin{align*}
 c_1(\tau) = -\frac{1}{(k_2-k_1)^{*}}\{((\epsilon^{eff}_1)^*-(k_2)^{*}) G^M_{11}(\tau,0^+) +
 (G^M_{11}*\Sigma^M_{11})_{(\tau,0^+)} + G^M_{12}(\tau,0^+)t_{21} \}, \\ 
 c_2(\tau) = \frac{1}{(k_2-k_1)^{*}}\{((\epsilon^{eff}_1)^*-(k_1)^{*}) G^M_{11}(\tau,0^+) +
 (G^M_{11}*\Sigma^M_{11})_{(\tau,0^+)} + G^M_{12}(\tau,0^+)t_{21} \},
\end{align*}
\begin{multline}
 G^{\rceil}_{12} (t,\tau) = d_1(\tau) e^{-ik_1(t-t_0)} + d_2(\tau)
  e^{-ik_2(t-t_0)} \\ + \frac{i}{k_2-k_1} \left[\int^t_{t_0} ds
 \{ (\tilde{\Sigma}^{\rceil}_{11}*G^M_{12})_{(s,\tau)} \} (e^{-ik_1(t-s)}-e^{-ik_2(t-s)})
 \right], \label{eq:59}
\end{multline}
\begin{align*}
 d_1(\tau) = -\frac{1}{k_2-k_1}\{(\epsilon^{eff}_1-k_2) G^M_{12}(0^+,\tau) +
 (\Sigma^M_{11}*G^M_{12})_{(0^+,\tau)} + t_{12}G^M_{22}(0^+,\tau) \}, \\ 
 d_2(\tau) = \frac{1}{k_2-k_1}\{(\epsilon^{eff}_1-k_1) G^M_{12}(0^+,\tau) +
 (\Sigma^M_{11}*G^M_{12})_{(0^+,\tau)} + t_{12}G^M_{22}(0^+,\tau) \}.
\end{align*}
The right Green functions between the lead and the system are determined from the
solutions (\ref{eq:30}) and (\ref{eq:31}) as
\begin{equation}
 \mathbf{G}^{\rceil}_{1L}(t,\tau) = (G^r_{11}\cdot
  \mathrm{h}_{1L}\mathbf{g}^{\rceil}_{LL} +
  G^{\rceil}_{11}*\mathrm{h}_{1L}\mathbf{g}^{M}_{LL})_{(t,\tau)}, \label{eq:60}
\end{equation}
\begin{equation}
 \mathbf{G}^{\rceil}_{L1}(t,\tau) = (\mathbf{g}^{r}_{LL}\mathrm{h}_{L1}
  \cdot G^{\rceil}_{11} +
  \mathbf{g}^{\rceil}_{LL}\mathrm{h}_{L1}*G^{\rceil}_{11})_{(t,\tau)} \label{eq:61}.
\end{equation}
 \subsubsection{Lesser Green function}
The second-order differential equation for the lesser Green
function is written in the following form:
\begin{multline}
G^{<}_{11} (t_1,t_2) \bigg[-\frac{\overleftarrow{d}^2}{dt^2_2} + i
 ((\epsilon_1^{eff} )^{*}+ (\epsilon^{eff}_2)^{*} ) \frac{\overleftarrow{d}}{dt_2} +
((\epsilon^{eff}_1)^{*} (\epsilon^{eff}_2)^{*} -t_{12}
 t_{21} )\bigg]  \\ =
\left( G^r_{11} \cdot \bar{\Sigma}^{<}_{11} +
 G^{\rceil}_{11}*\bar{\Sigma}^{\lceil}_{11} \right)_{(t_1,t_2)}
+ i G^{r}_{11}(t_1,t_0) \Sigma^{<}_{22}(t_0,t_2) -
 \Sigma^{<}_{22}(t_1,t_2), \label{eq:62}
\end{multline}
where we apply the modified Langreth rule to Eq. (\ref{eq:35}). We can solve this equation in
the same way as for the right/left Green functions and the solution is
expressed as
\begin{multline}
 G^{<}_{11} (t_1,t_2) = l_1(t_1) e^{ik^{*}_1(t_2-t_0)} + l_2(t_1)
  e^{ik^{*}_2(t_2-t_0)}  - \frac{i}{(k_2-k_1)^{*}}
  \int^{t_2}_{t_0} ds \\ \times \{\left( G^r_{11} \cdot \bar{\Sigma}^{<}_{11} +
 G^{\rceil}_{11}*\bar{\Sigma}^{\lceil}_{11} \right)_{(t_1,s)}
+ i G^{r}_{11}(t_1,t_0) \Sigma^{<}_{22}(t_0,s) -
 \Sigma^{<}_{22}(t_1,s) \} 
 (e^{ik^{*}_1(t_2-s)}-e^{ik^{*}_2(t_2-s)}). \label{eq:63}
\end{multline}
These coefficients $l_1(t), l_2(t)$ are determined from two boundary conditions, $ G^{<}_{11}
(t_1,t_0)=G^{\rceil}_{11} (t_1,0^+)$ and $G^{<}_{11}
(t_1,t_2)(-i\frac{\overleftarrow{\partial}}{\partial t_2})|_{t_2=t_0}= (\epsilon^{eff}_1)^*
G^{\rceil}_{11} (t_1,0^+) + \mathbf{G}^{\rceil}_{1L}(t_1,0^+)\cdot
\mathbf{h}_{L1}+G^{\rceil}_{12}(t_1,0^+)t_{21}$ as
\begin{align*}
 l_1(t) = -\frac{1}{(k_2-k_1)^{*}} \{ ((\epsilon^{eff}_1)^* -
 k^{*}_2) G^{\rceil}_{11} (t,0^+) + \left( G^r_{11} \cdot \Sigma^{\rceil}_{11} +
G^{\rceil}_{11}*\Sigma^M_{11} \right)_{(t,0^+)} + G^{\rceil}_{12}(t,0^+)t_{21} \}, \\
l_2(t) = \frac{1}{(k_2-k_1)^{*}} \{ ((\epsilon^{eff}_1)^* -
 k^{*}_1) G^{\rceil}_{11} (t,0^+) + \left( G^r_{11} \cdot \Sigma^{\rceil}_{11} +
G^{\rceil}_{11}*\Sigma^M_{11} \right)_{(t,0^+)} + G^{\rceil}_{12}(t,0^+)t_{21} \},
\end{align*}
where we use the expression $\mathbf{G}^{\rceil}_{1L}(t,\tau)\cdot
\mathbf{h}_{L1}=\left( G^r_{11} \cdot \Sigma^{\rceil}_{11} +
G^{\rceil}_{11}*\Sigma^M_{11} \right)_{(t,\tau)}$, which is derived from
(\ref{eq:60}) and the definition of the self-energy.

\subsection{Calculation of physical quantities}
In Sect. 4.4, we obtained the concrete expressions of the
nonequilibrium Green functions. In this subsection, we explain how the
physical quantities are expressed in terms of the nonequilibrium Green functions.
A definition of the electron density at dot $i$ is
$\rho_{i}(t):=\braket{d^{\dagger}_{H,i}(t) d_{H,i}(t)}$. Using the
nonequilibrium Green function, the density is expressed as
\begin{equation*}
 \rho_{i}(t)= -i G^{<}_{ii}(t,t),
\end{equation*}
where we use the definition of the lesser Green function
$G^{<}_{ii}(t,t')=i \braket{d^{\dagger}_{H,i}(t) d_{H,i}(t)}$. This
relation implies that the lesser Green function is directly related to the electron density.
We define the current from dot $i$ as
\begin{equation}
 J_{i}(t) := - \frac{d}{dt} \rho_{i}(t), \label{eq:64} 
\end{equation}
and that from reservoir $\alpha$ as 
\begin{equation*}
 J_{\alpha}(t) := - \frac{d}{dt} \rho_{\alpha}(t),
\end{equation*}
where we define the electron density of reservoir $\alpha$ in the same way as
that of dot$i$: $\rho_{\alpha}(t):= \sum_{k}
\braket{d^{\dagger}_{H,k \alpha}(t) d_{H,k \alpha}(t)}$. We can find a representation of the current from the left reservoir in
terms of the nonequilibrium Green function as
\begin{align}
 J_{L}(t) &= - \frac{d}{dt} \rho_{L}(t) \notag \\
 &= -2 \mathrm{Re}[\mathbf{G}^<_{1L} \mathbf{h}_{L1}], \notag 
\end{align}
where we use the Heisenberg equation of the Hamiltonian
(\ref{eq:18}) for the derivative of the operators. By substituting the
expression of the lesser Green function (\ref{eq:61}) and using the Langreth rule, we can express the
current in terms of the Green function as
\begin{align}
J_{L}(t) &= -2 {\rm Re} [(\Sigma^{\rceil}_{11} \cdot
 G^{\lceil}_{11})_{(t,t)} + (\Sigma^r_{11} \cdot G^<_{11})_{(t,t)} + (\Sigma^<_{11}
 \cdot G^a_{11})_{(t,t)}]. \label{eq:65}
\end{align}

 We can obtain the current between the dots as follows. Because the particles
in the left reservoir only move to dot 1, the current between the left
reservoir and dot 1 is equal to the current from the left reservoir:
$J_{L}(t)=J_{L\rightarrow 1}(t)$. The current from dot1 equals the sum of the current from dot 1 to reservoir $L$ and that from dot 1 to dot 2: $J_1(t) = J_{1\rightarrow L} + J_{1\rightarrow
2}$. Therefore, by summing these two relations, the current
from the dot1 to the dot2 is expressed as 
\begin{equation}
J_{1\rightarrow2}(t) = J_{1}(t) + J_{L}(t),
\label{eq:66}
\end{equation}
where we use $J_{L\rightarrow 1}(t) = -J_{1\rightarrow
L}(t)$. Using this relation, we can obtain an expression of the current between the dots.

\section{Numerical Results}
Using the expressions obtained in Sect. 4.5, we numerically calculate the
electron density of dot 1, the current of dot 1, and the current between the dots. Here, we consider the quenched case, where a constant
bias voltage is suddenly applied to the left reservoir $V_L(t)=V_L$ at an initial
time $t_0$ and $V_R=0$. We
consider the cases: $V_L=6$ as a high bias voltage and $V_L=2$ as a low
bias voltage. In both cases, the inverse temperature is taken to be
$\beta=100$. The energy levels of the dots are
$\epsilon_1=\epsilon_2=1$. The energies of the two subsystems must take
the same value because of energy conservation\cite{ziegler2000}. The
couplings between the subsystems and the reservoirs are taken to be
symmetric, $\Gamma_L=\Gamma_R=1/2$, and the couplings between the
subsystems are $t_{12}=t_{21}=1$. Throughout our numerical computations,
we use the representation of each physical quantity \cite{Ridley2015}. In this representation, the physical quantities are
expressed in terms of the integration with respect to frequency $\omega$
such as in Eq. (\ref{eq:17}). As an example, we derive the formula for the electron density of dot 1 in Appendix F.

The results are shown in  Fig. \ref{fig:4}-\ref{fig:7}. First we discuss
the graph of the electron density of dot 1 in Fig. \ref{fig:4}.
\begin{figure}[ht]
 \begin{center}
  \includegraphics[width=0.5\hsize]{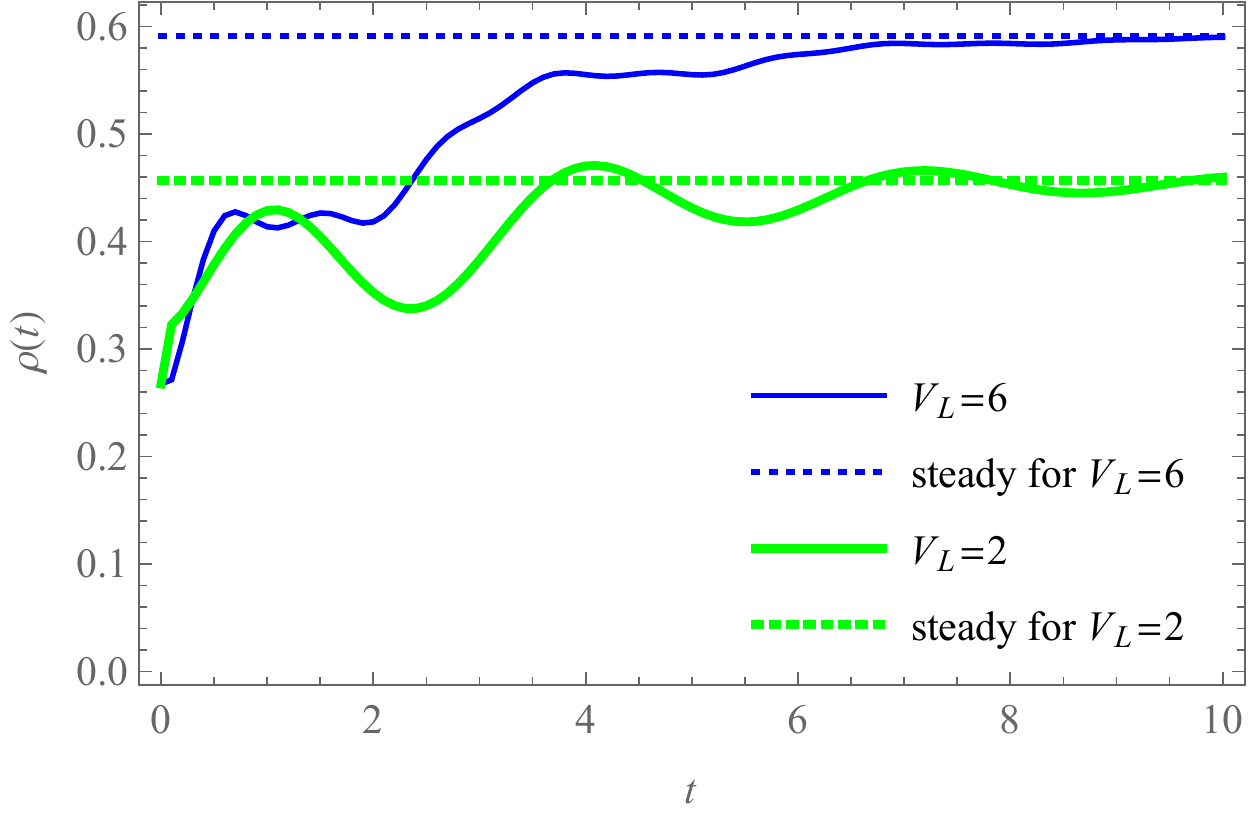}
\end{center}
\caption{(Color Online) Electron density of dot 1 as a function of time for $V_{L}=2$ and
 $V_L=6$. The straight
 line is the steady-state value of the electron density that is
 calculated from the analytical expression of the density.}
\label{fig:4}
\end{figure}
The electron density oscillates in our DQD case. This cannot be seen in
the case of the single dot (Fig. \ref{fig:2}). Since the electron density is directly related
to the lesser Green function $\rho_{11}(t)=-iG^<_{11}(t,t)$, the cause
of the difference between the cases is the different behavior of the
lesser Green function. In the single-dot case, the expression of the
lesser Green function is Eq. (\ref{eq:17}). If we take the two times to be the same, $t_1=t_2=t$, then
the imaginary part of the exponential $e^{-ih^{eff}_{11}(t_1-t_0)}$
vanishes because of the factor $e^{i(h^{eff}_{11})^{*} (t_2-t_0)}$. Therefore,
the oscillation does not appear in the single-dot case. In the double-dot
case, the expression of the lesser Green function is Eq. (\ref{eq:63}). Unlike the singe-dot case, there are two exponential functions with different exponents $k_1$ and
$k_2$. Therefore, the imaginary parts of these exponential
functions do not vanish and the oscillation appears.

\begin{figure}[ht]
 \begin{center}
  \includegraphics[width=0.5\hsize]{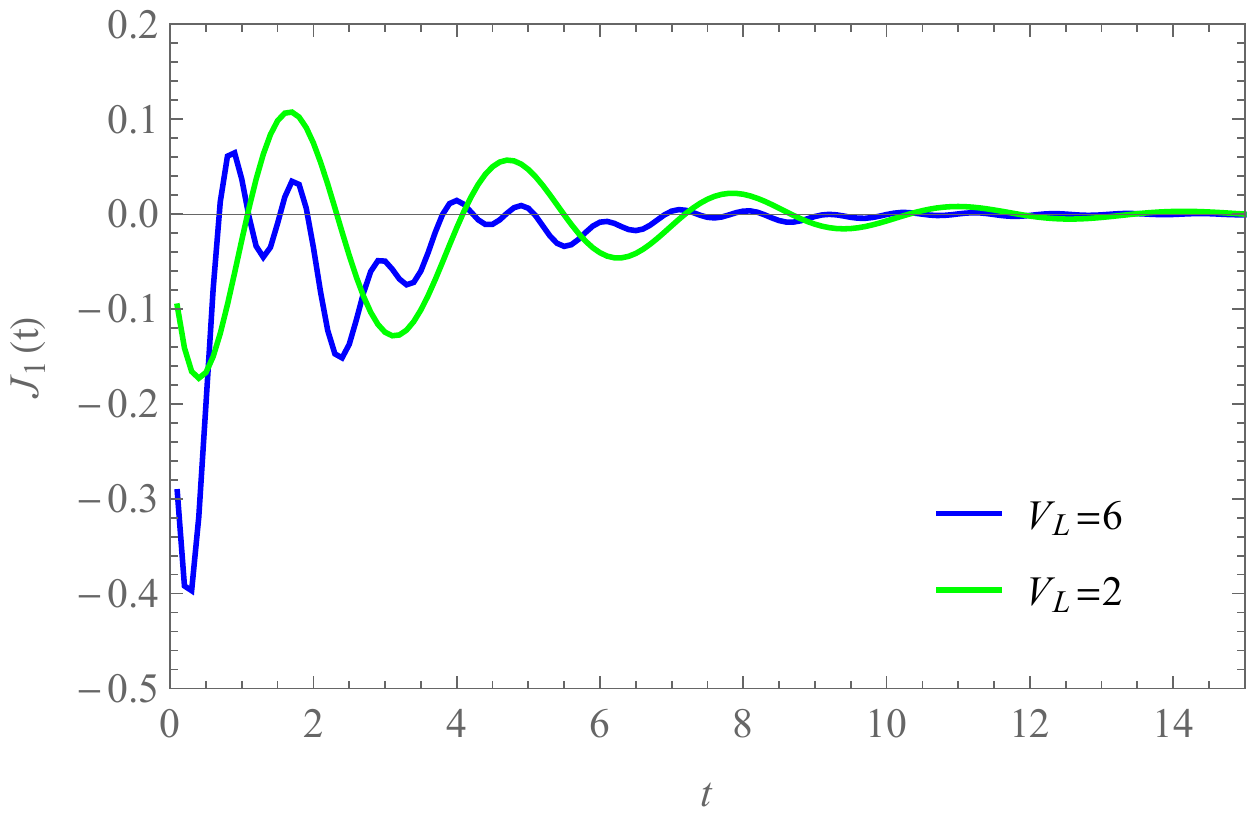}
\end{center}
\caption{(Color Online) Current from dot 1 for $V_{L}=2$ and
 $V_L=5$. }
\label{fig:5}
\end{figure}
We can obtain the graph of the current from dot 1 by
differentiating the electron density of dot 1. The result is shown in
Fig. \ref{fig:5}. The behavior of the current depends on the bias voltage. For the low bias of
$V_L=2$, we can see that there is a single crest in a period. In
contrast, for the high bias of $V_L=6$, there are two crests in a
period. Because the current from dot 1 is obtained by differentiating the electron density of dot 1, this difference arises from the existence of $k_1$ and $k_2$, which are the solutions of the characteristic equation of the second-order differential equation for the retarded Green function
 (\ref{eq:51}). Since the necessity to investigate the second-order
 partial differential equations arises from the existence of the
 coupling between the dots, we conclude that the two wave crests in the
 current between the dots is a manifestation of the fact that the system consists of two subsystems.

\begin{figure}[ht]
 \begin{center}
  \includegraphics[width=0.5\hsize]{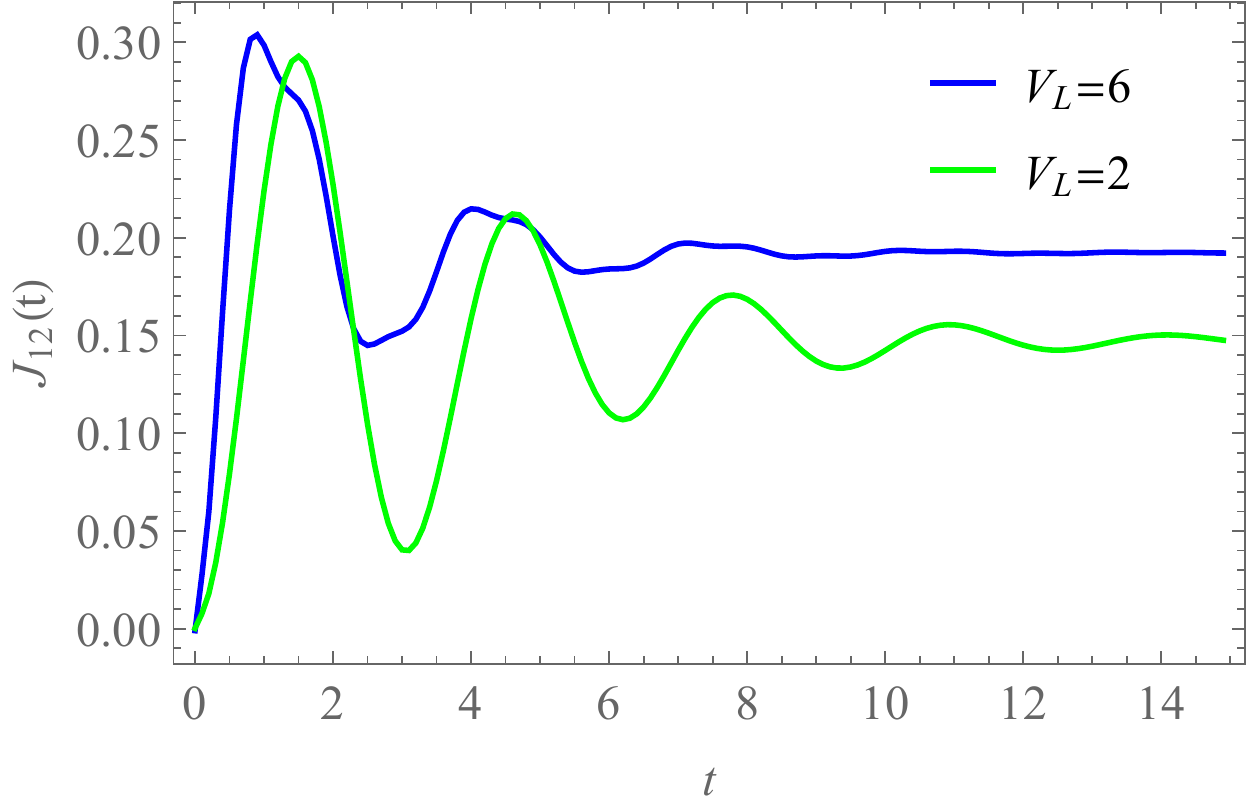}
\end{center}
\caption{(Color Online) Current between the dots for $V_{L}=2$ and $V_L=6$.}
\label{fig:6}
\end{figure}
We also calculate the graph of the current between the
dots based on Eq. (\ref{eq:66}). Fig. \ref{fig:6} shows the
result. From the graph, we can see that the bias voltage changes the frequency of the waves. The change occurs because the bias voltage causes a phase shift of the current meaning the term $e^{-i\psi_{L}(t,t_0)}$. We can understand this by
considering the fact that the lesser Green function exists in the expression
of the current from the left reservoir (\ref{eq:65}), which is a part of
the current between the dots (\ref{eq:66}). In the lesser Green function
(\ref{eq:63}), the pseudo self-energy and the self-energy exist, which
have the term $e^{-i\psi_{L}(t,t_0)}$. The concrete expressions of these
self-energies are in Appendices A and C. Because of the term, the
frequency becomes high as the bias voltage increases. In addition, we
can see that the difference in the number of peaks of the waves
vanishes, which we saw in the behavior of the current from dot 1(Fig. \ref{fig:5}). This is because the current between the dots is expressed as the sum of the currents
(\ref{eq:66}), and the current from the left reservoir cancels the behavior of the current from dot 1.

We also investigate the relaxation time. When a physical quantity $A(t)$
behaves as $A(t) \eqsim e^{-\frac{t}{\tau}}$, $\tau$ is called the
relaxation time. The notation $\eqsim$ means that we ignore algebraic
functions multiplied by the exponential.
From expression (\ref{eq:63}), we can see that the relaxation
time is determined by $\mathrm{min}_{{i,j}={1,2}} \mathrm{Im}[k_i-(k_j)^*]$. Let us explain
this with Eq. (\ref{eq:63}). From the first term in Eq. (\ref{eq:63}), the terms
$e^{-i(k_1-(k_1)^*)t}$ and $e^{-i(k_2-(k_1)^*)t}$ appear as the
exponential functions $e^{-ik_1 t }$ and $e^{-ik_2 t}$ exist in the
right Green function (\ref{eq:57}), which is a part of $l_1(t)$. Similarly, $e^{-i(k_1-(k_2)^*)t}$ and $e^{-i(k_2-(k_2)^*)t}$ appear from
the second term in Eq. (\ref{eq:63}). Therefore, the smallest term of $\mathrm{Im}[k_i-(k_j)^*](i,j={1,2})$ determines the relaxation time. 
For the case $\epsilon_1=\epsilon_2=\epsilon$,
the exponent takes the following forms:
\begin{equation}
\argmin_{{i,j}={1,2}} \mathrm{Im}[k_i-(k_j)^*] =
 \begin{cases}
    - \frac{\Gamma_L + \Gamma_R}{2} +  \sqrt{ \frac{1}{16} (\Gamma_L 
  - \Gamma_R)^2 - |t_{12}|^2}  & |t_{12}| \leq 1/4|\Gamma_L-\Gamma_R| \\
  - \frac{\Gamma_L + \Gamma_R}{2} & |t_{12}|>1/4|\Gamma_L-\Gamma_R|.
 \end{cases}
\end{equation}
The division arises because the
inside of the square root in the expression for $k_1$ or $k_2$ can be both negative or positive. When we consider the case
$|t_{12}| \leq 1/4|\Gamma_L-\Gamma_R|$, the relaxation
time $\tau$ is $2/ \left( \Gamma_L + \Gamma_R - 2 \sqrt{ \frac{1}{16} (\Gamma_L 
  - \Gamma_R)^2 - |t_{12}|^2} \right)$. Therefore, the relaxation time becomes smaller as $|t_{12}|$ increases. This is intuitively a natural result. For the case
$|t_{12}|>1/4|\Gamma_L-\Gamma_R|$, however, we can see an interesting
fact. In this case, $\tau$ is $2/(\Gamma_L
+ \Gamma_R)$. This expression reveals that the relaxation time
$\tau$ does not depend on $|t_{12}|$. This is against our
intuition. Since the coupling strength between the dots determines the probability that an electron in a dot is transported to the other dot, we can
 expect that the relaxation time becomes smaller as the coupling strength between the
 dots increases. To test this hypothesis, we calculate the
relaxation time from the numerical results and compare it with the
theoretically expected value. We consider the case where
the parameters take the values $\Gamma_L=\Gamma_R=1/2$, $\epsilon=1$,
$V_{L}=6$, and $\beta=1$. In this situation, the relaxation time does not
depend on the coupling strength $|t_{12}|>0$. In our analysis, we obtain
the relaxation time numerically by fitting the data of the current from dot 1, $J_{1}$, to an exponential function. 

\begin{figure}[ht]
 \begin{center}
  \includegraphics[width=0.5\hsize]{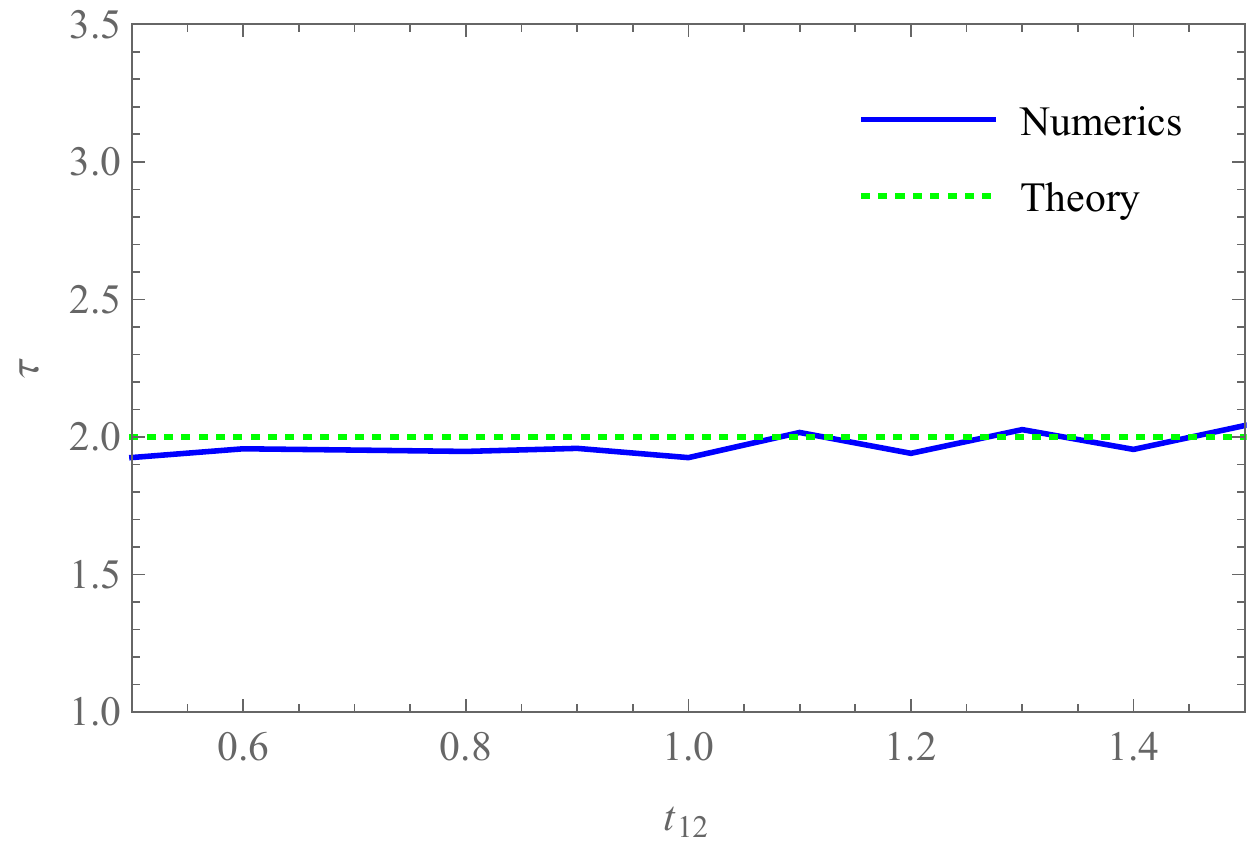}
\end{center}
\caption{(Color Online) Relaxation time for various couplings between the dots. The
 relaxation time takes almost the same value as the theoretically
 expected value of $\tau=2$.}
\label{fig:7}
\end{figure}
We calculate the relaxation time for various coupling strengths of the
dots $|t_{12}|$, $0.5 \le |t_{12}| \le 1.5$. The result is shown in
Fig. \ref{fig:7}. In this case, the theoretical value is $\tau =
2$. Therefore, we conclude that the relaxation time is actually constant
and matches the theoretical value. This phenomenon can be understood as
follows. From the representation of the retarded Green function
(\ref{eq:53}), we see that the imaginary parts of $k_1$ and $k_2$
determine the lifetime of quasiparticles\cite{zagoskin}. The forms of the
imaginary parts of $k_1$ and $k_2$ depend on the square root in the expression for $k_1$ and $k_2$. Since the sign of the inside of the square root is
determined by $|t_{12}|$ and $1/4|\Gamma_L-\Gamma_R|$, the lifetime
of quasiparticles, or the relaxation time, depends on the
parameters. Note that this behavior of the relaxation time can be
seen in any temperature regime 
because the temperature does not appear in the above discussion. This result may be useful for manipulating the coherence time such as in quantum computation.


\section{Conclusion}
We have investigated the transient dynamics of double quantum dots with
the nonequilibrium Green function method in the wide-band-limit
approximation. As a result, we obtained the analytic expressions of
the electron density of a dot, the current from dot1, and the current between the dots. Based on
these results, we calculated the physical quantities numerically for the
quenched case, where a constant bias voltage is suddenly applied to a reservoir. From the numerical computation, we found that a quantum fluctuation
appears in the electron
density and the currents. In particular, the qualitative behavior of the current from dot1 depends on the bias voltage. For a low
bias voltage, there is one crest in a period of oscillation in the current. However, for a high bias voltage, the number of crests in a period changes to
two. This difference arises from the existence of the coupling between
the dots. In addition, we calculated the relaxation time and found that the relaxation time becomes constant when the coupling constant between the dots is sufficiently large in comparison with the difference in coupling strength between the dots and the reservoirs. This implies that if a system is an open system, we have to take the effects from the edges into consideration
even when we consider a quantity in the scatterers.

Until now, we have only considered the regime where the Coulomb interaction is
irrelevant. Therefore, our next goal is to investigate the transient
dynamics of the double quantum dots including the Coulomb interaction.
Since many interesting studies have been reported, even for the steady case, we
can expect that much richer phenomena will arise from the competition of
the initial correlation effects and the Coulomb interaction in the transient
dynamics of double quantum dots.

\section*{Acknowledgements}
Part of this work was performed during a stay at ICTS Bangalore. 
The work of T.S. is supported by JSPS KAKENHI Grant Numbers 
JP25103004, JP14510499, JP15K05203, and JP16H06338.


\appendix
\section{Non-perturbative Green function and the self-energy in the
 WBLA}
  \subsection{Non-perturbative Green function}
 The non-perturbative Green function obeys the equation of motion
\begin{equation*}
  \left[  i \frac{d}{dz_{1}} - \mathbf{h}_{\alpha \alpha}(z_1) \right]
  \mathbf{g}_{\alpha \alpha}(z_1,z_2) = \mathbf{1}\delta(z_1,z_2).
\end{equation*}
This means that the non-perturbative Green function is the Green
function for the case where reservoir $\alpha$ evolves independently, or with no interaction. Then we consider the expression of the operator $d_{k\alpha}(z)$
under the Hamiltonian $\mathbf{h}_{\alpha\alpha}$. Because the
time-dependent part of the Hamiltonian
$V_{\alpha}(t) d^{\dagger}_{k\alpha} d_{k \alpha}$ and the remaining part
$\epsilon_{k \alpha} d^{\dagger}_{k \alpha} d_{k \alpha}$ are
interchangeable, the operator
$d_{k\alpha}(z)$ for $z \in C_{\pm}$ is represented as
\begin{equation*}
 d_{k \alpha}(t) =  d_{k \alpha} e^{-i \phi(t,t_0)} = (d^{\dagger}_{k \alpha}(t))^{\dagger},
\end{equation*}
where we define the function $\phi_{k \alpha}(t,t_0) = \epsilon_{k \alpha}(t-t_0) +
\int^t_{t_0} ds V_{\alpha}(s)$ for convenience. In the same way, the operator
$d_{k\alpha}(z)$ for $z=t_0-i\tau \in C_{M}$ is expressed as 
\begin{gather*}
 d_{k \alpha}(t_0-i\tau)  = d_{k \alpha} e^{-(\epsilon_{k \alpha}-\mu)
 \tau}, \\
d^{\dagger}_{k \alpha}(t_0-i\tau)  = d^{\dagger}_{k \alpha} e^{(\epsilon_{k \alpha}-\mu)
 \tau}.
\end{gather*}
Note that $ (d_{k \alpha}(t_0-i\tau))^{\dagger} \neq
d^{\dagger}_{k \alpha}(t_0-i\tau)$. From these operators, the
non-perturbative Green functions are calculated as
\begin{gather}
 [\mathbf{g}^r_{\alpha \alpha}(t_1,t_2)]_{k,k'} = -i \theta(t_1-t_2)
 \delta_{kk'} e^{-i \phi_{k \alpha}(t_1,t_2)}, \tag{A.1}
 \label{A.1} \\
 [\mathbf{g}^a_{\alpha \alpha}(t_1,t_2)]_{k,k'} = i \theta(t_2-t_1)
 \delta_{kk'} e^{-i \phi_{k \alpha}(t_1,t_2)}, \tag{A.2}
 \label{A.2} \\
 [\mathbf{g}^>_{\alpha \alpha}(t_1,t_2)]_{k,k'} = -i \delta_{k k'} [1-
 f(\epsilon_{k\alpha} - \mu)] e^{-i \phi_{k \alpha}(t_1,t_2)}, \tag{A.3}
 \label{A.3} \\
[\mathbf{g}^<_{\alpha \alpha}(t_1,t_2)]_{k,k'} = i \delta_{k k'} f(\epsilon_{k\alpha} - \mu) e^{-i \phi_{k \alpha}(t_1,t_2)}. \tag{A.4}
 \label{A.4} 
\end{gather}
For the Matsubara component, it is useful to use the representation of
the Matsubara sum, which arises from the boundary conditions
$\mathbf{g}_{\alpha \alpha}(z_1,t_0) = - \mathbf{g}_{\alpha
\alpha}(z_1,\beta)$, $\mathbf{g}_{\alpha \alpha}(t_0,z_2) = - \mathbf{g}_{\alpha
\alpha}(\beta,z_2)$,
\begin{equation}
 [\mathbf{g}^M_{\alpha \alpha}(\tau_1,\tau_2)]_{k,k'} = \delta_{k k'}
  \frac{i}{\beta} \sum_q \frac{e^{-\omega_q(\tau_1-\tau_2)}}{\omega_q -
  \epsilon_{k \alpha} + \mu}, \tag{A.5} \label{A.5}
\end{equation}
where $\omega_q= i \pi (2q+1)/\beta$ is the Matsubara frequency and the
summation over $q$ runs over all integers. The derivation is as
follows. From the definition of the Matsubara Green function,
\begin{align*}
 \left[\mathbf{g}^M_{\alpha\alpha} (\tau_1,\tau_2) \right]_{kk'} &= -i
 \braket{\mathcal{T}_{\tau} (d^M_{k\alpha}(\tau_1)
 (d^{\dagger}_{k'\alpha})^M)} \\
 &= -i \left[ \theta(\tau_1-\tau_2) (1-f(\epsilon_{k\alpha}-\mu)) -
 \theta(\tau_2-\tau_1) f(\epsilon_{k \alpha}-\mu) \right]
 e^{-(\epsilon_{k\alpha}-\mu)(\tau_1-\tau_2)} \delta_{kk'} \\
 &(\because
 \mathrm{the\ expectation\ is\ taken\ in\ the\ equilibrium\ state}\\
 &= -i \left[ \theta(\tau) (1-f(\epsilon_{k\alpha}-\mu)) -
 \theta(-\tau) f(\epsilon_{k \alpha}-\mu) \right]
 e^{-(\epsilon_{k\alpha}-\mu)\tau} \delta_{kk'}. (\because \tau_1-\tau_2=\tau 
\end{align*} 
Then we use these expressions in the Fourier transform related to the Matsubara frequency:
\begin{align*}
 \frac{1}{i} \int^{\beta}_0 d\tau \theta(\tau)
 e^{(\omega_q-\epsilon_{k\alpha}+\mu)\tau} &= i
 \frac{1+e^{-\beta(\epsilon_{k \alpha}-\mu)}}{\omega_q - \epsilon_{k
 \alpha} + \mu}, \\
 \frac{1}{i} \int^{\beta}_0 d\tau \theta(-\tau)
 e^{(\omega_q-\epsilon_{k\alpha}+\mu)\tau} &= 0.
\end{align*}
By substituting these expressions in the definition, we obtain
\begin{align*}
 \left[\mathbf{g}^M_{\alpha\alpha} (\tau_1,\tau_2) \right]_{kk'} &=
  i \frac{1}{\beta} \sum_q e^{-\omega_q(\tau_1-\tau_2)} (1-f(\epsilon_{k
 \alpha}-\mu)) \left(
 \frac{1+e^{-\beta(\epsilon_{k \alpha}-\mu)}}{\omega_q - \epsilon_{k
 \alpha} + \mu} \right) \delta_{k k'} \\
 &= \frac{i}{\beta} \sum_q \frac{e^{-\omega_q(\tau_1-\tau_2)}}{\omega_q -
 \epsilon_{k \alpha} + \mu}.
\end{align*}
The right/left non-perturbative Green functions are calculated with these
results, 
\begin{gather}
 [\mathbf{g}^{\rceil}_{\alpha \alpha}(t,\tau)]_{k,k'} =
 [\mathbf{g}^{M}_{\alpha \alpha}(0,\tau)]_{k,k'} e^{-i \phi_{k
 \alpha}(t,t_0)} = \delta_{k k'}
  \frac{i}{\beta} e^{-i \phi_{k
 \alpha}(t,t_0)} \sum_q \frac{e^{\omega_q \tau}}{\omega_q -
  \epsilon_{k \alpha} + \mu}, \tag{A.6} \label{A.6} \\
  [\mathbf{g}^{\lceil}_{\alpha \alpha}(\tau,t)]_{k,k'} =
 [\mathbf{g}^{M}_{\alpha \alpha}(\tau,0)]_{k,k'} e^{i \phi_{k
 \alpha}(t,t_0)} = \delta_{k k'}
  \frac{i}{\beta} e^{i \phi_{k
 \alpha}(t,t_0)} \sum_q \frac{e^{-\omega_q \tau}}{\omega_q -
  \epsilon_{k \alpha} + \mu}. \tag{A.7} \label{A.7}
\end{gather}
These are the same as in \cite{Ridley2015}.

 \subsection{Self-energy in the WBLA}
From the definition of the self-energy (\ref{eq:34}) and the expressions of
the non-perturbative Green function obtained above, we can calculate
the self-energy. For the retarded part, by the Fourier transform, we obtain 
\begin{align*}
 \Sigma^r_{11}(t_1,t_2) &= \theta(t_1-t_2) \left[ \Sigma^>_{11}(t_1,t_2)
 - \Sigma^<_{11}(t_1,t_2) \right] \notag \\
 &= \theta(t_1-t_2) \sum_{k,k'} T_{1,kL}
 ([\mathbf{g}^>_{LL}(t_1,t_2)]_{k,k'}-[\mathbf{g}^<_{LL}(t_1,t_2)]_{k,k'})
 T_{k' L,1}\ (\because \mathrm{By\ definition(\ref{eq:34})}\\
 &= \theta(t_1-t_2) \sum_{k} T_{1,k L}  (-i e^{-i
 \phi_{k L}(t_1,t_2)}) T_{kL,1}\ (\because \mathrm{(\ref{B.3})\ and\ (\ref{B.4})}\\
 &= -i e^{-i\psi_{L}(t_1,t_2)} \sum_k  T_{1,kL} T_{kL,1} \theta(t_1-t_2)
 e^{-i\epsilon_{kL}(t_1-t_2)} \\
 &= e^{-i \psi_{L}(t_1,t_2)} \int \frac{d \omega}{2\pi} e^{-i \omega
 (t_1-t_2)} \sum_k \frac{T_{1,k\alpha} T_{k L,1}}{\omega + i 0 -
 \epsilon_{k L}} \notag \ (\because \mathrm{From\ the\ formula\ below}\\
 &= e^{-i \psi_{\alpha}(t_1,t_2)} \int \frac{d \omega}{2\pi} e^{-i \omega
 (t_1-t_2)} \left[ \Lambda_{L}(\omega) - \frac{i}{2}
 \Gamma_{L}(\omega) \right]\ (\because \frac{1}{x\pm i0}=
\mathcal{P}\left( \frac{1}{x}\right) \mp i \pi \delta(x),
\end{align*}
where we define
\begin{gather*}
 \Lambda_{L}(\omega) = \mathcal{P} \int \frac{d{\omega}'}{2 \pi}
 \frac{\Gamma_{L}(\omega')}{\omega-\omega'}, \\
  \Gamma_{L}(\omega) = 2 \pi \sum_{k} T_{1,kL} T_{kL,1}
 \delta(\epsilon_{kL}- \omega),
\end{gather*}
and $\mathcal{P}$ represents the Cauchy principal part. In
the derivation, we use the Fourier transformation which is defined as
$F(\omega)= \int^{\infty}_{-\infty} dt f(t) e^{i\omega t}$. The inverse
Fourier transform is expressed as $f(t)=\int^{\infty}_{-\infty}
\frac{d \omega}{2 \pi} F(\omega) e^{-i\omega t}$. In this
definition, the Fourier transformation of $\theta(t) e^{-i a t}, a\in \mathbf{R}$ is
expressed as
\begin{align*}
 \int^{\infty}_{-\infty} dt \theta(t) e^{-i a t} e^{ i\omega t} &=
 \lim_{\delta \rightarrow 0^+} \int^{\infty}_{-\infty} dt \theta(t)
 e^{-(\delta - i(\omega-a))t}\ (\because \omega\rightarrow \omega +
 i\delta \\
 &= \lim_{\delta \rightarrow 0^+} \int^{\infty}_{0} dt e^{-\delta t+
 i(\omega-a)t} \\
 &= \lim_{\delta \rightarrow 0^+}  \left[ \frac{1}{-\delta+i(\omega-a)} e^{-\delta t + i(\omega-a)t}  \right]^{\infty}_0 \\
 &= \lim_{\delta \rightarrow 0^+}  \frac{-1}{-\delta+i(\omega-a)} \\
 &= \lim_{\delta \rightarrow 0^+} - \frac{1}{i} \frac{1}{\omega + i \delta
 - a} \\
 &=: - \frac{1}{i} \frac{1}{\omega + i 0- a}.
\end{align*}
Therefore, we obtain the formula $\theta(t_1-t_2) e^{-i \epsilon_{k L} (t_1-t_2)} = \int^{\infty}_{-\infty}
\frac{d \omega}{2 \pi} e^{-i \omega(t_1-t_2)}\frac{i}{\omega + i 0-
\epsilon_{kL}}$. In the WBLA, the line width $\Gamma_{L}(\omega)$ does not depend on the energy $\omega$ and
$\Lambda_{L}(\omega)$ becomes $0$: $\Gamma_{L}(\omega) \sim \Gamma_{L}$,
$\Lambda_{L}=0$. Then, the
retarded self-energy in the WBLA takes the form
\begin{equation*}
  \Sigma^r_{11}(t_1,t_2) = - \frac{i}{2} \Gamma_{L} \delta(t_1-t_2), 
\end{equation*} 
which is Eq. (\ref{eq:39}). Other self-energies are calculated similarly
and the results are
\begin{align}
 \Sigma^{M}_{11}(\tau_1,\tau_2)&=-\frac{\Gamma_{L}}{2\beta} \sum_q
  \xi_q e^{-\omega_q(\tau_1-\tau_2)}, \tag{A.8} \label{A.8} \\ 
 \Sigma^{M}_{22}(\tau_1,\tau_2)&=-\frac{\Gamma_{R}}{2\beta} \sum_q
  \xi_q e^{-\omega_q(\tau_1-\tau_2)}, \tag{A.9} \label{A.9}
\end{align}
\begin{align}
  \Sigma^{<}_{11}(t_1,t_2) &= i \Gamma_{L} e^{-i \psi_L(t_1,t_2)} \int
 \frac{d \omega}{ 2 \pi} f(\omega-\mu) e^{-i\omega(t_1-t_2)},
 \tag{A.10} \label{A.10} \\
\Sigma^{<}_{22}(t_1,t_2) &= i \Gamma_{R} e^{-i \psi_R(t_1,t_2)} \int
 \frac{d \omega}{ 2 \pi} f(\omega-\mu) e^{-i\omega(t_1-t_2)}, \tag{A.11} \label{A.11}
\end{align}
\begin{align}
\Sigma^{>}_{11}(t_1,t_2) &= -i \Gamma_{L} e^{-i \psi_L(t_1,t_2)} \int
 \frac{d \omega}{ 2 \pi} [1-f(\omega-\mu)] e^{-i\omega(t_1-t_2)},
 \tag{A.12} \label{A.12} \\
 \Sigma^{>}_{22}(t_1,t_2) &= -i \Gamma_{R} e^{-i \psi_R(t_1,t_2)} \int
 \frac{d \omega}{ 2 \pi} [1-f(\omega-\mu)] e^{-i\omega(t_1-t_2)}, \tag{A.13} \label{A.13}
\end{align}
\begin{align}
\Sigma^{\lceil}_{11}(\tau,t) &= \frac{i}{\beta} \Gamma_{L} e^{i
 \psi_L(t,t_0)} \sum_q e^{-{\omega}_q \tau} \int
 \frac{d \omega}{ 2 \pi} \frac{e^{i\omega(t-t_0)}}{\omega_q-\omega+\mu}, \tag{A.14} \label{A.14} \\
\Sigma^{\lceil}_{22}(\tau,t) &= \frac{i}{\beta} \Gamma_{R} e^{i
 \psi_R(t,t_0)} \sum_q e^{-{\omega}_q \tau} \int
 \frac{d \omega}{ 2 \pi} \frac{e^{i\omega(t-t_0)}}{\omega_q-\omega+\mu}, \tag{A.15} \label{A.15}
\end{align}
\begin{align}
\Sigma^{\rceil}_{11}(\tau,t) &= \frac{i}{\beta} \Gamma_{L} e^{-i
 \psi_L(t,t_0)} \sum_q e^{{\omega}_q \tau} \int
 \frac{d \omega}{ 2 \pi}
 \frac{e^{-i\omega(t-t_0)}}{\omega_q-\omega+\mu}, \tag{A.16} \label{A.16} \\
\Sigma^{\rceil}_{22}(\tau,t) &= \frac{i}{\beta} \Gamma_{R} e^{-i
 \psi_R(t,t_0)} \sum_q e^{{\omega}_q \tau} \int
 \frac{d \omega}{ 2 \pi} \frac{e^{-i\omega(t-t_0)}}{\omega_q-\omega+\mu}. \tag{A.17} \label{A.17}
\end{align}

\section{Modified Langreth rule for the pseudo self-energy in
 the WBLA}
 The Langreth rule\cite{langreth1976} are written as 
\begin{align}
 \Sigma^r_{11}(t_1,t_2) = \Sigma^{--}_{11}(t_1,t_2) -
  \Sigma^<_{11}(t_1,t_2),
 \tag{B.1} \label{B.1} \\
\Sigma^a_{11}(t_1,t_2) = \Sigma^{--}_{11}(t_1,t_2) -
  \Sigma^>_{11}(t_1,t_2).
 \tag{B.2} \label{B.2}
\end{align}
The rules express the relations between the retarded or the advanced part of the
 self-energy and other parts of the self-energy. The rules are derived straightforwardly from the fact that the
 self-energies are defined on contour $\gamma$. Therefore, the rules
 also hold for other quantities defined on the
 contour such as the nonequilibrium Green functions. Since the pseudo self-energies $\tilde{\Sigma}_{11}$ are expressed in terms of the
 quantities on the contour, one may expect the pseudo self-energy
 to follow the Langreth rule. However, we cannot apply the Langreth rule to the
 pseudo self-energies. This is because there is a product of quantities that
 depends on contour $\gamma$ in its definition (\ref{eq:37}). Instead,
 the following modified Langreth rules hold for the pseudo self-energy: 
\begin{align}
 \tilde{\Sigma}^r_{11} = \tilde{\Sigma}^{--}_{11}(t_1,t_2) -
  \tilde{\Sigma}^<_{11}(t_1,t_2) + \left( \Sigma^{r}_{11} \cdot
 \Sigma^{r}_{22} \right)_{(t_1,t_2)} - \delta(t_1-t_2) (\mathbf{h}_{1L}
 \mathbf{h}_{L1} + \mathbf{h}_{2R} \mathbf{h}_{R2}), \tag{B.3}
 \label{B.3} \\
\tilde{\Sigma}^a_{11} = \tilde{\Sigma}^{--}_{11}(t_1,t_2) -
  \tilde{\Sigma}^>_{11}(t_1,t_2) + \left( \Sigma^{r}_{22} \cdot
 \Sigma^{r}_{11} \right)_{(t_1,t_2)} - \delta(t_1-t_2) (\mathbf{h}_{1L}
 \mathbf{h}_{L1} + \mathbf{h}_{2R} \mathbf{h}_{R2}). \tag{B.4} \label{B.4}
\end{align}
When we calculate the quantities including the pseudo self-energies, we
use the modified Langreth rules.

Their derivation is as follows. To derive the modified Langreth rule
(\ref{B.3}), we
compare the expression for the retarded part of the pseudo self-energy
$\tilde{\Sigma}^r_{11}:=\theta(t_1-t_2)
(\tilde{\Sigma}^{>}_{11}-\tilde{\Sigma}^{<}_{11})$ and these for other parts of
the pseudo self-energy $\tilde{\Sigma}^{--}_{11}-\tilde{\Sigma}^{<}_{11}$. By substituting the equation of motion for the non-perturbative Green
function into the derivative of the self-energies in the definition of
the pseudo self-energy (\ref{eq:37}), we obtain the expression 
\begin{multline}
 \tilde{\Sigma}_{11}(z_1,z_2) = \mathbf{h}_{1L} \mathbf{h}_{LL}
  \mathbf{g}_{LL}(z_1,z_2) \mathbf{h}_{L1} + \mathbf{h}_{2R} \mathbf{h}_{RR}
  \mathbf{g}_{RR}(z_1,z_2) \mathbf{h}_{R2} - \epsilon_1
  \Sigma_{11}(z_1,z_2) - \epsilon_2 \Sigma_{22}(z_1,z_2) \\ - \int_{\gamma}
  d \bar{z}  \Sigma_{22}(z_1,\bar{z})  \Sigma_{11}(\bar{z},z_2) +
  \delta(z_1,z_2) (\mathbf{h}_{1L}
 \mathbf{h}_{L1} + \mathbf{h}_{2R} \mathbf{h}_{R2}). \tag{B.5} \label{B.5}
\end{multline}
From Eq. (\ref{B.5}), we can obtain expressions of $\tilde{\Sigma}^r_{11}$
and $\tilde{\Sigma}^{--}_{11}-\tilde{\Sigma}^{<}_{11}$. If there were only the first four terms in Eq. (\ref{B.5}), we could apply the
ordinary Langreth rule because the non-perturbative Green function and
the self-energy satisfy the standard Langreth rule. Howecer, a modification is
necessary due to the existence of the fifth term $\int_{\gamma}
  d \bar{z}  \Sigma_{22}(z_1,\bar{z})  \Sigma_{11}(\bar{z},z_2)$ and the
  sixth term $\delta(z_1,z_2) (\mathbf{h}_{1L}
 \mathbf{h}_{L1} + \mathbf{h}_{2R} \mathbf{h}_{R2})$. Therefore, we only consider the expressions of the fifth and the sixth terms of
 $\tilde{\Sigma}^r_{11}$ and
 $\tilde{\Sigma}^{--}_{11}-\tilde{\Sigma}^{<}_{11}$ and compare them. In
 fact, the fifth term for $\tilde{\Sigma}^r_{11}$ is 0 in the
WBLA,
\begin{equation}
 \theta(t_1-t_2) \int_{\gamma}
  d \bar{z} ( \Sigma_{22}^{+\ }(t_1,\bar{z})  \Sigma_{11}^{\
  -}(\bar{z},t_2) - \Sigma_{22}^{-}(t_1,\bar{z})  \Sigma_{11}^{\
  +}(\bar{z},t_2) = 0, \tag{B.6} \label{B.6}
\end{equation}
where we use the expression of the self-energy in the WBLA, which is
calculated in Appendix A. The sixth term vanishes for the pseudo
retarded self-energy due to the Heaviside function in the definition. For
$\tilde{\Sigma}^{--}_{11}-\tilde{\Sigma}^{<}_{11}$, the fifth term is written as
\begin{equation}
 \int_{\gamma}
  d \bar{z} ( \Sigma_{22}^{-}(t_1,\bar{z})  \Sigma_{11}^{\
  -}(\bar{z},t_2) - \Sigma_{22}^{-}(t_1,\bar{z})  \Sigma_{11}^{\
  +}(\bar{z},t_2) = \left( \Sigma^{r}_{22} \cdot
 \Sigma^{r}_{11} \right)_{(t_1,t_2)}, \tag{B.7} \label{B.7} 
\end{equation}
where we use the Langreth rule. The notation
$\Sigma_{22}^{-}(t_1,\bar{z})$ means that the argument $z_1$ is on contour $C_{-}$ and the argument $z_2$ is not identified where it is on contour $\gamma$. The sixth term remains unchanged.
Therefore, we can conclude that the difference between
$\tilde{\Sigma}^{--}_{11}-\tilde{\Sigma}^{<}_{11}$ and
$\tilde{\Sigma}^r_{11}$ is $\left( \Sigma^{r}_{22} \cdot
 \Sigma^{r}_{11} \right)_{(t_1,t_2)} - \delta(t_1-t_2) (\mathbf{h}_{1L}
 \mathbf{h}_{L1} + \mathbf{h}_{2R} \mathbf{h}_{R2})$. This means that the
 modified Langreth rule (\ref{B.3}) holds. The modified Langreth rule
 for the advanced pseudo self-energy (\ref{B.4}) is derived in the same way.

\section{Pseudo self-energy in the WBLA}

 \subsection{Matsubara pseudo self-energy}
From the definition of the pseudo self-energy (\ref{eq:37}), we obtain 
\begin{align}
 \tilde{\Sigma}^M_{11}(\tau_1,\tau_2) &= \mathbf{h}_{1L}
 \left(-\frac{d}{d\tau_1} \right) \mathbf{g}^M_{LL}(\tau_1,\tau_2)
 \mathbf{h}_{L1} +  \mathbf{h}_{2R} \mathbf{g}^M_{LL}(\tau_1,\tau_2)
 \left(\frac{\overleftarrow{d}}{d\tau_2} \right) 
 \mathbf{h}_{R2} \notag \\
 &\  -(\epsilon_1-\mu) \Sigma^M_{22}(\tau_1,\tau_2) -
 (\epsilon_2-\mu) \Sigma^M_{11}(\tau_1,\tau_2) -
 \left(\Sigma_{22}*\Sigma_{11} \right)_{(\tau_1,\tau_2)} \notag \\
&= (\mathrm{i}) + (\mathrm{i\hspace{-.1em} i}) +
  (\mathrm{i\hspace{-.1em}i \hspace{-.1em} i}) +
  (\mathrm{i\hspace{-.1em}v}) + (\mathrm{v}). \tag{C.1} \label{C.1}
\end{align}
The first term is computed as
\begin{align}
 (\mathrm{i}) &= \Gamma_{L} \frac{i}{\beta} \sum_q \int
  \frac{d\omega}{2\pi} \frac{\omega_q}{\omega_q-\omega+\mu}
  e^{-\omega_q (\tau_1-\tau_2)} \notag \\
 &= -\frac{\Gamma_{L}}{2\beta} \sum_q \omega_q \xi_q e^{-\omega_q
  (\tau_1-\tau_2)}, \tag{C.2} \label{C.2}
\end{align}
where we use the expression
\begin{equation*}
 \int \frac{d\omega}{2\pi} \frac{1}{\omega_q-\omega+\mu} = \frac{i}{2}
  \xi_q.
\end{equation*}
The function $\xi_q$ is defined as $\xi_q=1$ (Im $\omega_q<0$), $-1$ (Im
$\omega_q>0$).
In the same way, the second term is written as 
\begin{equation}
 (\mathrm{i\hspace{-.1em} i}) =  -\frac{\Gamma_{R}}{2\beta} \sum_q \omega_q \xi_q e^{-\omega_q
  (\tau_1-\tau_2)}. \tag{C.3} \label{C.3}
\end{equation}
Using the expressions for the Matsubara self-energies (\ref{A.8}) and (\ref{A.9}) calculated in Appendix A, we obtain
\begin{align}
  (\mathrm{i\hspace{-.1em}i \hspace{-.1em} i})  &= (\epsilon_1 - \mu)
 \frac{\Gamma_{R}}{2\beta} \sum_q \omega_q
 e^{-\omega_q(\tau_1-\tau_2)}, \tag{C.4} \label{C.4} \\
(\mathrm{i\hspace{-.1em}v})  &= (\epsilon_2 - \mu)
 \frac{\Gamma_{L}}{2\beta} \sum_q \omega_q
 e^{-\omega_q(\tau_1-\tau_2)}. \tag{C.5} \label{C.5}
\end{align} 
The fifth term is calculated using the relation $\int^\beta_0 d \tau
e^{-(\omega_q-\omega_q')\tau} = \beta \delta_{q,q'}$ and the result is
\begin{equation}
 (\mathrm{v}) =  i \frac{\Gamma_{L} \Gamma_{R}}{4\beta} \sum_q
  e^{-\omega_q(\tau_1-\tau_2)}. \tag{C.6} \label{C.6}
\end{equation}
Using expressions from Eq. (\ref{C.2})-(\ref{C.6}), we obtain the Matsubara pseudo self-energy (\ref{eq:52}) as
\begin{equation}
  \tilde{\Sigma}^M_{11}(\tau_1,\tau_2) = \sum_q
  e^{-\omega_q(\tau_1-\tau_2)}  \big[ -\frac{\xi_q}{2\beta} \{
 (\omega_q -\epsilon_1 + \mu) \Gamma_{R}   
+ (\omega_q -\epsilon_2 +\mu)
  \Gamma_{L} \} + i \frac{\Gamma_{L} \Gamma_{R}}{4 \beta} \big].
  \tag{C.7} \label{C.7}
\end{equation}
 \subsection{Retarded/advanced pseudo self-energy}
The retarded pseudo self-energy is defined as
$\tilde{\Sigma}^r_{11}(t_1,t_2) =
\theta(t_1-t_2)(\tilde{\Sigma}^>_{11}(t_1,t_2) -
\tilde{\Sigma}^<_{11}(t_1,t_2))$. Using the definition of the pseudo
self-energy (\ref{eq:37}), this is written in the form
\begin{align}
 \tilde{\Sigma}^r_{11}(t_1,t_2) &= \theta(t_1-t_2) \bigg[ \left(i \frac{d}{dt_1} - \epsilon_2 \right)
  \left( \Sigma^>_{11}(t_1,t_2) - \Sigma^<_{11}(t_1,t_2) \right) \notag \\
 & \quad + \left( \Sigma^>_{22}(t_1,t_2) - \Sigma^<_{22}(t_1,t_2)
		 \right) \left(-i \frac{\overleftarrow{d}}{dt_2} -
 \epsilon_1 \right) \notag \notag \notag \\
 & \qquad - \int_{\gamma}
  d \bar{z} ( \Sigma_{22}^{+\ }(t_1,\bar{z})  \Sigma_{11}^{\
  -}(\bar{z},t_2) - \Sigma_{22}^{-}(t_1,\bar{z})  \Sigma_{11}^{\
  +}(\bar{z},t_2) \bigg] \notag \\
 &= (\mathrm{v\hspace{-.1em}i}) +
 (\mathrm{v\hspace{-.1em}i\hspace{-.1em}i}) +
 (\mathrm{v\hspace{-.1em}i\hspace{-.1em}i\hspace{-.1em}i}). \tag{C.8} \label{C.8}
\end{align}
Using the definition of the retarded self-energy, we rewrite the first term $(\mathrm{v\hspace{-.1em}i})$ as
\begin{multline}
 \theta(t_1-t_2) \left(i \frac{d}{dt_1} - \epsilon_2 \right)
  \left( \Sigma^>_{11}(t_1,t_2) - \Sigma^<_{11}(t_1,t_2) \right) =
  \left(i \frac{d}{dt_1} - \epsilon_2 \right) \Sigma^r_{11}(t_1,t_2) \\ -
  i\delta(t_1 -t_2) \left( \Sigma^>_{11}(t_1,t_2) -
		     \Sigma^<_{11}(t_1,t_2) \right). \tag{C.9} \label{C.9}
\end{multline}
Using the expressions for the self-energies, (\ref{A.10}) and (\ref{A.12}), we obtain
\begin{equation*}
 \left( \Sigma^>_{11}(t_1,t_2) - \Sigma^<_{11}(t_1,t_2) \right) = -i
\mathbf{h}_{1L} \mathbf{h}_{L1} e^{-i \psi_{kL}(t_1,t_2)}.
\end{equation*}
Therefore, Eq. (\ref{C.9}) is expressed in the form
\begin{equation}
 \theta(t_1-t_2) \left(i \frac{d}{dt_1} - \epsilon_2 \right)
  \left( \Sigma^>_{11}(t_1,t_2) - \Sigma^<_{11}(t_1,t_2) \right) =
  \left(i \frac{d}{dt_1} - \epsilon_2 \right) \Sigma^r_{11}(t_1,t_2) -
  \mathbf{h}_{1L} \mathbf{h}_{L1} \delta(t_1-t_2). \tag{C.10} \label{C.10}
\end{equation}
In the same way, $(\mathrm{v\hspace{-.1em}i\hspace{-.1em}i})$ is represented as 
\begin{equation}
\theta(t_1-t_2)  \left( \Sigma^>_{22}(t_1,t_2) - \Sigma^<_{22}(t_1,t_2)
		 \right) \left(-i \frac{\overleftarrow{d}}{dt_2} - \epsilon_1 \right) =
   \Sigma^r_{22}(t_1,t_2) \left(-i \frac{\overleftarrow{d}}{dt_2} - \epsilon_1 \right) -
  \mathbf{h}_{2R} \mathbf{h}_{R2} \delta(t_1-t_2). \tag{C.11} \label{C.11}
\end{equation}
Then we use the expression for the retarded self-energy in the WBLA
(\ref{eq:39}). By substituting the expression into Eqs. (\ref{C.10}) and (\ref{C.11}), we obtain
\begin{align*}
 & \theta(t_1-t_2) \bigg[ \left(i \frac{d}{dt_1} - \epsilon_2 \right)
  \left( \Sigma^>_{11}(t_1,t_2) - \Sigma^<_{11}(t_1,t_2) \right) +
 \left( \Sigma^>_{22}(t_1,t_2) - \Sigma^<_{22}(t_1,t_2) \right) \left(-i
 \frac{\overleftarrow{d}}{dt_2} - \epsilon_1 \right) \bigg] \\ 
 & = \frac{1}{2}(\Gamma_{L} +\Gamma_{R})
 \frac{d}{d(t_1-t_2)} \delta(t_1-t_2) + \frac{i}{2} (\epsilon_2
 \Gamma_{L} +\epsilon_1 \Gamma_{R}) \delta(t_1-t_2) \\
 & \qquad - (\mathbf{h}_{1L} \mathbf{h}_{L1} +\mathbf{h}_{2R} \mathbf{h}_{R2})
 \delta(t_1-t_2). \tag{C.12} \label{C.12}
\end{align*} 
By the substitution of Eqs. (\ref{C.12}) and (\ref{B.7}) in Eq. (\ref{C.8}),
we obtain the expression of the retarded pseudo self-energy
(\ref{eq:40}). The advanced pseudo self-energy is calculated in the same
way and the result is
\begin{align}
\tilde{\Sigma}^a_{11}(t_1,t_2) &:= -\theta(t_2-t_1)
  (\tilde{\Sigma}^{>}_{11}(t_1,t_2)-\tilde{\Sigma}^{<}_{11}(t_1,t_2) \notag \\
 &= -\frac{1}{2}(\Gamma_{L}  + \Gamma_{R} ) \frac{d}{d(t_1-t_2)} \delta(t_1-t_2) - \frac{i}{2}
(\epsilon_2 \Gamma_{L} + \epsilon_1 \Gamma_{R})
\delta(t_1-t_2)  \tag{C.13} \label{C.13} \\
& - (\mathbf{h}_{1L}\mathbf{h}_{L1} + \mathbf{h}_{2R}\mathbf{h}_{R2})
 \delta(t_1-t_2). \notag
\end{align}
 \subsection{Left/right pseudo self-energy}
From the definition (\ref{eq:37}), the left pseudo self-energy is represented as
\begin{align}
 \tilde{{\Sigma}^{\lceil}}_{11}(\tau,t) & = \left(-\frac{d}{d\tau} -
 \epsilon_2 + \mu \right) \Sigma^{\lceil}_{11}(\tau,t) +
\Sigma^{\lceil}_{22}(\tau,t) \left(-i\frac{\overleftarrow{d}}{dt}
 -\epsilon_1 \right) \notag \\
 & - \int_{\gamma} d \bar{z} \Sigma_{22}(\tau,\bar{z})
 \Sigma_{11}(\bar{z},t)  \notag \\
 & = \mathrm{(i\hspace{-.1em}x) + (x) + (x\hspace{-.1em}i)}. \tag{C.14} \label{C.14}
\end{align}
From the expressions (\ref{A.14}) and (\ref{A.15}) for the left self-energy
in the WBLA, the first and second terms are written as
\begin{align}
\mathrm{(i\hspace{-.1em}x)} &= \frac{i}{\beta} \Gamma_{L} e^{i \psi_{L}(t,t_0)} \sum_q
  (\omega_q-\epsilon_2 + \mu) e^{-\omega_q \tau} \int
  \frac{d\omega}{2\pi} \frac{e^{i\omega(t-t_0)}}{\omega_q -\omega + \mu}
 \tag{C.15}, \label{C.15} \\
 \mathrm{(x)} &= \frac{i}{\beta} \Gamma_{R,11} e^{i \psi_{R}(t,t_0)} \sum_q
  e^{-\omega_q \tau} \int
  \frac{d\omega}{2\pi} \frac{\omega+V_{R}(t) -\epsilon_1}{\omega_q
 -\omega + \mu} e^{i\omega(t-t_0)}.
 \tag{C.16} \label{C.16}
\end{align}
Using the Langreth rule and the expressions for the retarded and advanced
self-energy in the WBLA (\ref{eq:39}), we obtain
\begin{equation}
 \mathrm{(x\hspace{-.1em}i)} = - \frac{i}{2} \Gamma_{L}(\tau,t)
  \Sigma^{\lceil}_{22} - \left( \Sigma^M_{22} * \Sigma^{\lceil}_{11}
			 \right)_{(\tau,t)}. \tag{C.17} \label{C.17}
\end{equation}
We can derive the expression of the left pseudo self-energy from Eqs. (\ref{C.14})-(\ref{C.16}) as
\begin{multline}
 \tilde{\Sigma}^{\lceil}_{11}(\tau_1,t_2) = \frac{i}{\beta}
 \Gamma_{L} e^{i \psi_{L}(t_1,t_0)} \sum_{q}
 (\omega_q-\epsilon_2+\mu) e^{-\omega_{q} \tau}
 \int \frac{d \omega}{2 \pi} \frac{e^{i\omega(t-t_0)}}{\omega_{q} -\omega + \mu}   + \frac{i}{\beta}
 \Gamma_{R} e^{i \psi_{R}(t_1,t_0)} \\
 \times  \sum_q 
 e^{-\omega_{q} \tau} \int \frac{d \omega}{2 \pi} \frac{\omega+V_R(t) -
 \epsilon_1}{\omega_{q} -\omega + \mu} e^{i\omega(t-t_0)} 
 - \frac{i}{2} \Gamma_{L} \Sigma^{\lceil}_{22}(\tau_1,t_2) - \left(
 \Sigma^{M}_{22} * \Sigma^{\lceil}_{11} \right)_{(\tau_1,t_2)}. \tag{C.18} \label{C.18}
\end{multline}
The right pseudo self-energy is derived in the same way
as the left pseudo self-energy and it is expressed as
\begin{multline}
 \tilde{\Sigma}^{\rceil}_{11}(t_1,\tau_2) = \frac{i}{\beta}
 \Gamma_{L} e^{-i \psi_{L}(t_1,t_0)} \sum_{q} e^{\omega_{q} \tau}
 \int \frac{d \omega}{2 \pi} \frac{\omega+V_L(t) -
 \epsilon_2}{\omega_{q} -\omega + \mu} e^{-i\omega(t-t_0)}  + \frac{i}{\beta}
 \Gamma_{R} e^{-i \psi_{R}(t_1,t_0)} \\
 \times  \sum_q (\omega_q-\epsilon_1+\mu)
 e^{\omega_{q} \tau} \int \frac{d \omega}{2 \pi} \frac{e^{-i\omega(t-t_0)}}{\omega_{q}
 -\omega + \mu}  + \frac{i}{2} \Gamma_{R} \Sigma^{\rceil}_{11}(t_1,\tau_2) - \left(
 \Sigma^{\rceil}_{22} * \Sigma^{M}_{11} \right)_{(t_1,\tau_2)}.  \tag{C.19} \label{C.19}
\end{multline}

 \subsection{Lesser/greater pseudo self-energy}
From definition (\ref{eq:37}), the lesser pseudo self-energy is represented as 
\begin{align}
 \tilde{\Sigma}^{<}_{11}(\tau,t) & = \left(i\frac{d}{dt} -
 \epsilon_2  \right) \Sigma^{<}_{11}(t_1,t_2) +
\Sigma^{<}_{22}(t_1,t_2) \left(-i\frac{\overleftarrow{d}}{dt_2}
 -\epsilon_1 \right) \notag \\
 & - \int_{\gamma} d \bar{z} \Sigma^{-\ }_{22}(t_1,\bar{z})
 \Sigma^{\ +}_{11}(\bar{z},t_2)  \notag \\
 & = \mathrm{(x\hspace{-.1em}i\hspace{-.1em}i) +
 (x\hspace{-.1em}i\hspace{-.1em}i\hspace{-.1em}i) + (x\hspace{-.1em}i\hspace{-.1em}v)}. \tag{C.20} \label{C.20}
\end{align}
By differentiating the representations, (\ref{A.10}) and (\ref{A.11}), of the left self-energy in the WBLA, we obtain
\begin{align}
 \mathrm{(x\hspace{-.1em}i\hspace{-.1em}i)} &= i \Gamma_{L}
 e^{-i\psi_L(t_1,t_2)} \int \frac{d\omega}{2\pi} (\omega + V_L(t_1) -
 \epsilon_2) f(\omega-\mu) e^{-i\omega(t_1-t_2)}, \tag{C.21}
 \label{C.21} \\
\mathrm{(x\hspace{-.1em}i\hspace{-.1em}i\hspace{-.1em}i)} &= i \Gamma_{R}
 e^{-i\psi_R(t_1,t_2)} \int \frac{d\omega}{2\pi} (\omega + V_R(t_2) -
 \epsilon_1) f(\omega-\mu) e^{-i\omega(t_1-t_2)}. \tag{C.22} \label{C.22}
\end{align}
Using the expressions for the retarded/advanced self-energies in the WBLA
(\ref{eq:39}) and the Langreth rule, the third term is written as 
\begin{equation*}
 \mathrm{(x\hspace{-.1em}i\hspace{-.1em}v)} = \frac{i}{2}
  \left(\Gamma_{R} \Sigma^<_{11} - \Gamma_{L} \Sigma^<_{22}
  \right) - \left(\Sigma^{\rceil}_{22} * \Sigma^{\lceil}_{11} \right)_{(t_1,t_2)}.
\end{equation*}
Actually, the second term in the representation above equals to 0:
\begin{align}
 \left(\Sigma^{\rceil}_{22} * \Sigma^{\lceil}_{11} \right)_{(t_1,t_2)}
 &= -i \int^{\beta}_0 d \tau \frac{i}{\beta} \Gamma_{R} e^{-i
 \psi_R(t_1,t_0)} \sum_q e^{\omega_q \tau} \frac{d\omega}{2\pi}
 \frac{e^{-i\omega(t_1-t_2)}}{\omega_q - \omega + \mu} \notag \\
 & \qquad \times \frac{i}{\beta} \Gamma_{L} e^{i
 \psi_L(t_2,t_0)} \sum_{q'} e^{- \omega_{q'} \tau} \frac{d\omega}{2\pi}
 \frac{e^{-i\omega(t_1-t_2)}}{\omega_{q'} - \omega + \mu} \notag \\
 &= 0, \notag
\end{align}
where we use the expressions for the left/right self-energies in the WBLA
(\ref{A.14}) and (\ref{A.17}). From these above results, we
derive the following expression for the lesser pseudo self-energy:
\begin{multline}
 \tilde{\Sigma}^{<}_{11}(t_1,t_2) = i
 \Gamma_{L} e^{-i \psi_{L}(t_1,t_0)} 
 \int \frac{d \omega}{2 \pi} (\omega +V_L(t_1) -\epsilon_2)
 f(\omega-\mu) e^{-i\omega(t-t_0)} + 
 i \Gamma_{R} e^{-i \psi_{R}(t_1,t_0)} \\  \int \frac{d \omega}{2 \pi}
 (\omega +V_R(t_2) -\epsilon_1) f(\omega-\mu ) e^{-i\omega(t-t_0)} 
 + \frac{i}{2} \left(\Gamma_{R} \Sigma^{<}_{11}(t_1,t_2) -
 \Gamma_{L} \Sigma^{<}_{22}(t_1,t_2) \right). \tag{C.23} \label{C.23}
\end{multline}
The greater pseudo self-energy is calculated in the same way as
\begin{multline}
 \tilde{\Sigma}^{>}_{11}(t_1,t_2) = -i
 \Gamma_{L} e^{-i \psi_{L}(t_1,t_0)} 
 \int \frac{d \omega}{2 \pi} (\omega +V_L(t_1) -\epsilon_2)
 [1-f(\omega-\mu)] e^{-i\omega(t-t_0)} \\ -i \Gamma_{R} e^{-i
 \psi_{R}(t_1,t_0)} \int \frac{d \omega}{2 \pi}
 (\omega +V_R(t_2) -\epsilon_1) [1-f(\omega-\mu )] e^{-i\omega(t-t_0)} \\
 + \frac{i}{2} \left(\Gamma_{R} \Sigma^{>}_{11}(t_1,t_2) -
 \Gamma_{L} \Sigma^{>}_{22}(t_1,t_2) \right). \tag{C.24} \label{C.24}
\end{multline}

\section{Derivation of the second-order partial differential equation}

 \subsection{Lesser Green function}
Here, we derive the second-order partial differential
equation for the lesser Green function (\ref{eq:35}).
The derivation starts from the equations of motion for $G_{11}$ and
$G_{21}$ of the lesser part:
\begin{align}
 \left[ i \frac{d}{dt_1} - \epsilon_1 \right] G^<_{11}(t_1,t_2) &=
 \int_{\gamma} d \bar{z} \Sigma^{-\ }_{11}(t_1,\bar{z}) G^{\
 +}_{11}(\bar{z},t_2) + t_{12} G^<_{21}(t_1,t_2), \tag{D.1} \label{D.1} \\
 \left[ i \frac{d}{dt_1} - \epsilon_2 \right] G^<_{21}(t_1,t_2) &=
 \int_{\gamma} d \bar{z} \Sigma^{-\ }_{22}(t_1,\bar{z}) G^{\
 +}_{21}(\bar{z},t_2) + t_{21} G^<_{11}(t_1,t_2). \tag{D.2} \label{D.2}
\end{align}
To substitute the differential equation for $G_{21}$ into the equation for
$G_{11}$, we rewrite Eq. (\ref{D.1}) in the form
\begin{align}
 G^<_{21}(t_1,t_2) &= \frac{1}{t_{12}} \bigg{\{}  \left[ i \frac{d}{dt_1} -
 \epsilon_1 \right] G^<_{11}(t_1,t_2) -  \int_{\gamma} d \bar{z} \Sigma^{-\ }_{11}(t_1,\bar{z}) G^{\
 +}_{11}(\bar{z},t_2) \bigg{\}}, \tag{D.3} \label{D.3} \\
 i \frac{d}{dt_1}G^<_{21}(t_1,t_2) &= \frac{1}{t_{12}} \bigg{\{}  \left[ -
 \frac{d^2}{dt^2_1} - i 
 \epsilon_1 \frac{d}{dt_1} \right] G^<_{11}(t_1,t_2) -i  \int_{\gamma} d
 \bar{z}  \frac{d}{dt_1} \Sigma^{-\ }_{11}(t_1,\bar{z}) G^{\
 +}_{11}(\bar{z},t_2) \bigg{\}}. \tag{D.4} \label{D.4}
\end{align}
By substituting Eqs. (\ref{D.3}) and (\ref{D.4}) into Eq. (\ref{D.2})
and multiplying by $t_{12}$, we obtain
\begin{multline}
   \bigg[ - \frac{d^2}{dt^2_1}  - i  
 (\epsilon_1 + \epsilon_2) \frac{d}{dt_1} + (\epsilon_1 \epsilon_2 -
 t_{12} t_{21} ) \bigg] G^<_{11}(t_1,t_2) \\
 = \int_{\gamma} d \bar{z}  (i \frac{d}{dt_1} - \epsilon_2 )  \Sigma^{-\ }_{11}(t_1,\bar{z}) G^{\
 +}_{11}(\bar{z},t_2) + t_{12} \int_{\gamma} d \bar{z}
 \Sigma^{-\ }_{22}(t_1, \bar{z}) G^{\ +}_{21}(\bar{z},t_2). \tag{D.5} \label{D.5}
\end{multline}
With the Langreth rule, the rightmost term is expressed as
\begin{equation}
 t_{12} \int_{\gamma} d \bar{z}
 \Sigma^{-\ }_{22}(t_1, \bar{z}) G^{\ +}_{21}(\bar{z},t_2) = t_{12}
 \left( \Sigma^{--}_{22} \cdot G^<_{21} - \Sigma^<_{22} \cdot
  G^{++}_{21} + \Sigma^{\rceil}_{12} * G^{\lceil}_{21}
 \right)_{(t_1,t_2)}. \tag{D.6} \label{D.6}
\end{equation}
Using the equation of motion for $G_{11}$ (\ref{eq:32}), we rewrite $G_{21}$ in Eq. (\ref{D.6}) in terms of $G_{11}$. From Eq. (\ref{eq:32}), we obtain 
\begin{gather*}
 G^<_{21}(t_1,t_2) = \frac{1}{t_{12}}  \bigg{\{}  \left[ i \frac{d}{dt_1} -
 \epsilon_1 \right] G^<_{11}(t_1,t_2) -  \int_{\gamma} d \bar{z} \Sigma^{-\ }_{11}(t_1,\bar{z}) G^{\
 +}_{11}(\bar{z},t_2) \bigg{\}}, \\
 G^{++}_{21}(t_1,t_2) = \frac{1}{t_{12}}  \bigg{\{}  \left[ i \frac{d}{dt_1} -
 \epsilon_1 \right] G^{++}_{11}(t_1,t_2) -  \int_{\gamma} d \bar{z} \Sigma^{+\ }_{11}(t_1,\bar{z}) G^{\
 +}_{11}(\bar{z},t_2) - \delta(t_1-t_2) \bigg{\}}, \\
 G^{\rceil}_{21}(\tau_1,t_2) = \frac{1}{t_{12}}  \bigg{\{}  \left[ - \frac{d}{d\tau_1} -
 \epsilon_1+\mu \right] G^{\rceil}_{11}(\tau_1,t_2) -  \int_{\gamma} d \bar{z} \Sigma_{11}(\tau_1,\bar{z}) G^{\ +}_{11}(\bar{z},t_2) \bigg{\}}.
\end{gather*}
Using these equations, we obtain
\begin{equation*}
 t_{12} \left( \Sigma^{--}_{22} \cdot G^<_{21} \right)_{(t_1,t_2)} =
 \int^{\infty}_{t_0} dt \Sigma^{--}_{22}(t_1,t)  \bigg{\{}  \left[ i
 \frac{d}{dt} - \epsilon_1 \right] G^<_{11}(t,t_2) -  \int_{\gamma} d \bar{z} \Sigma^{-\ }_{11}(t,\bar{z}) G^{\
 +}_{11}(\bar{z},t_2) \bigg{\}}, \tag{D.7} \label{D.7} 
\end{equation*}
\begin{multline*}
 - t_{12} \left( \Sigma^<_{22} \cdot G^{++}_{21}\right)_{(t_1,t_2)} = -
 \int^{\infty}_{t_0} dt \Sigma^{<}_{22}(t_1,t) \bigg{\{}  \left[ i
 \frac{d}{dt} - \epsilon_1 \right] G^{++}_{11}(t,t_2) \\
 -  \int_{\gamma} d \bar{z} \Sigma^{+\ }_{11}(t,\bar{z}) G^{\
 +}_{11}(\bar{z},t_2) \bigg{\}} + \Sigma^<_{22},  \tag{D.8}
 \label{D.8}
\end{multline*}
\begin{multline*}
 t_{12} \left( \Sigma^{\rceil}_{22} * G^{\lceil}_{21}
 \right)_{(t_1,t_2)} = -i \int^{\beta}_0 d \tau
 \Sigma^{\rceil}_{22}(t_1,\tau) \bigg{\{}  \left[ -
 \frac{d}{d\tau} - \epsilon_1 + \mu \right] G^{\lceil}_{11}(\tau,t_2) \\
 -  \int_{\gamma} d \bar{z} \Sigma_{11}(\tau,\bar{z}) G^{\
 +}_{11}(\bar{z},t_2) \bigg{\}}. \tag{D.9} \label{D.9}
\end{multline*}
Integrating by parts, the first terms of these expressions are written as 
\begin{gather}
  \int^{\infty}_{t_0} dt \Sigma^{--}_{22}(t_1,t) i
 \frac{d}{dt} G^<_{11}(t,t_2) = i \left[ \Sigma^{--}_{22}(t_1,t)
 G^<_{11}(t,t_2) \right]^{\infty}_{t_0} +  \int^{\infty}_{t_0} dt
 \left( -i \frac{d}{dt} \right) \Sigma^{--}_{22}(t_1,t) G^<_{11}(t,t_2), \notag \\
  \int^{\infty}_{t_0} dt \Sigma^{<}_{22}(t_1,t) i
 \frac{d}{dt} G^{++}_{11}(t,t_2) = i \left[ \Sigma^{<}_{22}(t_1,t)
 G^{++}_{11}(t,t_2) \right]^{\infty}_{t_0} +  \int^{\infty}_{t_0} dt
 \left( -i \frac{d}{dt} \right) \Sigma^{<}_{22}(t_1,t) G^{++}_{11}(t,t_2), \notag \\
\int^{\beta}_0 d \tau
 \Sigma^{\rceil}_{22}(t_1,\tau) \left( -
 \frac{d}{d\tau} \right) G^{\lceil}_{11}(\tau,t_2) = \int^{\beta}_0 d \tau
 \frac{d}{d\tau} \Sigma^{\rceil}_{22}(t_1,\tau)
 G^{\lceil}_{11}(\tau,t_2).  \tag{D.10} \label{D.10}
\end{gather}
The boundary term $\left[ \Sigma^{\rceil}_{22}(t_1,\tau)
G^{\lceil}_{11}(\tau,t_2) \right]^{\beta}_0$ in Eq. (\ref{D.10}) vanishes
due to the KMS condition (\ref{eq:29}). Using these expressions and Eqs. (\ref{D.7})-(\ref{D.9}), Eq. (\ref{D.6}) is written as 
\begin{align*}
& t_{12} \int_{\gamma} d \bar{z}
 \Sigma^{-\ }_{22}(t_1, \bar{z}) G^{\ +}_{21}(\bar{z},t_2) \\
 & \quad =
 \int_{\gamma} d \bar{z} \bigg[ \Sigma^{-\ }_{22}(t_1, \bar{z}) \left( -i
 \frac{\overleftarrow{d}}{d\bar{z}} - h_1(\bar{z}) \right) 
  - \int_{\gamma} d\bar{\bar{z}} \Sigma^{-\ }_{22}(t_1, \bar{\bar{z}})
 \Sigma_{11} (\bar{\bar{z}}, \bar{z}) \bigg] G^{\ +}_{11}(\bar{z},t_2) \\
 & \qquad + i \left[ \Sigma^{--}_{22}(t_1,t)
 G^<_{11}(t,t_2) - \Sigma^{<}_{22}(t_1,t)
 G^{++}_{11}(t,t_2) \right]^{\infty}_{t_0} + \Sigma^<_{22}(t_1,t_2).
 \tag{D.11} \label{D.11}
\end{align*}
By substituting Eq. (\ref{D.11}) into Eq. (\ref{D.5}) and using the definition of
the pseudo self-energy, we obtain the second-order partial differential
equation for the lesser Green function (\ref{eq:35}).

\subsection{Retarded Green function}
By definition, the retarded Green function is expressed as
\begin{equation*}
 G^r_{11}(t_1,t_2) = \theta(t_1-t_2) (G^>_{11}(t_1,t_2) - G^<(t_1,t_2)).
\end{equation*}
We  can obtain the following relations by differentiating the above definition:
\begin{align*}
 i \frac{d}{dt_1} G^r_{11}(t_1,t_2) &= i \delta (t_1-t_2)
 (G^>_{11}(t_1,t_2) - G^<_{11}(t_1,t_2)) +i \theta(t_1-t_2)
 \frac{d}{dt_1} (G^>_{11}(t_1,t_2) - G^<_{11}(t_1,t_2) ) \notag \\
 &= \delta(t_1-t_2) +i \theta(t_1-t_2)
 \frac{d}{dt_1} (G^>_{11}(t_1,t_2) - G^<_{11}(t_1,t_2) ), \tag{D.12} \label{D.12}
\end{align*}
\begin{align*}
 - \frac{d^2}{dt^2_1} G^r_{11}(t_1,t_2) &= i \frac{d}{dt_1} \left( \delta(t_1-t_2) +i \theta(t_1-t_2)
 \frac{d}{dt_1} (G^>_{11}(t_1,t_2) - G^<_{11}(t_1,t_2) ) \right)
 \notag \\
 &= i \frac{d}{d(t_1-t_2)} \delta(t_1-t_2) - \delta(t_1-t_2)
 \frac{d}{dt_1} (G^>_{11}(t_1,t_2) - G^<_{11}(t_1,t_2)) \notag \\
 &\qquad - \theta(t_1-t_2) \frac{d^2}{dt^2_1} (G^>_{11}(t_1,t_2) -
 G^<_{11}(t_1,t_2)) \notag \\
 &= i \frac{d}{d(t_1-t_2)} \delta(t_1-t_2) + \epsilon_1 \delta(t_1-t_2) -
 \theta(t_1-t_2) \frac{d^2}{dt^2_1} (G^>_{11}(t_1,t_2) -
 G^<_{11}(t_1,t_2)), \tag{D.13} \label{D.13}
\end{align*}
where we use the relations
\begin{gather*}
 \left. (G^>_{11}(t_1,t_2) -
 G^<_{11}(t_1,t_2)) \right|_{t_1=t_2}=-i, \\
 i\frac{d}{dt_1} \left. (G^>_{11}(t_1,t_2) -
 G^<_{11}(t_1,t_2)) \right|_{t_1=t_2}= -i \epsilon_1.
\end{gather*} 
The first relation is derived by substituting the definitions of the
 lesser Green function $G^{<}_{11}(t_1,t_2)= i \braket{
 d^{\dagger}_{H,1}(t_2) d_{H,1}(t_1)}$ and greater Green function
 $G^{>}_{11}(t_1,t_2)= -i \braket{d_{H,1}(t_1) d^{\dagger}_{H,1}(t_2)}$
 and using the commutation relation for fermions. The second relation follows from the equation of motion for the lesser and greater Green
 functions derived from Eq. (\ref{eq:19}) and using the commutation relation
 again. Using these relations, we obtain
\begin{multline*}
\left[ - \frac{d^2}{dt^2_1} -i (\epsilon_1+ \epsilon_2) \frac{d}{dt_1}
 + (\epsilon_1 \epsilon_2 - t_{12} t_{21}) \right] G^r_{11}(t_1,t_2) =  
 i \frac{d}{d(t_1-t_2)} \delta(t_1-t_2) 
 = - \epsilon_2 \delta(t_1-t_2) \\ +
 \theta(t_1-t_2) \left[ - \frac{d^2}{dt^2_1} -i (\epsilon_1+ \epsilon_2) \frac{d}{dt_1}
 + (\epsilon_1 \epsilon_2 - t_{12} t_{21}) \right] (G^>_{11}(t_1,t_2) -
 G^<_{11}(t_1,t_2)). \tag{D.14} \label{D.14}
\end{multline*}
Since we want to obtain a differential equation for $G^r_{11}$ in a closed
form, next we rewrite the last term of the RHS in Eq. (\ref{D.14}) in a form, that
is expressed using $G^{r}_{11}$. To do this, first we apply the Langreth
rule and the modified
Langreth rules (\ref{B.3}) and (\ref{B.4}), and use the second-order partial
differential equations for the lesser ({\ref{eq:35}}) and greater
Green functions for the last term. Then it is finally decomposed into the following five terms:
\begin{align}
&\theta(t_1-t_2) \left[ - \frac{d^2}{dt^2_1} -i (\epsilon_1+ \epsilon_2) \frac{d}{dt_1}
 + (\epsilon_1 \epsilon_2 - t_{12} t_{21}) \right] (G^>_{11}(t_1,t_2) -
 G^<_{11}(t_1,t_2)) \notag \\
&= \theta(t_1-t_2) \int_\gamma ( \tilde{\Sigma}^{+\
 }_{11}(t_1,z) G^{\ -}_{11}(z,t_2) - \tilde{\Sigma}^{-\
 }_{11}(t_1,z) G^{\ +}_{11}(z,t_2)) - \Sigma^r_{22}(t_1,t_2) \notag \\
&= \theta(t_1-t_2) \left[ (\tilde{\Sigma}^{>}_{11} \cdot G^{--}_{11} -
 \tilde{\Sigma}^{--}_{11} \cdot G^{<}_{11} ) - (\tilde{\Sigma}^{++}_{11}
 \cdot G^{>}_{11} - \tilde{\Sigma}^{<}_{11} \cdot
 G^{++}_{11}) \right]_{(t_1,t_2)} - \Sigma^r_{22}(t_1,t_2) \notag \\
&= \theta(t_1-t_2) \bigg[  
(\tilde{\Sigma}^{>}_{11} \cdot (G^a_{11} + G^>_{11}) 
 - \big{\}} \tilde{\Sigma}^{r}_{11} + \tilde{\Sigma}^{<}_{11} 
 - ( \Sigma^r_{11} \cdot \Sigma^r_{22})_{(t_1,t)} \notag \\
& \quad + \delta (t_1-t) (\mathbf{h}_{1L} \mathbf{h}_{L1} +  \mathbf{h}_{2R}
 \mathbf{h}_{R2} ) \big{\}} \cdot G^<_{11} - \big{\{} \tilde{\Sigma}^{>}_{11} - \tilde{\Sigma}^{r}_{11} 
 + ( \Sigma^r_{11} \cdot \Sigma^r_{22})_{(t_1,t)} \notag \\
& \qquad - \delta (t_1-t) (\mathbf{h}_{1L} \mathbf{h}_{L1} +  \mathbf{h}_{2R}
 \mathbf{h}_{R2} ) \big{\}} \cdot G^>_{11} + \tilde{\Sigma}^{<}_{11} \cdot (G^<_{11} + G^a_{11})
 \bigg]_{(t_1,t_2)} - \Sigma^r_{22}(t_1,t_2) \notag \\
&= \theta(t_1-t_2) \bigg[ (\tilde{\Sigma}^{>}_{11} - \tilde{\Sigma}^{<}_{11} )
 \cdot G^a_{11} \bigg]_{(t_1,t_2)} + \theta(t_1-t_2) \bigg[ \tilde{\Sigma}^{r}_{11} \cdot
 (G^>_{11}-G^<_{11}) \bigg]_{(t_1,t_2)}  \notag \\
&\quad - \theta(t_1-t_2) \bigg[ ( \Sigma^r_{11} \cdot
 \Sigma^r_{22})_{(t_1,t)} \cdot (G^>_{11} - G^<_{11}) \bigg]_{(t_1,t_2)}
 + (\mathbf{h}_{1L}
 \mathbf{h}_{L1} +  \mathbf{h}_{2R} \mathbf{h}_{R2} )G^r_{11}(t_1,t_2) -
 \Sigma^r_{22}(t_1,t_2) \notag \\
&= R_1 + R_2 + R_3 + R_4 - \Sigma^r_{22}(t_1,t_2). \tag{D.15} \label{D.15}
\end{align}
From the third line to the fourth line, we use the Langreth rule and the
modified Langreth rule. Using the concrete
expressions for the self-energy and the pseudo self-energy, we can express
$R_1$ as
\begin{align*}
 R_1 &= \theta (t_1-t_2) \int^{\infty}_{t_0} dt (\tilde{\Sigma}^{>}_{11} -
 \tilde{\Sigma}^{<}_{11}) G^a_{11}(t,t_2) \notag \\
 &= \theta (t_1-t_2) \int^{\infty}_{t_0} dt r_1(t_1,t) \frac{d}{d(t_1-t)}
 \delta(t_1-t) G^a_{11}(t_1,t_2),
\end{align*}
where we use $\theta(t_1-t_2)G^a_{11}(t_1,t_2)=0$ and define
$r_1(t,s):=\Gamma_{L} e^{-i \psi_L(t,s)} + \Gamma_{R}
e^{-i\psi_R(t,s) }$. By integrating by parts, we obtain
\begin{align}
 R_1 &= \theta(t_1-t_2) \left[r(t_1,t) G^a_{11}(t,t_2) \delta(t_1-t)
 \right]^{\infty}_{t_0} \notag \\
 & \quad - \theta(t_1-t_2) \int^{\infty}_{t_0} dt 
 \frac{d}{d(t_1-t)} \{r_1(t_1,t) G^a_{11}(t,t_2) \} \delta(t_1-t)
 \notag \\
 &= - \theta(t_1-t_2) \int^{\infty}_{t_0} - \frac{d}{dt}
 \{G^a_{11}(t,t_2) \} r_1(t_1,t) \delta(t_1-t) dt \notag \\
 &= \theta(t_1-t) r(t_1,t_1) \frac{d}{dt_1} G^a_{11}(t_1,t_2). \notag 
\end{align} 
By differentiating the definition of $G^a_{11}(t_1,t_2)$ and substituting it into the expression
above, we finally obtain
\begin{align}
 R_1 &= -i \theta(t_1-t_2) \delta(t_2-t_1) r_1(t_1,t_1) \notag \\
 &= - \frac{i}{2} (\Gamma_{L} + \Gamma_{R}) \delta(t_1-t_2),
 \tag{D.16} \label{D.16} 
\end{align}
where we use the fact that $\int dt \theta(t)\delta(t) f(t) = \int dt
\frac{1}{2}\delta(t) f(t)$ for an arbitrary test function $f$. For
$R_2$, using the expression of the retarded pseudo self-energy
(\ref{eq:40}), we can see that it consists of three terms:
\begin{align*}
 R_2 &= \theta(t_1-t_2) \bigg[ \tilde{\Sigma}^{r}_{11} \cdot
 (G^>_{11}-G^<_{11}) \bigg]_{(t_1,t_2)} \notag \\
 &= \theta(t_1-t_2) \int^{\infty}_{t_0} \bigg[ \frac{1}{2}
 (\Gamma_{L} +  \Gamma_{R} ) \frac{d}{d(t_1-t)} \delta(t_1-t) \notag \\
 &\qquad + \frac{i}{2}	  (\epsilon_2 \Gamma_{L} +\epsilon_1 \Gamma_{R})  \delta(t_1-t) - (\mathbf{h}_{1L}
 \mathbf{h}_{L1} +  \mathbf{h}_{2R} \mathbf{h}_{R2} ) \delta(t_1-t)
 \bigg] (G^>_{11}(t_1,t_2) - G^<_{11}(t_1,t_2)) \notag \\
 &= r_{2,1} + r_{2,2} + r_{2,3}. 
\end{align*}
Each term is calculated as follows:
\begin{align*}
 r_{2,1} &= \theta(t_1-t_2) \int^{\infty}_{t_0} dt
 \frac{1}{2}(\Gamma_{L} + \Gamma_{R}) \frac{d}{d(t_1-t)}
 \delta(t_1-t) (G^>_{11}(t,t_2) - G^<_{11}(t,t_2)) \notag \\
 &= \theta(t_1-t_2) \left[ \frac{1}{2} (\Gamma_{L} + \Gamma_{R})
 \delta(t_1-t) (G^>_{11}(t,t_2) - G^<_{11}(t,t_2))
 \right]^{\infty}_{t_0} \notag \\
 &\qquad - \theta(t_1-t_2) \frac{1}{2}
 (\Gamma_{L}+\Gamma_{R} ) \int^{\infty}_{t_0} \frac{d}{d(t_1-t)}
 \{ G^>_{11}(t,t_2) - G^<_{11}(t,t_2) \} \delta(t_1-t) \notag \\
 &= \frac{1}{2} (\Gamma_{L}+ \Gamma_{R}) \theta(t_1-t_2)
 \frac{d}{dt_1} (G^>_{11}(t_1,t_2) - G^<_{11}(t_1,t_2)), \tag{D.17} \label{D.17}
\end{align*}
\begin{align*}
 r_{2,2} &= \theta(t_1-t_2) \int^{\infty}_{t_0} dt \frac{i}{2}
 (\epsilon_2 \Gamma_{L} + \epsilon_1 \Gamma_{R}) \delta(t_1-t)
 (G^>_{11}(t,t_2) - G^<_{11}(t,t_2)) \notag \\
 &= \frac{i}{2}(\epsilon_2 \Gamma_{L} + \epsilon_1 \Gamma_{R})
 G^r_{11}(t_1,t_2), \tag{D.18} \label{D.18}
\end{align*}
\begin{align}
 r_{2,3} &= - \theta(t_1-t_2) \int^{\infty}_{t_0} dt (\mathbf{h}_{1L}
 \mathbf{h}_{L1} +  \mathbf{h}_{2R} \mathbf{h}_{R2}) \delta(t_1-t)
 (G^>_{11}(t,t_2) - G^<_{11}(t,t_2)) \notag \\
 &= - (\mathbf{h}_{1L}
 \mathbf{h}_{L1} +  \mathbf{h}_{2R} \mathbf{h}_{R2}) G^r_{11}(t_1,t_2). \tag{D.19} \label{D.19}
\end{align}
By taking the sum of the terms (\ref{D.17})-(\ref{D.19}), we
obtain the expression of $R_2$ as
\begin{align*}
 R_2 &= \frac{1}{2} (\Gamma_{L}+ \Gamma_{R}) \theta(t_1-t_2)
 \frac{d}{dt_1} (G^>_{11}(t_1,t_2) - G^<_{11}(t_1,t_2)) \notag \\
 &\qquad + \frac{i}{2}(\epsilon_2 \Gamma_{L} + \epsilon_1 \Gamma_{R})
 G^r_{11}(t_1,t_2) - (\mathbf{h}_{1L}
 \mathbf{h}_{L1} +  \mathbf{h}_{2R} \mathbf{h}_{R2}) G^r_{11}(t_1,t_2)
 \notag \\
 &= \frac{1}{2} (\Gamma_{L}+ \Gamma_{R}) \frac{d}{dt_1} G^r_{11}(t_1,t_2) +
 \frac{i}{2} (\Gamma_{L}+ \Gamma_{R}) \delta(t_1-t_2) \notag \\
 &\qquad + \frac{i}{2}(\epsilon_2 \Gamma_{L} + \epsilon_1 \Gamma_{R})
 G^r_{11}(t_1,t_2) - (\mathbf{h}_{1L}
 \mathbf{h}_{L1} +  \mathbf{h}_{2R} \mathbf{h}_{R2}) G^r_{11}(t_1,t_2). \tag{D.20} \label{D.20}
\end{align*}
In the WBLA, the term in the expression for $R_3$ is represented as $( \Sigma^r_{11} \cdot
\Sigma^r_{22})_{(t,s)}=-1/4 \Gamma_{L} \Gamma_{R}
\delta(t-s)$. Then we obtain the following expression for $R_3$:
\begin{align*}
 R_3 &= - \theta(t_1-t_2) \int^{\infty}_{t_0} dt ( \Sigma^r_{11} \cdot
\Sigma^r_{22})_{(t_1,t)} (G^>_{11}(t,t_2)-G^<_{11}(t,t_2)) \notag \\
 &= \frac{1}{4} \Gamma_{L} \Gamma_{R} G^r_{11}(t_1,t_2). \tag{D.21}
 \label{D.21} 
\end{align*}
By substituting Eqs. (\ref{D.16}), (\ref{D.20}), and (\ref{D.21})
into Eq. (\ref{D.15}), we obtain 
\begin{align}
&\theta(t_1-t_2) \left[ - \frac{d^2}{dt^2_1} -i (\epsilon_1+ \epsilon_2) \frac{d}{dt_1}
 + (\epsilon_1 \epsilon_2 - t_{12} t_{21}) \right] (G^>_{11}(t_1,t_2) -
 G^<_{11}(t_1,t_2)) \notag \\
 &= \frac{i}{2}(\epsilon_2 \Gamma_{L} + \epsilon_1 \Gamma_{R} -
 \frac{i}{2} \Gamma_{L} \Gamma_{R}) G^r_{11}(t_1,t_2) \notag \\
 & \qquad + \frac{1}{2} (\Gamma_{L}+
 \Gamma_{R}) \frac{d}{dt_1} G^r_{11}(t_1,t_2) + \frac{i}{2}
 \Gamma_{R} \delta(t_1-t_2) \notag \\ 
 &= (\epsilon_1 \epsilon_2 - \epsilon^{eff}_{1} \epsilon^{eff}_{2})
 G^r_{11}(t_1,t_2) + \frac{1}{2} (\Gamma_{L}+
 \Gamma_{R}) \frac{d}{dt_1} G^r_{11}(t_1,t_2) + \frac{i}{2}
 \Gamma_{R} \delta(t_1-t_2), \tag{D.22} \label{D.22}
\end{align}
where we define $\epsilon^{eff}_1:= \epsilon_1 - i/2 \Gamma_{L}$ and
$\epsilon^{eff}_2:= \epsilon_2 - i/2 \Gamma_{R}$. Using Eq. (\ref{D.22}),
we can rewrite Eq. (\ref{D.14}) as a closed-form differential equation for
$G^r_{11}(t_1,t_2)$. This is the second-order partial
differential equation for $G^r_{11}(t_1,t_2)$, or Eq. (\ref{eq:51}).

\section{Problem of calculating the retarded pseudo self-energy}
The definition of the pseudo self-energy is Eq. (\ref{eq:37})
\begin{multline}
 \tilde{\Sigma}_{11}(z_1,z_2) := \left( i \frac{d}{dz_1} - h_2(z_1)
 \right) \Sigma_{11}(z_1,z_2) + \Sigma_{22}(z_1,z_2)
 \left( -i\frac{\overleftarrow{d}}{dz_2} -h_1(z_2) \right) \\
 - \int_{\gamma} d \bar{z} \Sigma_{22}(z_1,\bar{z})
 \Sigma_{11}(\bar{z},z_2), \notag
\end{multline}
where the self-energies are defined as $\Sigma_{11}(z_1,\bar{z}) = \mathbf{h}_{1L} \cdot
 \mathbf{g}_{LL}(z_1,\bar{z}) \mathbf{h}_{L1}, \  \Sigma_{22}(z_1,\bar{z})
 = \mathbf{h}_{2R} \cdot \mathbf{g}_{RR}(z_1,\bar{z})
 \mathbf{h}_{R2}$. From this definition, we can calculate the retarded part
 of the pseudo self-energy straightforwardly in the following
 way. However, as we will see soon, there is a problem of
 divergence because the term $\theta(t) \frac{d}{dt}\delta
 (t)$ appears in this calculation. The difficulty can be avoided by changing the method of calculation as in Appendix D. 

 Based on the definition of the self-energy, the
 retarded part of the pseudo self-energy is calculated as
\begin{align}
 \tilde{\Sigma}^r_{11}(t_1,t_2) &:= \theta(t_1-t_2)
  (\tilde{\Sigma}^{>}_{11}(t_1,t_2)-\tilde{\Sigma}^{<}_{11}(t_1,t_2)
 \notag \\
 &= \theta(t_1-t_2) \left( i \frac{d}{dt_1} - h_2(t_1)
 \right) \mathbf{h}_{1L} \cdot
 (\mathbf{g}^>_{LL}(z_1,\bar{z})-\mathbf{g}^<_{LL}(z_1,\bar{z}))
 \mathbf{h}_{L1} \notag \\
 &\quad + \theta(t_1-t_2) \mathbf{h}_{2R} \cdot
 (\mathbf{g}^>_{RR}(z_1,\bar{z}) - \mathbf{g}^<_{RR}(z_1,\bar{z}) )
 \mathbf{h}_{R2} \left( -i\frac{\overleftarrow{d}}{dt_2} -h_1(t_2)
 \right)  \notag \\
 &\quad \quad - \theta(t_1-t_2) \int_{\gamma} d \bar{z} \left( \Sigma^{+\ }_{22}(t_1,\bar{z})
 \Sigma^{\ -}_{11}(\bar{z},t_2) - \Sigma^{-\ }_{22}(t_1,\bar{z})
 \Sigma^{\ +}_{11}(\bar{z},t_2) \right) \notag \\
 &= \theta(t_1-t_2) \left( i \frac{d}{dt_1} - h_2(z_1)
 \right) \sum_{k,k'} T_{1,kL} (-i \delta_{kk'})
 e^{-i\phi_{kL}(t_1,t_2)} T_{kL,1} \notag \\
 &\quad +  \theta(t_1-t_2) \sum_{k,k'} T_{kR,2} (-i \delta_{kk'})
 e^{-i\phi_{kR}(t_1,t_2)} T_{kR,2} \left(
 -i\frac{\overleftarrow{d}}{dt_2} -h_1(z_2) \right), \notag
\end{align}
where we use the concrete expressions of the non-perturbative Green function
(\ref{A.1}) and (\ref{A.3}) and the relation (\ref{B.3}). Next we rewrite the expression as follows so
that it is convenient for applying the WBLA:
\begin{align}
  \tilde{\Sigma}^r_{11}(t_1,t_2) &= -i \theta(t_1-t_2) \left( i \frac{d}{dt_1} - h_2(z_1)
 \right) \sum_k |T_{1,kL}|^2 e^{-i\phi_{kL}(t_1,t_2)} \notag \\
 & \quad -i \theta(t_1-t_2) \sum_k |T_{2,kR}|^2 e^{-i\phi_{kR}(t_1,t_2)} \left(
 -i\frac{\overleftarrow{d}}{dt_2} -h_1(z_2) \right) \notag \\
 &= -i \theta(t_1-t_2) \sum_k |T_{1,kL}|^2 (V_L(t_1)-\epsilon_{kL} -
 \epsilon_2) e^{-i\phi_{kL}(t_1,t_2)} \notag \\
 &\quad -i \theta(t_1-t_2) \sum_k |T_{2,kR}|^2 (V_R(t_2)-\epsilon_{kR} -
 \epsilon_1) e^{-i\phi_{kR}(t_1,t_2)} \notag \\
 &=-i \theta(t_1-t_2) e^{-i\psi_{L}(t_1,t_2)} \sum_k |T_{1,kL}|^2 (V_L(t_1)-\epsilon_{kL} -
 \epsilon_2) e^{-i\epsilon_{kL}(t_1-t_2)} \notag \\
 &\quad -i \theta(t_1-t_2) e^{-i \psi_{R}(t_1,t_2)} \sum_k |T_{2,kR}|^2 (V_R(t_2)-\epsilon_{kR} -
 \epsilon_1) e^{-i \epsilon_{kR}(t_1-t_2)} \notag \\
 &=-i \theta(t_1-t_2) e^{-i\psi_{L}(t_1,t_2)} \int \frac{d \omega}{ 2
 \pi}  \sum_k 2 \pi |T_{1,kL}|^2 \delta(\omega - \epsilon_{kL}) (V_L(t_1)- \omega -
 \epsilon_2) e^{-i \omega (t_1-t_2)}  \notag \\
 & \quad -i \theta(t_1-t_2) e^{-i \psi_{R}(t_1,t_2)} \int \frac{d\omega}{2\pi}
 \sum_k  2\pi |T_{2,kR}|^2 \delta(\omega-\epsilon_{kR}) (V_R(t_2)-\omega -
 \epsilon_1)  e^{-i \omega(t_1-t_2)}.  \notag 
\end{align}
At this point, we use the WBLA, which is equivalent to the substitution of
$\Gamma_{L}$ and $\Gamma_{R}$ by $\sum_k 2
\pi |T_{1,kL}|^2 \delta(\omega - \epsilon_{kL})$ and $\sum_k
2\pi |T_{2,kR}|^2 \delta(\omega-\epsilon_{kR})$, respectively. Then, the
retarded self-energy is written as
\begin{align}
 \tilde{\Sigma}^r_{11}(t_1,t_2) & =-i \theta(t_1-t_2) e^{-i\psi_{L}(t_1,t_2)} \int \frac{d \omega}{ 2
 \pi}  \Gamma_{L} (V_L(t_1)- \omega -
 \epsilon_2) e^{-i \omega (t_1-t_2)}  \notag \\
 & \qquad -i \theta(t_1-t_2) e^{-i \psi_{R}(t_1,t_2)} \int \frac{d\omega}{2\pi}
   \Gamma_{R} (V_R(t_2)-\omega -
 \epsilon_1)  e^{-i \omega(t_1-t_2)} \notag \\
 &=-i \theta(t_1-t_2) \Gamma_{L} e^{-i\psi_{L}(t_1,t_2)}
 \left((V_L(t_1)- \epsilon_2) \delta(t_1-t_2) + i \frac{d}{d(t_1-t_2)}
 \delta(t_1-t_2) \right) \notag \\
 &\qquad -i \theta(t_1-t_2) \Gamma_{R} e^{-i \psi_{R}(t_1,t_2)}  \left( (V_R(t_2)-
 \epsilon_1) \delta(t_1-t_2) +i \frac{d}{d(t_1-t_2)}
 \delta(t_1-t_2) \right) \notag \\
 &= - \frac{i}{2} \big( \Gamma_{L}(V_L(t_1) - \epsilon_2) + \Gamma_{R}
 (V_R(t_2) - \epsilon_1) \big) \delta (t_1-t_2) \notag \\
 & \qquad + \theta(t_1-t_2) (\Gamma_{L}e^{-i\psi_{L}(t_1,t_2)} +
 \Gamma_{R} e^{-i \psi_{R}(t_1,t_2)} ) \frac{d}{d(t_1-t_2)}
 \delta(t_1-t_2), \tag{E.1} \label{E.1}
\end{align}
where we use the relations $\int \frac{d\omega}{2\pi}
 e^{-i\omega t}=\delta(t)$ and $\int dt \theta(t)\delta(t) f(t) = \int dt
\frac{1}{2}\delta(t) f(t)$ for an arbitrary test function $f$. In this
 expression, the term $\theta(t) \frac{d}{dt}\delta
 (t)$ appears. In expression
 (\ref{eq:40}), which we use in the calculation of the retarded
 self-energy, the term does not appear. The problem is that a diverging term
 appears if one uses this expression including $\theta(t) \frac{d}{dt}\delta
 (t)$ for the derivation of the second-order
 differential equation for the retarded Green function. We can see this fact
 as follows. In Appendix D, we derived
 the second-order partial differential equation for the retarded Green
 function. In the derivation, we needed an expression of the retarded
 pseudo self-energy for the calculation of the term $r_{2,1}$ in the expression for $R_2$,
 where we integrate by parts a function that is multiplied by
 $\frac{d}{dt}\delta(t)$. Therefore, if we use expression (\ref{E.1}) including the term $\theta(t) \frac{d}{dt}
 \delta (t)$ for the calculation of $r_{2,1}$, the square of the delta function
 $\delta^2(t)$ appears in the integration by parts as $ \theta(t)
 \frac{d}{dt} \delta(t) = \frac{d}{dt}(\theta(t) \delta(t)) -
 \delta^2(t)$. This is why the
 diverging term emerges when we use expression (\ref{E.1}), and we
 have to use expression (\ref{eq:40}) to calculate the
 retarded pseudo self-energy. 

The essential difference between the two methods of calculation in  Appendices C and E appears when one employs the WBLA. To understand the
 difference, we first simplify the problem. Let us take a continuous and
 differentiable function
 $f_{N}(t)$, which becomes $\delta(t)$ as $N \rightarrow \infty$. We suppose
 that the $N \rightarrow \infty$ limit corresponds to the WBLA. Let us
 consider the limit of $\theta(t) \frac{d}{dt} f_{N}(t)$ as $N \rightarrow
 \infty$. Naturally, we calculate the limit as 
\begin{align*}
 \lim_{N \rightarrow \infty} \theta(t) \frac{d}{dt} f_{N}(t) &= \theta(t)
 \lim_{N \rightarrow \infty} \frac{d}{dt} f_{N}(t) \\
 &= \theta(t) \frac{d}{dt} \delta(t).
\end{align*} 
This corresponds to the method of calculation carried out in Appendix E. By
contrast, we calculate the same quantity in Appendix C as follows:
\begin{align*}
 \lim_{N \rightarrow \infty} \theta(t) \frac{d}{dt} f_{N}(t) &= \lim_{N
 \rightarrow \infty} \left[ \frac{d}{dt} (\theta (t) f_{N}(t)) - \delta(t)
 \frac{d}{dt} f_{N}(t) \right] \\
 &= \lim_{N
 \rightarrow \infty} \left[ \frac{d}{dt} (\theta (t) f_{N}(t)) \right] - \delta(t)
 \left. \frac{d}{dt} f_{N}(t) \right|_{t=0} \\ 
 &= \frac{d}{dt} (\theta (t) \delta(t)) - \left. \frac{d}{dt} f_{N}(t) \right|_{t=0} \delta(t), \\
 &= \frac{d}{dt} (\frac{1}{2} \delta(t)) - \left. \frac{d}{dt} f_{N}(t) \right|_{t=0} \delta(t), \\
 &= \frac{1}{2} \frac{d}{dt} \delta(t) - \left. \frac{d}{dt} f_{N}(t) \right|_{t=0} \delta(t). \tag{E.2} \label{E.2}
\end{align*}
(\ref{E.2}) corresponds to the expression (\ref{C.9}) in Appendix C. Therefore, $f_N(t)$ corresponds to the term
$\Sigma^>_{11}(t_1,t_2) - \Sigma^<_{11}(t_1,t_2)$ in Appendix D. The important points in this calculation are that we replace
$\theta(t) \frac{d}{dt} f_{N}(t)$ by $\frac{d}{dt} (\theta (t) f_{N}(t)) - \delta(t)
 \frac{d}{dt} f_{N}(t)$ and use the relation $\theta(t) \delta (t) =
 \frac{1}{2}\delta(t)$. In this way, we avoid the problem of
 divergence.

 We have not yet found any reasons to justify this method of calculation
 physically or mathematically, but the expressions for the Matsubara,
 retarded, and advanced Green functions obtained from Eq. (\ref{eq:40})
 reproduce the same results as in the previous study of a single quantum
 dot when we take the limit in which the coupling constant between the
 dots is zero.

\section{Representation of physical quantities as integrals
with respect to frequency}

In this appendix, we explain how we rewrite the electron density of dot 1 in terms of an integration with respect to frequency. This expression
is useful for comparing the expression of the electron
density of the double dots with that of the single dot (\ref{eq:17}). In
addition, it is convenient for numerical computation. We only
explain the rewriting of $G^{\rceil}_{11}(t,0^+)$, $(G^{\rceil}_{11} *
\Sigma^{M}_{11})_{(t,0^+)}$, and $G^{\rceil}_{12}(t,0^+)$, which appear in
the two coefficients $l_1(t)$ and $l_2(t)$ of the lesser Green function
(\ref{eq:65}).

Throughout the calculations below, we use the following relation for
the Matsubara frequency:
\begin{equation*}
 \frac{i}{\beta} \sum^{q=\infty}_{q=-\infty} Q(\omega_q) e^{\omega_q 0^+} =
  \int^{\infty}_{-\infty} \frac{d\omega}{2 \pi} f(\omega) [Q(\omega^{-})
  - Q(\omega^+)], \tag{F.1} \label{F.1}
\end{equation*}
where $Q(\omega)$ is a function that satisfies $\lim_{\omega
\rightarrow \infty} Q(\omega)e^{-\omega}=0$ and $f(\omega)$ is the Fermi
distribution. This relation can be proved by the fact that the Matsubara
frequency is the residue of the function $1/(e^{\beta z}+1)$. 
We define the following functions that include the effects between the reservoirs
and the system:
\begin{align*}
 K^i_{\alpha}(t,t':\omega) &:= e^{-ik_i(t-t_0)} \left[\int^t_{t_0} ds
 e^{-i (\omega-k_2)(s-t')} e^{-i \psi_{\alpha}(s,t')} \right], \\
 \bar{K}^i_{\alpha}(t,t':\omega) &:= e^{-ik_i(t-t_0)} \left[\int^t_{t_0} ds
 e^{-i (\omega-k_2)(s-t')} V_{\alpha}(s) e^{-i \psi_{\alpha}(s,t')} \right].
\end{align*}
In the following, we express $K^i_{\alpha}(t,t_0:\omega)$ as
$K^i_{\alpha}(\omega)$ and represent $G^M_{11}(\omega^-)$ as
the Matsubara Green function (\ref{eq:42}) with the condition
that the imaginary part of its variable is
negative. $G^M_{11}(\omega^+)$ is also defined similarly.

Using the expression of the right Green function (\ref{eq:57}),
$G^{\rceil}_{11}(t,0^+)$ is written as 
\begin{multline*}
 G^{\rceil}_{11} (t,0^+) = b_1(0^+) e^{-ik_1(t-t_0)} + b_2(0^+)
  e^{-ik_2(t-t_0)}  \\ - \frac{1}{k_2-k_1} \left[\int^t_{t_0} ds g(s,0^+) (e^{-ik_1(t-s)}-e^{-ik_2(t-s)})
 \right], \label{F.2} \tag{F.2}
\end{multline*}
where coefficients $b_1(0^+)$, $b_2(0^+)$ and the function
$g(s,0^+)$ are defined as 
\begin{align*}
  b_1(0^+) &= -\frac{1}{k_2-k_1}\{(\epsilon^{eff}_1-k_2) G^M_{11}(0^+,0^+) +
 (\Sigma^M_{11}*G^M_{11})_{(0^+,0^+)} + t_{12}G^M_{12}(0^+,0^+) \}, \\ 
 b_2(0^+) &= \frac{1}{k_2-k_1}\{(\epsilon^{eff}_1-k_1) G^M_{11}(0^+,0^+) +
 (\Sigma^M_{11}*G^M_{11})_{(0^+,0^+)} + t_{12}G^M_{12}(0^+,0^+) \}, \\
 g(s,0^+) &= -i \{ (\tilde{\Sigma}^{\rceil}_{11}*G^M_{11})_{(s,0^+)}
 -\Sigma^{\rceil}_{22}(s,0^+) \}.
\end{align*}
Now we explain how we obtain a representation expressed as an integration over frequency using the relation (\ref{F.1}). The constant
$G^M_{11}(0^+,0^+)$ is rewritten as 
\begin{align*}
G^M_{11}(0^+,0^+) &= \lim_{\tau_1,\tau_2 \rightarrow 0}
 G^M_{11}(\tau_1,\tau_2) \\
 &=  \lim_{\tau_2 \rightarrow 0} \frac{i}{\beta} \sum_q e^{\omega_q
 \tau} G^M_{11}(\omega_q) \\
 &= \int \frac{d\omega}{ 2 \pi} f(\omega-\mu) \left[
 G^M_{11}(\omega^{-}-\mu) -  G^M_{11}(\omega^{+}-\mu) \right]. (\because
 \mathrm{(\ref{F.1})} \tag{F.3} \label{F.3}
\end{align*}
Similarly, $(\Sigma^M_{11}*G^M_{11})_{(0^+,0^+)}$ is rearranged as
\begin{align*}
 (\Sigma^M_{11}*G^M_{11})_{(0^+,0^+)} &= -i \int^{\beta}_0 d \tau
 \Sigma^M_{11}(0^+,\tau) G^M_{11}(\tau,0^+) \\
 &= \lim_{\tau' \rightarrow 0} -i \int^{\beta}_0 d \tau
 \Sigma^M_{11}(\tau',\tau) G^M_{11}(\tau,\tau') \\
 &= \lim_{\tau' \rightarrow 0} -i \int^{\beta}_0 d \tau
 \left(-\frac{\Gamma_{L}}{2\beta} \sum_q \xi_q
 e^{-\omega_q(\tau'-\tau)} \right) \left(\frac{i}{\beta} \sum_{q'}
 e^{-\omega_{q'} (\tau-\tau')} G^M_{11}(\omega_{q'}) \right) \\
 &= \frac{i}{2} \Gamma_{L} \frac{i}{\beta} \sum_q \left( e^{\omega_q 0^+}
 \xi_q G^M_{11}(\omega_q) \right ). \ (\because \int^\beta_0 d\tau
 e^{-(\omega_q-\omega_{q'}) \tau} = \beta \delta_{qq'}  
\end{align*}
Here we use the relation $\xi_q= 2 U \left( \mathrm{Im}(\omega_q)
\right) -1 $, where $U(t)$ is the unit step function: $U(t)=1$ for $t \ge 1$,
$U(t)=0$ for $t < 0$. Using this relation and Eq. (\ref{F.1}), we obtain the
expression
\begin{align*}
 (\Sigma^M_{11}*G^M_{11})_{(0^+,0^+)} &= \frac{i}{2} \Gamma_{L} \int
 \frac{d \omega}{2 \pi} f(\omega-\mu) \bigg[ \{ 2 U \left(
 \mathrm{Im}(\omega^{-} - \mu) \right) -1 \} G^M_{11}(\omega^{-} - \mu)) \\
 &\qquad - \{ 2 U \left(
 \mathrm{Im}(\omega^{+} - \mu) \right) -1 \} G^M_{11}(\omega^{+} - \mu))
 \bigg] \ (\because (\mathrm{\ref{E.1}}) \\
 &= - \frac{i}{2} \Gamma_{L} \int
 \frac{d \omega}{2 \pi} f(\omega-\mu) \left[
 G^M_{11}(\omega^{-}-\mu) +  G^M_{11}(\omega^{+}-\mu) \right]. \tag{F.4} \label{F.4}
\end{align*}
These calculations are the basis of our rearrangement of the lesser Green function. The expression of
$G_{21}(0^+,0^+)$ is obtained in the same way as that for
$G_{11}(0^+,0^+)$ and the result is
\begin{equation*}
 G^M_{21}(0^+,0^+) = \int \frac{d\omega}{ 2 \pi} f(\omega-\mu) \left[
 G^M_{21}(\omega^{-}-\mu) -  G^M_{21}(\omega^{+}-\mu) \right]. \tag{F.5} \label{F.5}
\end{equation*}
 By substituting Eqs. (\ref{F.3})-(\ref{F.5}) in $b_1(0^+)$ and $b_2(0^+)$, we obtain
\begin{align*}
 b_1(0^+) &= - \frac{1}{k_2-k_1} \{ (\epsilon^{eff}_1 - k_2) \int \frac{d \omega }{2
 \pi } f(\omega -\mu) [G^M_{11}(\omega^- - \mu) - G^M_{11} (\omega^+ -
 \mu) ] \\
  &\quad - \frac{i}{2} \Gamma_{L} \int \frac{d \omega }{2
 \pi } f(\omega -\mu) [G^M_{11}(\omega^- - \mu) + G^M_{11} (\omega^+ -
 \mu) ] \\
 & \qquad+ t_{12} \int \frac{d \omega }{2
 \pi } f(\omega -\mu) [ G^M_{21}(\omega^- - \mu) - G^M_{21} (\omega^+ -
 \mu) ] \}, \label{F.6} \tag{F.6}
\end{align*}
\begin{align*}
 b_2(0^+) &=  \frac{1}{k_2-k_1} \{ (\epsilon^{eff}_1 - k_1) \int \frac{d \omega }{2
 \pi } f(\omega -\mu) [G^M_{11}(\omega^- - \mu) - G^M_{11} (\omega^+ -
 \mu) ] \\
  &\quad - \frac{i}{2} \Gamma_{L}  \int \frac{d \omega }{2
 \pi } f(\omega -\mu) [G^M_{11}(\omega^- - \mu) + G^M_{11} (\omega^+ -
 \mu) ] \\
 &\qquad + t_{12} \int \frac{d \omega }{2
 \pi } f(\omega -\mu) [ G^M_{21}(\omega^- - \mu) - G^M_{21} (\omega^+ -
 \mu) ] \}. \label{F.7} \tag{F.7}
\end{align*}
Next we explain the rewriting of $\int^t_{t_0} ds
g(s,0^+) (e^{-ik_1(t-s)}-e^{-ik_2(t-s)})$ in Eq. (\ref{F.2}). We represent
the term as $b_3(t)$. In the same way as in the rearrangement of $b_1(0^+)$ and
$b_2(0^+)$, $b_3(t)$ is expressed as
\begin{align*}
 b_3(t) &= \frac{i}{2} \Gamma_{L} \int \frac{d \omega}{2 \pi} f(\omega
 -\mu)  [G^M_{11}(\omega^- - \mu) - G^M_{11} (\omega^+ -
 \mu) ] (e^{- i k_1 (t-t_0)} - e^{-i k_2(t-t_0)}) \\
 &\quad - \int \frac{d \omega}{2 \pi}  f(\omega
 -\mu) \omega G^M_{11}(\omega^- - \mu) [\Gamma_{L}( K^1_L(\omega) -
 K^2_L(\omega)) +\Gamma_{R}( K^1_R(\omega) -
 K^2_R(\omega))] \\
 &\quad - \int \frac{d \omega}{2 \pi} f(\omega
 -\mu) G^M_{11}(\omega^- - \mu) [ \Gamma_{L}( \bar{K}^1_L(\omega) -
 \bar{K}^2_L(\omega)) \\ & \qquad - \epsilon^{eff}_2 \Gamma_{L}( K^1_L(\omega) - K^2_L(\omega)) -\epsilon^{eff}_1 \Gamma_{R}( K^1_R(\omega) -
 K^2_R(\omega)) ] \\
 &\quad + \Gamma_{R} \int \frac{d \omega}{2 \pi}
  f(\omega  -\mu) (K^1_R(\omega) - K^2_R(\omega)). \tag{F.8} \label{F.8}
\end{align*}
Using Eqs. (\ref{F.6})-(\ref{F.8}), the function
$G^{\rceil}_{11}(t,0^+)$ is expressed as
\begin{equation*}
 G^{\rceil}_{11}(t,0^+) = b_1 e^{-i k_1(t-t_0)} + b_2 e^{-ik_2 (t-t_0)}
  - \frac{1}{k_2-k_1} b_3(t).
\end{equation*}
The other functions $(G^{\rceil}_{11} *
\Sigma^{M}_{11})_{(t,0^+)}$ and $G^{\rceil}_{12}(t,0^+)$ are
rewritten in the same way as
\begin{equation*}
 (G^r_{11} \cdot \Sigma^{\rceil}_{11})_{(t,0^+)} = - \Gamma_{L} \int \frac{d \omega}{2 \pi}
  f(\omega  -\mu) (C_1 K^1_L(\omega) + C_2 K^2_L(\omega)),
\end{equation*}
\begin{align*}
 c_1 &= - \frac{1}{k_2-k_1} \{ (\epsilon{eff}_1 - k_2) \int \frac{d \omega }{2
 \pi } f(\omega -\mu) [G^M_{12}(\omega^- - \mu) - G^M_{12} (\omega^+ -
 \mu) ] \\
  &- \frac{i}{2} \Gamma_{L}  \int \frac{d \omega }{2
 \pi } f(\omega -\mu) [G^M_{12}(\omega^- - \mu) + G^M_{12} (\omega^+ -
 \mu) ] \\
 &+ t_{12} \int \frac{d \omega }{2
 \pi } f(\omega -\mu) [ G^M_{22}(\omega^- - \mu) - G^M_{22} (\omega^+ -
 \mu) ] \},
\end{align*}
\begin{align*}
 c_2 &=  \frac{1}{k_2-k_1} \{ (\epsilon^[eff]_1 - k_1) \int \frac{d \omega }{2
 \pi } f(\omega -\mu) [G^M_{12}(\omega^- - \mu) - G^M_{12} (\omega^+ -
 \mu) ] \\
  & - \frac{i}{2} \Gamma_{L}  \int \frac{d \omega }{2
 \pi } f(\omega -\mu) [G^M_{12}(\omega^- - \mu) + G^M_{12} (\omega^+ -
 \mu) ] \\
 &+ t_{12} \int \frac{d \omega }{2
 \pi } f(\omega -\mu) [ G^M_{22}(\omega^- - \mu) - G^M_{22} (\omega^+ -
 \mu) ] \},
\end{align*}
\begin{align*}
 c_3(t) &= \frac{i}{2} \Gamma_{L} \int \frac{d \omega}{2 \pi} f(\omega
 -\mu)  [G^M_{12}(\omega^- - \mu) - G^M_{12} (\omega^+ -
 \mu) ] (e^{- i k_1 (t-t_0)} - e^{-i k_2(t-t_0)}) \\
 &+ \int \frac{d \omega}{2 \pi}  f(\omega
 -\mu) \omega G^M_{12}(\omega^- - \mu) [\Gamma_{L}( K^1_L(\omega) -
 K^2_L(\omega)) +\Gamma_{R}( K^1_R(\omega) -
 K^2_R(\omega))] \\
 & + \int \frac{d \omega}{2 \pi} f(\omega
 -\mu) G^M_{12}(\omega^- - \mu) [ \Gamma_{L}( \bar{K}^1_L(\omega) -
 \bar{K}^2_L(\omega)) \\ & \qquad - \epsilon^{eff}_2 \Gamma_{L}( K^1_L(\omega) -
 K^2_L(\omega)) -\epsilon^{eff}_1 \Gamma_{R}( K^1_R(\omega) -
 K^2_R(\omega)) ], \\
\end{align*}
\begin{equation*}
 G^{\rceil}_{12}(t,0^+) = c_1 e^{-i k_1(t-t_0)} + c_2 e^{-ik_2 (t-t_0)}
  - \frac{1}{k_2-k_1} c_3(t).
\end{equation*}

\end{document}